# Local nuclear energy density functional at next-to-next-to-next-to-leading order

B.G. Carlsson,[1] J. Dobaczewski,[1,2] and M. Kortelainen[1]

[1]Department of Physics, P.O. Box 35 (YFL), FI-40014 University of Jyväskylä, Finland
[2]Institute of Theoretical Physics, University of Warsaw, ul. Hoża 69, 00-681 Warsaw, Poland.

We construct nuclear energy density functionals in terms of derivatives of densities up to sixth, next-to-next-to-next-to-leading order ($N^3LO$). A phenomenological functional built in this way conforms to the ideas of the density matrix expansion and is rooted in the expansions characteristic to effective theories. It builds on the standard functionals related to the contact and Skyrme forces, which constitute the zero-order (LO) and second-order (NLO) expansions, respectively. At $N^3LO$, the full functional with density-independent coupling constants, and with the isospin degree of freedom taken into account, contains 376 terms, while the functionals restricted by the Galilean and gauge symmetries contain 100 and 42 terms, respectively. For functionals additionally restricted by the spherical, space-inversion, and time-reversal symmetries, the corresponding numbers of terms are equal to 100, 60, and 22, respectively.

## I. INTRODUCTION

The Skyrme force was introduced into nuclear physics more than half a century ago [1, 2] and it is still a concept that is widely used in methods striving to determine properties of nuclei irrespective of their mass number and isospin. However, at present we understand this concept in a significantly different way than it was originally proposed. Indeed, instead of using this force as an effective interaction within the Hartree-Fock (HF) approximation, we rather focus on the underlying Skyrme energy density functional (EDF), without direct references to the effective interaction or HF approximation.

In electronic systems, the use of functionals of density is motivated by formal results originating from the Hohenberg-Kohn [3] and Kohn-Sham theorems [4], whereby exact ground-state energies of many-fermion systems can be obtained by minimizing a certain exact functional of the one-body density. This led to numerous extensions and applications, now collectively known under the name of density functional theory (DFT) [5–7].

The fact that properties of electronic systems are governed by the well-known Coulomb interaction allows for derivations of functionals from first principles, by which token this approach can proudly be called a theory. For nuclear systems, the luxury of knowing the exact interaction is not there, so the analogous approaches developed in this domain of physics carry the name of the EDF methods.

In this article we construct a phenomenological nuclear EDF based on strategies that are proper to effective theories [8]. There, guiding principles [9] are based on: (i) appropriate choice of effective fields, (ii) building effective Lagrangian or Hamiltonian densities restricted only by symmetry principles, (iii) employing ideas of power counting. In the low-energy nuclear structure, correct fields can probably be associated with nonlocal one-body nuclear densities. Then, functionals of densities acquire the meaning of effective Hamiltonian densities. Although a formal construction of power-counting schemes is not yet available, ideas based on the density matrix expansion (DME) [10–16] (see also a recent example of an application to electronic systems in Ref. [17]) can be used to propose expansions in terms of moments of effective nuclear interactions, or equivalently, in orders of derivatives acting on the one-body densities. This is precisely the strategy we are going to follow in the present study.

Effective field theories (EFT) were recently extensively applied in analyzing properties of nuclear systems. Here we are not able to give an even shortest possible review of this rapidly developing area of physics, but let us mention two specific examples.

Firstly, the nucleon-nucleon (NN) scattering properties were very successfully described by employing the effective nucleon-pion Lagrangian at next-to-next-to-next-to-leading order ($N^3LO$), see Ref. [18] and references cited therein. This showed that the EFT expansion is capable of grasping the main features of nuclear interactions at low energies, without explicitly invoking microscopic foundations in terms of, e.g., heavy meson exchanges.

Secondly, methods using the harmonic-oscillator effective operators have been developed up to $N^3LO$, to be employed within the shell-model approaches [19]. There, the $N^3LO$ expansion was explicitly expressed in the form of pseudopotentials that contain derivatives up to sixth order. Evidently, such pseudopotentials are exact equivalents of higher-order Skyrme-like forces. When averaged within the HF approximation, they would lead to EDFs depending on derivatives of densities up to sixth order. This allows us to label our approach with the traditional name of the $N^3LO$ expansion too.

There is also a recent significant effort in deriving nuclear EDFs directly from low-energy QCD within chiral perturbation theory, see, e.g., Refs. [20–22]. This may have a potential of providing new important insight into the precise structure of terms in the EDF, while at present we are bound to proceed phenomenologically, with only the symmetry constraints available, as is done in the present study.

In all rigorous EFT expansions, one strives to achieve convergence in describing physical observables by going to higher and higher orders of expansion. This is best illustrated by the so-called Lepage plots [19, 23], where for theories cut at different orders, relative errors of observables are plotted as functions of energy. In nuclear EDF methods, this kind of convergence tests were never performed—simply because the functionals beyond the second order (NLO) of the standard Skyrme type had never been constructed or studied. The present study constitutes the first step towards this goal.

Our paper is organized as follows. In Sec. II we define basic building blocks for our construction, and then we construct local densities up to N$^3$LO. In Sec. III we construct terms in the EDF up to N$^3$LO and evaluate constraints imposed by the Galilean and gauge symmetries. In Sec. IV we derive results for the case of conserved spherical, space-inversion, and time-reversal symmetries. After formulating conclusions of the present study in Sec. V, in Appendix A we discuss general symmetry properties of the energy density, in Appendix B we present details of the adopted choice of the phase convention, and in Appendix C we list results pertaining to the Galilean and gauge symmetries.

## II. CONSTRUCTION OF LOCAL DENSITIES

### A. Building blocks

Let $\rho\left(\boldsymbol{r}\sigma,\boldsymbol{r}'\sigma'\right)$ denote the one-body density matrix in space-spin coordinates. In what follows, in order to simplify the notation, we omit the isospin degree of freedom, because in the particle-hole channel all densities appear in the isoscalar and isovector forms [24], and generalization to proton-neutron systems does not present any problem. Within this assumption, the EDF we consider has the form:

$$\mathcal{E} = \int \mathrm{d}^3\boldsymbol{r}\mathcal{H}_E(\boldsymbol{r}), \tag{1}$$

where the energy density $\mathcal{H}_E(\boldsymbol{r})$ can be represented as a sum of the kinetic and potential energies,

$$\mathcal{H}_E(\boldsymbol{r}) = \frac{\hbar^2}{2m}\tau_0 + \mathcal{H}(\boldsymbol{r}). \tag{2}$$

In the present study, we focus on the potential energy density $\mathcal{H}(\boldsymbol{r})$ only.

First, using the Pauli matrices $\sigma_a$, where index $a=\{x,y,z\}$ enumerates the Cartesian components of a vector, the density matrix is separated into the standard scalar and vector parts [25],

$$\rho\left(\boldsymbol{r}\sigma,\boldsymbol{r}'\sigma'\right)=\tfrac{1}{2}\rho\left(\boldsymbol{r},\boldsymbol{r}'\right)\delta_{\sigma\sigma'}+\tfrac{1}{2}\sum_a\left\langle\sigma\left|\sigma_a\right|\sigma'\right\rangle s_a\left(\boldsymbol{r},\boldsymbol{r}'\right), \tag{3}$$

where

$$\rho\left(\boldsymbol{r},\boldsymbol{r}'\right) = \sum_\sigma \rho\left(\boldsymbol{r}\sigma,\boldsymbol{r}'\sigma\right), \tag{4}$$

$$\boldsymbol{s}\left(\boldsymbol{r},\boldsymbol{r}'\right) = \sum_{\sigma\sigma'} \rho\left(\boldsymbol{r}\sigma,\boldsymbol{r}'\sigma'\right)\left\langle\sigma'\left|\boldsymbol{\sigma}\right|\sigma\right\rangle. \tag{5}$$

These two nonlocal densities will be used as building blocks of the functional together with the derivative operator $\boldsymbol{\nabla}$ and the relative momentum operator $\boldsymbol{k}$,

$$\boldsymbol{k} = \frac{1}{2i}\left(\boldsymbol{\nabla}-\boldsymbol{\nabla}'\right). \tag{6}$$

To most easily satisfy the constraints imposed by the rotational invariance, in our method, the building blocks are represented as spherical tensor operators [26], i.e., $\rho_{\lambda\mu}\left(\boldsymbol{r},\boldsymbol{r}'\right)$ for $\lambda=0$ and $s_{\lambda\mu}\left(\boldsymbol{r},\boldsymbol{r}'\right)$, $\nabla_{\lambda\mu}$, and $k_{\lambda\mu}$ for $\lambda=1$. In this notation, $\lambda$ is the rank of the tensor, and $\mu=-\lambda,\ldots,+\lambda$ is its tensor component. In the present study we use the following definitions of the building blocks in the spherical representation:

$$\rho_{00}(\boldsymbol{r},\boldsymbol{r}') = \rho(\boldsymbol{r},\boldsymbol{r}'), \tag{7}$$

$$s_{1,\mu=\{-1,0,1\}}(\boldsymbol{r},\boldsymbol{r}') = -i\left\{\tfrac{1}{\sqrt{2}}\left(s_x(\boldsymbol{r},\boldsymbol{r}')-is_y(\boldsymbol{r},\boldsymbol{r}')\right), s_z(\boldsymbol{r},\boldsymbol{r}'), \tfrac{-1}{\sqrt{2}}\left(s_x(\boldsymbol{r},\boldsymbol{r}')+is_y(\boldsymbol{r},\boldsymbol{r}')\right)\right\}, \tag{8}$$

$$\nabla_{1,\mu=\{-1,0,1\}} = -i\left\{\tfrac{1}{\sqrt{2}}\left(\nabla_x-i\nabla_y\right), \nabla_z, \tfrac{-1}{\sqrt{2}}\left(\nabla_x+i\nabla_y\right)\right\}, \tag{9}$$

$$k_{1,\mu=\{-1,0,1\}} = -i\left\{\tfrac{1}{\sqrt{2}}\left(k_x-ik_y\right), k_z, \tfrac{-1}{\sqrt{2}}\left(k_x+ik_y\right)\right\}. \tag{10}$$

In what follows, we most often omit indices and arguments of these spherical tensors and we simply write $\rho$, $s$, $\nabla$, and $k$ to lighten the notation.

In principle, arbitrary phase factors could be used in front of the spherical tensors. In Appendix B, we discuss possible choices of such phase conventions, and determine the particular ones selected in Eqs. (7)–(10). These phase conventions, which are *not* the standard ones, are used throughout the paper and define the phase properties of all other objects that we construct by using the building blocks above.

### B. Higher-order derivative operators

We begin by constructing all possible higher-order and higher-rank tensor operators from powers of the derivative $\nabla_{1\mu}$, where $\mu = -1, 0, +1$ are the spin-projection components of the vector (rank-1) operator $\nabla$. It is obvious that all possible $n$th-order powers of the derivative can be written as sums of terms $\nabla_{1\mu_1}\ldots\nabla_{1\mu_n}$. Therefore, any $(n+1)$th-order power is simply obtained by multiplying some $n$th-order power by a sum of $\nabla_{1\mu}$ operators. Then, powers of a given rank can be obtained iteratively by vector coupling.

In the second order, the two nabla operators can be coupled to angular momenta 0 and 2. The coupling to angular momentum 0, $[\nabla\nabla]_0 = \Delta/\sqrt{3}$, corresponds to the Laplacian operator. Furthermore, the coupling to angular momentum 2, $[\nabla\nabla]_2$, gives the second-order, rank-2 derivative operator. The rank-1 coupling, $[\nabla\nabla]_1 = 0$, vanishes because the derivatives commute. Similarly, in each one higher order, a rank-$L$ symmetric operator can be coupled with $\nabla$ only to $L-1$ and $L+1$. Hence, all the $n$th-order powers have the form of $\Delta^{(n-L)/2}$ multiplied by the $L$th-order rank-$L$ (stretched) coupled operators for $L = n, n-2, \ldots, (1)0$. Then, up to N$^3$LO, one obtains 16 different operators $D_{nL}$ listed in Table I. Any arbitrary tensor formed by coupled operators $\nabla$ can always be rewritten as a sum of operators $D_{nL}$ through the repeated use of the $6j$ symbols.

Exactly in the same way, we define 16 different operators $K_{nL}$, which are spherical tensors built of the relative momentum operators $k$ coupled up to N$^3$LO, i.e., for $n \leq 6$ and $L \leq 6$. In the remainder of this section, we only discuss operators $D_{nL}$, while all the results *mutatis mutandis* also pertain to operators $K_{nL}$.

The stretched coupled operators $D_{LL}$ for $n = L$,

$$D_{LL} = [\nabla\ldots[\nabla[\nabla\nabla]_2]_3\ldots]_L, \qquad (11)$$

play a central role in our derivations below. They correspond to irreducible symmetric traceless Cartesian tensors built of the derivative $\nabla$. They have $2L+1$ tensor components $D_{LLM}$ numbered by the quantum number $M = -L, \ldots, L$ that we most often do not show below explicitly. Moreover, since terms in the EDF up to N$^3$LO depend only on operators $D_{LL}$ and $K_{LL}$ up to fourth order, $L \leq 4$, see Sec. III A, below we do not discuss stretched coupled operators of fifth or sixth orders.

Equivalently, derivative operators $D_{LL}$ can be written in the Cartesian representation, in which their components are numbered by $L$ Cartesian indices, $D_{LL,a_1\ldots a_L}$, $a_i = x, y, z$. The order of these indices does not matter

TABLE I: Derivative operators $D_{nL}$ up to N$^3$LO as expressed through spherical tensor representation of the operator $\nabla$.

| No. | tensor $D_{nL}$ | order $n$ | rank $L$ |
|---|---|---|---|
| 1 | $1$ | 0 | 0 |
| 2 | $\nabla$ | 1 | 1 |
| 3 | $[\nabla\nabla]_0$ | 2 | 0 |
| 4 | $[\nabla\nabla]_2$ | 2 | 2 |
| 5 | $[\nabla\nabla]_0\nabla$ | 3 | 1 |
| 6 | $[\nabla[\nabla\nabla]_2]_3$ | 3 | 3 |
| 7 | $[\nabla\nabla]_0^2$ | 4 | 0 |
| 8 | $[\nabla\nabla]_0[\nabla\nabla]_2$ | 4 | 2 |
| 9 | $[\nabla[\nabla[\nabla\nabla]_2]_3]_4$ | 4 | 4 |
| 10 | $[\nabla\nabla]_0^2\nabla$ | 5 | 1 |
| 11 | $[\nabla\nabla]_0[\nabla[\nabla\nabla]_2]_3$ | 5 | 3 |
| 12 | $[\nabla[\nabla[\nabla[\nabla\nabla]_2]_3]_4]_5$ | 5 | 5 |
| 13 | $[\nabla\nabla]_0^3$ | 6 | 0 |
| 14 | $[\nabla\nabla]_0^2[\nabla\nabla]_2$ | 6 | 2 |
| 15 | $[\nabla\nabla]_0[\nabla[\nabla[\nabla\nabla]_2]_3]_4$ | 6 | 4 |
| 16 | $[\nabla[\nabla[\nabla[\nabla[\nabla\nabla]_2]_3]_4]_5]_6$ | 6 | 6 |

(totally symmetric tensors) and all traces vanish,

$$\sum_a D_{LL,aaa_3\ldots a_L} = 0. \qquad (12)$$

The Cartesian components $D_{LL,a_1\ldots a_L}$ can be calculated by using the detracer operator defined in Sec. 5 of Ref. [27]. Up to fourth order they read:

$$D_{00} = 1, \qquad (13)$$

$$D_{11,a_1} = \nabla_{a_1}, \qquad (14)$$

$$D_{22,a_1a_2} = \nabla_{a_1}\nabla_{a_2} - \tfrac{1}{3}\delta_{a_1a_2}\Delta, \qquad (15)$$

$$D_{33,a_1a_2a_3} = \nabla_{a_1}\nabla_{a_2}\nabla_{a_3} - \tfrac{1}{5}\Delta\Big(\nabla_{a_1}\delta_{a_2a_3} + \nabla_{a_2}\delta_{a_1a_3} + \nabla_{a_3}\delta_{a_1a_2}\Big), \qquad (16)$$

$$D_{44,a_1a_2a_3a_4} = \nabla_{a_1}\nabla_{a_2}\nabla_{a_3}\nabla_{a_4} - \tfrac{1}{7}\Delta\Big(\nabla_{a_1}\nabla_{a_2}\delta_{a_3a_4} + \nabla_{a_1}\nabla_{a_3}\delta_{a_2a_4} + \nabla_{a_1}\nabla_{a_4}\delta_{a_2a_3} + \nabla_{a_2}\nabla_{a_3}\delta_{a_1a_4} + \nabla_{a_2}\nabla_{a_4}\delta_{a_1a_3} + \nabla_{a_3}\nabla_{a_4}\delta_{a_1a_2}\Big) + \tfrac{1}{5\cdot7}\Delta^2\Big(\delta_{a_1a_2}\delta_{a_3a_4} + \delta_{a_1a_3}\delta_{a_2a_4} + \delta_{a_1a_4}\delta_{a_2a_3}\Big). \qquad (17)$$

TABLE II: Spherical components of the derivative operators $D_{nLM}$ expressed through the Cartesian derivatives. Expressions for negative components can be obtained as $D_{nL,-M} = (-1)^{L-M} D^*_{nLM}$, see Eqs. (B20) and (B22).

| $D_{nLM}$ | | Cartesian derivatives |
|---|---|---|
| $D_{110}$ | $=$ | $-i\partial_z$ |
| $D_{111}$ | $=$ | $i\frac{1}{\sqrt{2}}\left(\partial_x + i\partial_y\right)$ |
| $D_{200}$ | $=$ | $\frac{1}{\sqrt{3}}\Delta$ |
| $D_{220}$ | $=$ | $\frac{1}{\sqrt{6}}\left(\partial_x^2 + \partial_y^2 - 2\partial_z^2\right)$ |
| $D_{221}$ | $=$ | $\left(\partial_x + i\partial_y\right)\partial_z$ |
| $D_{222}$ | $=$ | $-\frac{1}{2}\left(\partial_x + i\partial_y\right)^2$ |
| $D_{330}$ | $=$ | $i\frac{1}{\sqrt{10}}\partial_z\left(-3\partial_x^2 - 3\partial_y^2 + 2\partial_z^2\right)$ |
| $D_{331}$ | $=$ | $i\frac{1}{2}\sqrt{\frac{3}{10}}\left(\partial_x + i\partial_y\right)\left(\partial_x^2 + \partial_y^2 - 4\partial_z^2\right)$ |
| $D_{332}$ | $=$ | $i\frac{1}{2}\sqrt{3}\left(\partial_x + i\partial_y\right)^2\partial_z$ |
| $D_{333}$ | $=$ | $-i\frac{1}{2\sqrt{2}}\left(\partial_x + i\partial_y\right)^3$ |
| $D_{440}$ | $=$ | $\frac{1}{2\sqrt{70}}\left(3\partial_x^4 + 6\left(\partial_y^2 - 4\partial_z^2\right)\partial_x^2 + 3\partial_y^4 + 8\partial_z^4 - 24\partial_y^2\partial_z^2\right)$ |
| $D_{441}$ | $=$ | $\frac{1}{\sqrt{14}}\left(\partial_x + i\partial_y\right)\partial_z\left(3\partial_x^2 + 3\partial_y^2 - 4\partial_z^2\right)$ |
| $D_{442}$ | $=$ | $-\frac{1}{2\sqrt{7}}\left(\partial_x + i\partial_y\right)^2\left(\partial_x^2 + \partial_y^2 - 6\partial_z^2\right)$ |
| $D_{443}$ | $=$ | $-\frac{1}{\sqrt{2}}\left(\partial_x + i\partial_y\right)^3\partial_z$ |
| $D_{444}$ | $=$ | $\frac{1}{4}\left(\partial_x + i\partial_y\right)^4$ |

We note here in passing that we could have equally well used the Cartesian derivative operators with traces not subtracted out, i.e.,

$$\mathcal{D}_{00} = 1, \tag{18}$$

$$\mathcal{D}_{11,a_1} = \nabla_{a_1}, \tag{19}$$

$$\mathcal{D}_{22,a_1a_2} = \nabla_{a_1}\nabla_{a_2}, \tag{20}$$

$$\mathcal{D}_{33,a_1a_2a_3} = \nabla_{a_1}\nabla_{a_2}\nabla_{a_3}, \tag{21}$$

$$\mathcal{D}_{44,a_1a_2a_3a_4} = \nabla_{a_1}\nabla_{a_2}\nabla_{a_3}\nabla_{a_4}. \tag{22}$$

Representations (13)–(17) and (18)–(22) are equivalent in the sense that each operator $D_{LL,a_1\ldots a_L}$ is evidently a linear combination of operators $\Delta^{(L-L')/2}\mathcal{D}_{L'L',a_1\ldots a_{L'}}$ for $L' = L, L-2, \ldots, (1)0$.

In principle, below one could replace the spherical representations of derivative operators shown in Table I by their Cartesian counterparts (13)–(17) or (18)–(22), and work entirely in the Cartesian representation. However, in our opinion, the use of the spherical representation is superior and more economical. Moreover, whenever calculation of the Cartesian derivatives is more suitable, we may express spherical components of the derivative operators through the Cartesian derivatives, as shown in Table II. An example of using the Cartesian representation (18)–(22) is given in Sec. IV.

TABLE III: Local primary densities (23) up to N$^3$LO built from the scalar nonlocal density $\rho(\boldsymbol{r}, \boldsymbol{r}')$ ($v = 0$). To simplify the notation the limit of $\boldsymbol{r}' = \boldsymbol{r}$ is not shown explicitly. Stars ($\star$) mark densities that enter the EDF up to N$^3$LO. Bullets ($\bullet$) mark densities that enter the EDF up to N$^3$LO for conserved spherical, space-inversion, and time-reversal symmetries, see Sec. IV. The last two columns show the $T$ and $P$ parities defined in Eqs. (25) and (26), respectively. In addition, the time-even densities are marked by using the bold-face font.

| No. | | | $\rho_{nLvJ}$ | = density | $n$ | $L$ | $v$ | $J$ | $T$ | $P$ |
|---|---|---|---|---|---|---|---|---|---|---|
| 1 | $\star$ | $\bullet$ | $\boldsymbol{\rho}_{\mathbf{0000}}$ | $= [\rho]_0$ | 0 | 0 | 0 | 0 | 1 | 1 |
| 2 | $\star$ | | $\rho_{1101}$ | $= [k\rho]_1$ | 1 | 1 | 0 | 1 | −1 | −1 |
| 3 | $\star$ | $\bullet$ | $\boldsymbol{\rho}_{\mathbf{2000}}$ | $= [[kk]_0\rho]_0$ | 2 | 0 | 0 | 0 | 1 | 1 |
| 4 | $\star$ | $\bullet$ | $\boldsymbol{\rho}_{\mathbf{2202}}$ | $= [[kk]_2\rho]_2$ | 2 | 2 | 0 | 2 | 1 | 1 |
| 5 | $\star$ | | $\rho_{3101}$ | $= [[kk]_0k\rho]_1$ | 3 | 1 | 0 | 1 | −1 | −1 |
| 6 | $\star$ | | $\rho_{3303}$ | $= [[k[kk]_2]_3\rho]_3$ | 3 | 3 | 0 | 3 | −1 | −1 |
| 7 | $\star$ | $\bullet$ | $\boldsymbol{\rho}_{\mathbf{4000}}$ | $= [[kk]_0^2\rho]_0$ | 4 | 0 | 0 | 0 | 1 | 1 |
| 8 | $\star$ | $\bullet$ | $\boldsymbol{\rho}_{\mathbf{4202}}$ | $= [[kk]_0[kk]_2\rho]_2$ | 4 | 2 | 0 | 2 | 1 | 1 |
| 9 | | | $\boldsymbol{\rho}_{\mathbf{4404}}$ | $= [[k[k[kk]_2]_3]_4\rho]_4$ | 4 | 4 | 0 | 4 | 1 | 1 |
| 10 | $\star$ | | $\rho_{5101}$ | $= [[kk]_0^2k\rho]_1$ | 5 | 1 | 0 | 1 | −1 | −1 |
| 11 | | | $\rho_{5303}$ | $= [[kk]_0[k[kk]_2]_3\rho]_3$ | 5 | 3 | 0 | 3 | −1 | −1 |
| 12 | | | $\rho_{5505}$ | $= [[k[k[k[kk]_2]_3]_4]_5\rho]_5$ | 5 | 5 | 0 | 5 | −1 | −1 |
| 13 | $\star$ | $\bullet$ | $\boldsymbol{\rho}_{\mathbf{6000}}$ | $= [[kk]_0^3\rho]_0$ | 6 | 0 | 0 | 0 | 1 | 1 |
| 14 | | | $\boldsymbol{\rho}_{\mathbf{6202}}$ | $= [[kk]_0^2[kk]_2\rho]_2$ | 6 | 2 | 0 | 2 | 1 | 1 |
| 15 | | | $\boldsymbol{\rho}_{\mathbf{6404}}$ | $= [[kk]_0[k[k[kk]_2]_3]_4\rho]_4$ | 6 | 4 | 0 | 4 | 1 | 1 |
| 16 | | | $\boldsymbol{\rho}_{\mathbf{6606}}$ | $= [[k[k[k[k[kk]_2]_3]_4]_5]_6\rho]_6$ | 6 | 6 | 0 | 6 | 1 | 1 |

### C. Local densities

Local densities are formed by acting several times on the scalar and vector nonlocal densities with the relative momentum operator $\boldsymbol{k}$ and taking the limit of $\boldsymbol{r}' = \boldsymbol{r}$. Using the spherical representation, the possible coupled $k$-tensors (10) (up to sixth order in derivatives) $K_{nL}$ are those given in Table I (replacing $\nabla$ with $k$).

Acting with $K_{nL}$ on the scalar nonlocal density $\rho(\boldsymbol{r}, \boldsymbol{r}')$ gives 16 different local densities up to N$^3$LO (one for every term in Table I). They are listed in Table III. When acting with $K_{nL}$ on the vector nonlocal densities $s(\boldsymbol{r}, \boldsymbol{r}')$, one has to construct all possible ways of coupling the $k$-tensors with the vector density. Obviously, each of the 4 scalar ($L = 0$) derivative operators gives one local density, while each of the 12 non-scalar ($L > 0$) derivative operators gives three local densities. Altogether, from the vector density one obtains 40 local densities up to N$^3$LO. They are listed in Table IV.

Finally, all local densities can be denoted by four integers $nLvJ$ as

$$\rho_{nLvJ}(\boldsymbol{r}) = \left\{[K_{nL}\rho_v(\boldsymbol{r}, \boldsymbol{r}')]_J\right\}_{\boldsymbol{r}'=\boldsymbol{r}}, \tag{23}$$

where the $n$th-order and rank-$L$ relative derivative operator $K_{nL}$ acts on the scalar ($v = 0$) or vector ($v = 1$) nonlocal density, and ranks $L$ and $v$ are then vector coupled to $J$. We call these local densities primary densities.

TABLE IV: Same as in Table III but for densities built from the vector nonlocal density $\boldsymbol{s}(\boldsymbol{r},\boldsymbol{r}')$ ($v=1$).

| No. | | $\rho_{nLvJ}$ = density | $n$ | $L$ | $v$ | $J$ | $T$ | $P$ |
|---|---|---|---|---|---|---|---|---|
| 17 | ⋆ | $\rho_{0011} = [s]_1$ | 0 | 0 | 1 | 1 | −1 | 1 |
| 18 | ⋆ | $\boldsymbol{\rho_{1110}} = [ks]_0$ | 1 | 1 | 1 | 0 | 1 | −1 |
| 19 | ⋆ • | $\boldsymbol{\rho_{1111}} = [ks]_1$ | 1 | 1 | 1 | 1 | 1 | −1 |
| 20 | ⋆ | $\boldsymbol{\rho_{1112}} = [ks]_2$ | 1 | 1 | 1 | 2 | 1 | −1 |
| 21 | ⋆ | $\rho_{2011} = [[kk]_0 s]_1$ | 2 | 0 | 1 | 1 | −1 | 1 |
| 22 | ⋆ | $\rho_{2211} = [[kk]_2 s]_1$ | 2 | 2 | 1 | 1 | −1 | 1 |
| 23 | ⋆ | $\rho_{2212} = [[kk]_2 s]_2$ | 2 | 2 | 1 | 2 | −1 | 1 |
| 24 | ⋆ | $\rho_{2213} = [[kk]_2 s]_3$ | 2 | 2 | 1 | 3 | −1 | 1 |
| 25 | ⋆ | $\boldsymbol{\rho_{3110}} = [[kk]_0 ks]_0$ | 3 | 1 | 1 | 0 | 1 | −1 |
| 26 | ⋆ • | $\boldsymbol{\rho_{3111}} = [[kk]_0 ks]_1$ | 3 | 1 | 1 | 1 | 1 | −1 |
| 27 | ⋆ | $\boldsymbol{\rho_{3112}} = [[kk]_0 ks]_2$ | 3 | 1 | 1 | 2 | 1 | −1 |
| 28 | ⋆ | $\boldsymbol{\rho_{3312}} = [[k[kk]_2]_3 s]_2$ | 3 | 3 | 1 | 2 | 1 | −1 |
| 29 | ⋆ • | $\boldsymbol{\rho_{3313}} = [[k[kk]_2]_3 s]_3$ | 3 | 3 | 1 | 3 | 1 | −1 |
| 30 | ⋆ | $\boldsymbol{\rho_{3314}} = [[k[kk]_2]_3 s]_4$ | 3 | 3 | 1 | 4 | 1 | −1 |
| 31 | ⋆ | $\rho_{4011} = [[kk]_0^2 s]_1$ | 4 | 0 | 1 | 1 | −1 | 1 |
| 32 | ⋆ | $\rho_{4211} = [[kk]_0 [kk]_2 s]_1$ | 4 | 2 | 1 | 1 | −1 | 1 |
| 33 | ⋆ | $\rho_{4212} = [[kk]_0 [kk]_2 s]_2$ | 4 | 2 | 1 | 2 | −1 | 1 |
| 34 | ⋆ | $\rho_{4213} = [[kk]_0 [kk]_2 s]_3$ | 4 | 2 | 1 | 3 | −1 | 1 |
| 35 | ⋆ | $\rho_{4413} = [[k[k[kk]_2]_3]_4 s]_3$ | 4 | 4 | 1 | 3 | −1 | 1 |
| 36 | | $\rho_{4414} = [[k[k[kk]_2]_3]_4 s]_4$ | 4 | 4 | 1 | 4 | −1 | 1 |
| 37 | | $\rho_{4415} = [[k[k[kk]_2]_3]_4 s]_5$ | 4 | 4 | 1 | 5 | −1 | 1 |

TABLE IV: continued.

| No. | | $\rho_{nLvJ}$ = density | $n$ | $L$ | $v$ | $J$ | $T$ | $P$ |
|---|---|---|---|---|---|---|---|---|
| 38 | ⋆ | $\boldsymbol{\rho_{5110}} = [[kk]_0^2 ks]_0$ | 5 | 1 | 1 | 0 | 1 | −1 |
| 39 | ⋆ • | $\boldsymbol{\rho_{5111}} = [[kk]_0^2 ks]_1$ | 5 | 1 | 1 | 1 | 1 | −1 |
| 40 | ⋆ | $\boldsymbol{\rho_{5112}} = [[kk]_0^2 ks]_2$ | 5 | 1 | 1 | 2 | 1 | −1 |
| 41 | ⋆ | $\boldsymbol{\rho_{5312}} = [[kk]_0 [k[kk]_2]_3 s]_2$ | 5 | 3 | 1 | 2 | 1 | −1 |
| 42 | | $\boldsymbol{\rho_{5313}} = [[kk]_0 [k[kk]_2]_3 s]_3$ | 5 | 3 | 1 | 3 | 1 | −1 |
| 43 | | $\boldsymbol{\rho_{5314}} = [[kk]_0 [k[kk]_2]_3 s]_4$ | 5 | 3 | 1 | 4 | 1 | −1 |
| 44 | | $\boldsymbol{\rho_{5514}} = [[k[k[k[kk]_2]_3]_4]_5 s]_4$ | 5 | 5 | 1 | 4 | 1 | −1 |
| 45 | | $\boldsymbol{\rho_{5515}} = [[k[k[k[kk]_2]_3]_4]_5 s]_5$ | 5 | 5 | 1 | 5 | 1 | −1 |
| 46 | | $\boldsymbol{\rho_{5516}} = [[k[k[k[kk]_2]_3]_4]_5 s]_6$ | 5 | 5 | 1 | 6 | 1 | −1 |
| 47 | ⋆ | $\rho_{6011} = [[kk]_0^3 s]_1$ | 6 | 0 | 1 | 1 | −1 | 1 |
| 48 | ⋆ | $\rho_{6211} = [[kk]_0^2 [kk]_2 s]_1$ | 6 | 2 | 1 | 1 | −1 | 1 |
| 49 | | $\rho_{6212} = [[kk]_0^2 [kk]_2 s]_2$ | 6 | 2 | 1 | 2 | −1 | 1 |
| 50 | | $\rho_{6213} = [[kk]_0^2 [kk]_2 s]_3$ | 6 | 2 | 1 | 3 | −1 | 1 |
| 51 | | $\rho_{6413} = [[kk]_0 [k[k[kk]_2]_3]_4 s]_3$ | 6 | 4 | 1 | 3 | −1 | 1 |
| 52 | | $\rho_{6414} = [[kk]_0 [k[k[kk]_2]_3]_4 s]_4$ | 6 | 4 | 1 | 4 | −1 | 1 |
| 53 | | $\rho_{6415} = [[kk]_0 [k[k[kk]_2]_3]_4 s]_5$ | 6 | 4 | 1 | 5 | −1 | 1 |
| 54 | | $\rho_{6615} = [[k[k[k[k[kk]_2]_3]_4]_5]_6 s]_5$ | 6 | 6 | 1 | 5 | −1 | 1 |
| 55 | | $\rho_{6616} = [[k[k[k[k[kk]_2]_3]_4]_5]_6 s]_6$ | 6 | 6 | 1 | 6 | −1 | 1 |
| 56 | | $\rho_{6617} = [[k[k[k[k[kk]_2]_3]_4]_5]_6 s]_7$ | 6 | 6 | 1 | 7 | −1 | 1 |

The tensor components corresponding to the total rank $J$ are not explicitly shown.

One can also act on each of the local densities with derivative operators $D_{mI}$ of Table I, and then couple ranks $I$ and $J$ to the total rank $Q$, i.e.,

$$\rho_{mI,nLvJ,Q}(\boldsymbol{r}) = [D_{mI}\rho_{nLvJ}(\boldsymbol{r})]_Q . \qquad (24)$$

For $m > 0$, we call these local densities secondary densities. We do not explicitly list them, because they can be obtained in a straightforward way from the primary densities corresponding to $m = 0$, $\rho_{nLvJ} = \rho_{00,InLvJ,J}$, which are listed in Tables III and IV.

In Tables III and IV, for completeness we also show the time-reversal ($T$) and space-inversion ($P$) parities defined as,

$$T = (-1)^{n+v}, \qquad (25)$$

$$P = (-1)^n. \qquad (26)$$

These definitions are based on the analysis of symmetry properties, which we present in Appendix A. To better visualize the time-reversal properties of the local densities, in Tables III and IV the time-even densities are shown in bold face.

Local densities constructed above are complex. Taking the complex conjugations gives relations derived in Appendix B:

$$\rho^*_{mI,nLvJ,Q,M} = (-1)^{Q-M}\,\rho_{mI,nLvJ,Q,-M}, \qquad (27)$$

where the tensor components, denoted $M$, are shown explicitly. These relations allow for expressing positive tensor components through negative ones or *vice versa*. Therefore, complete information is contained in non-negative or non-positive tensor components only. The $M=0$ components are either real (for even $Q$) or imaginary (for odd $Q$), and hence $2Q+1$ real functions always suffice to describe a given local density of rank $Q$. Moreover, all scalar densities are real, which was the basis of choosing this particular phase convention, as described in Appendix B.

## III. CONSTRUCTION OF THE ENERGY DENSITY

### A. Terms in the energy density

Terms in the EDF we construct here are required to be quadratic in densities, invariant with respect to time reversal (Sec. A 1), and covariant with respect to space inversion and rotations (Sec. A 2). All terms up to the $\mathrm{N^3LO}$ order in derivatives fulfilling these restrictions are constructed below.

Using notation of Eq. (24), a general term in the energy density can be written in the following form,

$$T^{m'I',n'L'v'J'}_{mI,nLvJ,Q}(\boldsymbol{r}) = [\rho_{m'I',n'L'v'J',Q}(\boldsymbol{r})\rho_{mI,nLvJ,Q}(\boldsymbol{r})]_0, \qquad (28)$$

TABLE V: Numbers of terms defined in Eq. (28) of different orders in the EDF up to N$^3$LO. Numbers of terms depending on the time-even and time-odd densities are given separately. The last two columns give numbers of terms when the Galilean or gauge invariance is assumed, respectively, see Sec. III B. To take into account both isospin channels, the numbers of terms should be multiplied by a factor of two.

| order | T-even | T-odd | Total | Galilean | Gauge |
|---|---|---|---|---|---|
| 0 | 1 | 1 | 2 | 2 | 2 |
| 2 | 8 | 10 | 18 | 12 | 12 |
| 4 | 53 | 61 | 114 | 45 | 29 |
| 6 | 250 | 274 | 524 | 129 | 54 |
| N$^3$LO | 312 | 346 | 658 | 188 | 97 |

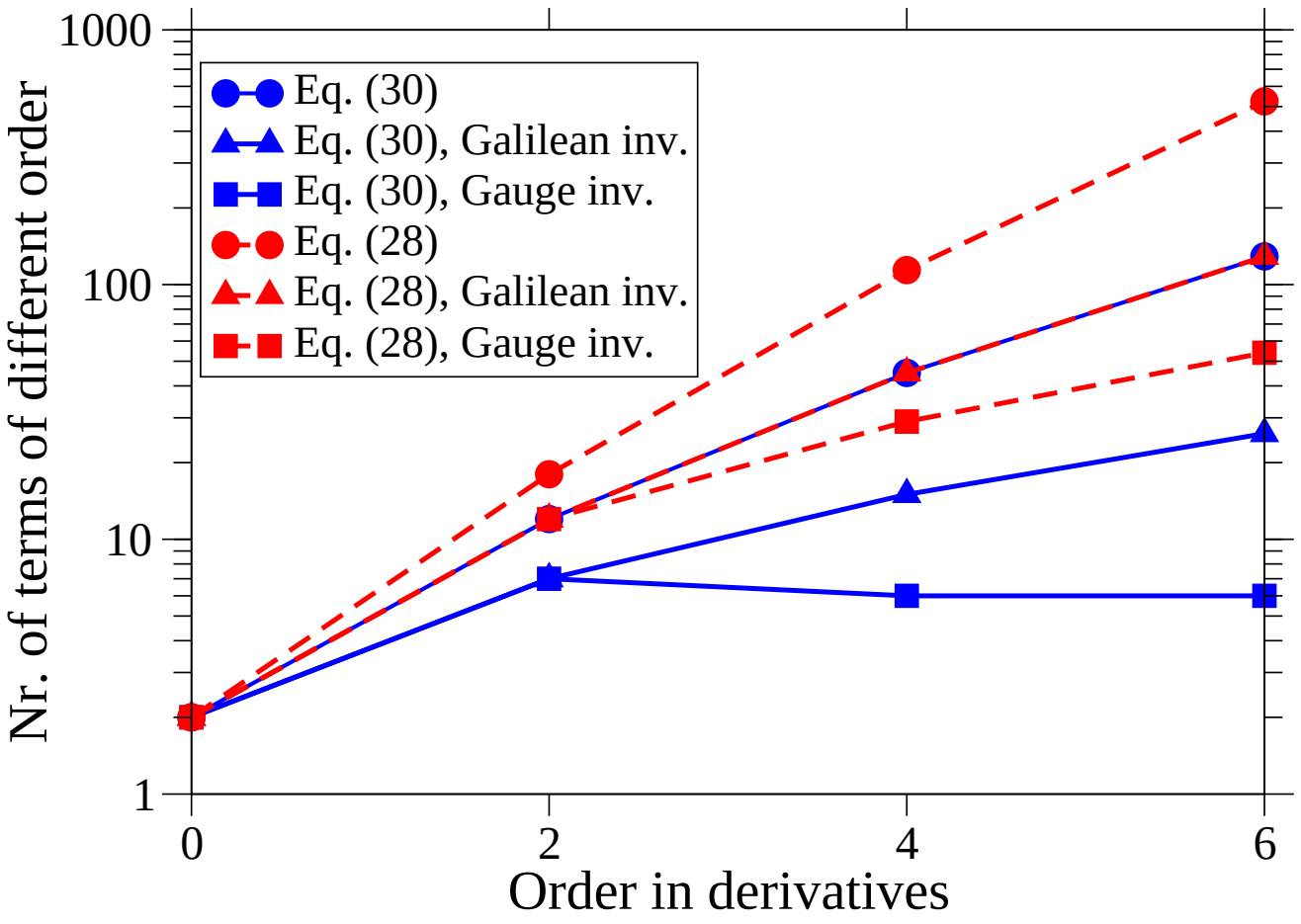

FIG. 1: (Color online) Numbers of terms (28) and (30) shown in Tables V and VI, respectively, plotted in logarithmic scale as a function of the order in derivatives.

where both densities must have the same rank $Q$ to be coupled to a scalar. Moreover, their time-reversal and space-inversion parities ($T$ and $P$) must be the same. Again, at N$^3$LO only terms with $m'+n'+m+n \leq 6$ are allowed. Then, the total energy density reads

$$\mathcal{H}(\boldsymbol{r}) = \sum_{\substack{m'I',n'L'v'J'\\ mI,nLvJ,Q}} C^{m'I',n'L'v'J'}_{mI,nLvJ,Q} T^{m'I',n'L'v'J'}_{mI,nLvJ,Q}(\boldsymbol{r}), \quad (29)$$

where $C^{m'I',n'L'v'J'}_{mI,nLvJ,Q}$ are coupling constants and the summation runs over all allowed indices.

Had we considered the case of coupling constants depending on density, all terms in Eq. (28) would have been independent of one another (up to a possible exchange of the two densities). Table V lists numbers of such independent terms, and they are also plotted in Fig. 1.

In the present study, we concentrate on the case of density-independent coupling constants, in which case one can perform integrations by parts, so that the derivative operators $D_{m'I'}$ are transferred from one density to the other. That this can always be done is obvious by the fact that the coupled derivative operators $D_{m'I'}$ can always be expressed as sums of products of uncoupled derivatives $\nabla_{1\mu}$ or $\nabla_a$. As a result of the integration by parts, the integral (28) can now be written as a sum of terms, where each term has the form:

$$T^{n'L'v'J'}_{mI,nLvJ}(\boldsymbol{r}) = [\rho_{n'L'v'J'}(\boldsymbol{r})[D_{mI}\rho_{nLvJ}(\boldsymbol{r})]_{J'}]_0, \quad (30)$$

where ranks $I$ and $J$ are coupled to $J'$. Here, at N$^3$LO (i) only terms with $n'+m+n \leq 6$ are allowed, (ii) both densities must have the same time-reversal parity $T$, and (iii) their space-inversion parities must differ by factors $(-1)^I$. Finally, in order to avoid double-counting one takes only terms with $n' < n$, and for $n' = n$ only those with $L' < L$, and for $L' = L$ only those with $v' < v$, and for $v' = v$ only those with $J' \leq J$. Then, the total energy density reads

$$\mathcal{H}(\boldsymbol{r}) = \sum_{\substack{n'L'v'J'\\ mI,nLvJ,J'}} C^{n'L'v'J'}_{mI,nLvJ} T^{n'L'v'J'}_{mI,nLvJ}(\boldsymbol{r}), \quad (31)$$

where $C^{n'L'v'J'}_{mI,nLvJ}$ are coupling constants and the summation again runs over all allowed indices. As we did for the local densities above, to better visualize the time-reversal characteristics of terms in the EDF, the coupling constants $C^{n'L'v'J'}_{mI,nLvJ}$ corresponding to terms that depend on time-even densities are shown in bold face.

TABLE VI: Same as in Table V, but for numbers of terms defined in Eq. (30).

| order | T-even | T-odd | Total | Galilean | Gauge |
|---|---|---|---|---|---|
| 0 | 1 | 1 | 2 | 2 | 2 |
| 2 | 6 | 6 | 12 | 7 | 7 |
| 4 | 22 | 23 | 45 | 15 | 6 |
| 6 | 64 | 65 | 129 | 26 | 6 |
| N$^3$LO | 93 | 95 | 188 | 50 | 21 |

Based on the results obtained in Secs. A 1 and A 2, and on Eqs. (25) and (26), we see that time-reversal invariance and space-inversion covariance require that

$$(-1)^{n'+v'+n+v} = 1, \quad (32)$$

$$(-1)^{n'+m+n} = 1, \quad (33)$$

respectively. This means that integers $v'+v$, $n'+n$, and $m$ must be simultaneously either even or odd. The numbers of all such allowed terms are given in Table VI and plotted in Fig. 1. The space-inversion covariance (33) requires that for all terms, the total orders in derivatives are even numbers, which defines our classification of the EDF up to LO (0), NLO (2), NNLO (4), and N$^3$LO (6).

In Appendix B, we presented terms in the EDF up to NLO, i.e., for zero and second orders, see Table XXII. The EDF at NLO is exactly equivalent to the standard

TABLE VII: Terms in the EDF (30) that are of fourth order, depend on time-even densities, and are built from the scalar nonlocal density $\rho(\boldsymbol{r},\boldsymbol{r}')$. Coupling constants corresponding to terms that depend on time-even densities are marked by using the bold-face font. Bullets (•) mark coupling constants corresponding to terms that do not vanish for conserved spherical, space-inversion, and time-reversal symmetries, see Sec. IV.

| No. | $C^{n'L'v'J'}_{mI,nLvJ}$ | $\rho_{n'L'v'J'}$ | $D_{mI}$ | $\rho_{nLvJ}$ |
|---|---|---|---|---|
| 1 • | $\boldsymbol{C^{0000}_{40,0000}}$ | $[\rho]_0$ | $[\nabla\nabla]^2_0$ | $[\rho]_0$ |
| 2 • | $\boldsymbol{C^{0000}_{20,2000}}$ | $[\rho]_0$ | $[\nabla\nabla]_0$ | $[[kk]_0\rho]_0$ |
| 3 • | $\boldsymbol{C^{0000}_{22,2202}}$ | $[\rho]_0$ | $[\nabla\nabla]_2$ | $[[kk]_2\rho]_2$ |
| 4 • | $\boldsymbol{C^{0000}_{00,4000}}$ | $[\rho]_0$ | 1 | $[[kk]^2_0\rho]_0$ |
| 5 • | $\boldsymbol{C^{2000}_{00,2000}}$ | $[[kk]_0\rho]_0$ | 1 | $[[kk]_0\rho]_0$ |
| 6 • | $\boldsymbol{C^{2202}_{00,2202}}$ | $[[kk]_2\rho]_2$ | 1 | $[[kk]_2\rho]_2$ |

TABLE VIII: Same as in Table VII but for terms that are built from the vector nonlocal density $\boldsymbol{s}(\boldsymbol{r},\boldsymbol{r}')$.

| No. | $C^{n'L'v'J'}_{mI,nLvJ}$ | $\rho_{n'L'v'J'}$ | $D_{mI}$ | $\rho_{nLvJ}$ |
|---|---|---|---|---|
| 7 | $\boldsymbol{C^{1110}_{20,1110}}$ | $[ks]_0$ | $[\nabla\nabla]_0$ | $[ks]_0$ |
| 8 | $\boldsymbol{C^{1110}_{22,1112}}$ | $[ks]_0$ | $[\nabla\nabla]_2$ | $[ks]_2$ |
| 9 | $\boldsymbol{C^{1110}_{00,3110}}$ | $[ks]_0$ | 1 | $[[kk]_0ks]_0$ |
| 10 • | $\boldsymbol{C^{1111}_{20,1111}}$ | $[ks]_1$ | $[\nabla\nabla]_0$ | $[ks]_1$ |
| 11 • | $\boldsymbol{C^{1111}_{22,1111}}$ | $[ks]_1$ | $[\nabla\nabla]_2$ | $[ks]_1$ |
| 12 | $\boldsymbol{C^{1111}_{22,1112}}$ | $[ks]_1$ | $[\nabla\nabla]_2$ | $[ks]_2$ |
| 13 • | $\boldsymbol{C^{1111}_{00,3111}}$ | $[ks]_1$ | 1 | $[[kk]_0ks]_1$ |
| 14 | $\boldsymbol{C^{1112}_{20,1112}}$ | $[ks]_2$ | $[\nabla\nabla]_0$ | $[ks]_2$ |
| 15 | $\boldsymbol{C^{1112}_{22,1112}}$ | $[ks]_2$ | $[\nabla\nabla]_2$ | $[ks]_2$ |
| 16 | $\boldsymbol{C^{1112}_{00,3112}}$ | $[ks]_2$ | 1 | $[[kk]_0ks]_2$ |
| 17 | $\boldsymbol{C^{1112}_{00,3312}}$ | $[ks]_2$ | 1 | $[[k[kk]_2]_3s]_2$ |

Skyrme functional [28, 29], generalized to include all time-odd terms [24, 25, 30]. In both representations the functional depends, in general, on 14 coupling constants, and both sets are related by simple expressions given in Eqs. (B6)–(B19).

In Tables VII–XVIII, we list all 45 and 129 terms in the EDF that are of fourth and sixth order, respectively. Together with 14 terms at NLO, listed in Table XXII, this constitutes the full list of 188 terms in the EDF at N$^3$LO.

After the complete list of terms in the EDF at N$^3$LO is constructed, one can check that not all of the local densities listed in Tables III and IV appear in the final EDF at N$^3$LO. This is so, because it is not possible to couple all these densities to scalars, and simultaneously fulfill conditions (32) and (33), without obtaining more than total sixth order in derivatives. It turns out that out of the 56 local densities at N$^3$LO, which are listed in Tables III and IV, only 35 occur in the final EDF at N$^3$LO. In Tables III and IV such densities are marked with stars ($\star$). Table XIX gives their numbers determined separately at each order.

TABLE IX: Same as in Table VII but for terms that are built from the scalar nonlocal density $\rho(\boldsymbol{r},\boldsymbol{r}')$ and vector nonlocal density $\boldsymbol{s}(\boldsymbol{r},\boldsymbol{r}')$.

| No. | $C^{n'L'v'J'}_{mI,nLvJ}$ | $\rho_{n'L'v'J'}$ | $D_{mI}$ | $\rho_{nLvJ}$ |
|---|---|---|---|---|
| 18 • | $\boldsymbol{C^{0000}_{31,1111}}$ | $[\rho]_0$ | $[\nabla\nabla]_0\nabla$ | $[ks]_1$ |
| 19 • | $\boldsymbol{C^{0000}_{11,3111}}$ | $[\rho]_0$ | $\nabla$ | $[[kk]_0ks]_1$ |
| 20 • | $\boldsymbol{C^{2000}_{11,1111}}$ | $[[kk]_0\rho]_0$ | $\nabla$ | $[ks]_1$ |
| 21 • | $\boldsymbol{C^{2202}_{11,1111}}$ | $[[kk]_2\rho]_2$ | $\nabla$ | $[ks]_1$ |
| 22 | $\boldsymbol{C^{2202}_{11,1112}}$ | $[[kk]_2\rho]_2$ | $\nabla$ | $[ks]_2$ |

TABLE X: Terms in the EDF (30) that are of fourth order, depend on time-odd densities, and are built from the scalar nonlocal density $\rho(\boldsymbol{r},\boldsymbol{r}')$.

| No. | $C^{n'L'v'J'}_{mI,nLvJ}$ | $\rho_{n'L'v'J'}$ | $D_{mI}$ | $\rho_{nLvJ}$ |
|---|---|---|---|---|
| 23 | $C^{1101}_{20,1101}$ | $[k\rho]_1$ | $[\nabla\nabla]_0$ | $[k\rho]_1$ |
| 24 | $C^{1101}_{22,1101}$ | $[k\rho]_1$ | $[\nabla\nabla]_2$ | $[k\rho]_1$ |
| 25 | $C^{1101}_{00,3101}$ | $[k\rho]_1$ | 1 | $[[kk]_0k\rho]_1$ |

TABLE XI: Same as in Table X but for terms that are built from the vector nonlocal density $\boldsymbol{s}(\boldsymbol{r},\boldsymbol{r}')$.

| No. | $C^{n'L'v'J'}_{mI,nLvJ}$ | $\rho_{n'L'v'J'}$ | $D_{mI}$ | $\rho_{nLvJ}$ |
|---|---|---|---|---|
| 26 | $C^{0011}_{40,0011}$ | $[s]_1$ | $[\nabla\nabla]^2_0$ | $[s]_1$ |
| 27 | $C^{0011}_{42,0011}$ | $[s]_1$ | $[\nabla\nabla]_0[\nabla\nabla]_2$ | $[s]_1$ |
| 28 | $C^{0011}_{20,2011}$ | $[s]_1$ | $[\nabla\nabla]_0$ | $[[kk]_0s]_1$ |
| 29 | $C^{0011}_{22,2011}$ | $[s]_1$ | $[\nabla\nabla]_2$ | $[[kk]_0s]_1$ |
| 30 | $C^{0011}_{20,2211}$ | $[s]_1$ | $[\nabla\nabla]_0$ | $[[kk]_2s]_1$ |
| 31 | $C^{0011}_{22,2211}$ | $[s]_1$ | $[\nabla\nabla]_2$ | $[[kk]_2s]_1$ |
| 32 | $C^{0011}_{22,2212}$ | $[s]_1$ | $[\nabla\nabla]_2$ | $[[kk]_2s]_2$ |
| 33 | $C^{0011}_{22,2213}$ | $[s]_1$ | $[\nabla\nabla]_2$ | $[[kk]_2s]_3$ |
| 34 | $C^{0011}_{00,4011}$ | $[s]_1$ | 1 | $[[kk]^2_0s]_1$ |
| 35 | $C^{0011}_{00,4211}$ | $[s]_1$ | 1 | $[[kk]_0[kk]_2s]_1$ |
| 36 | $C^{2011}_{00,2011}$ | $[[kk]_0s]_1$ | 1 | $[[kk]_0s]_1$ |
| 37 | $C^{2011}_{00,2211}$ | $[[kk]_0s]_1$ | 1 | $[[kk]_2s]_1$ |
| 38 | $C^{2211}_{00,2211}$ | $[[kk]_2s]_1$ | 1 | $[[kk]_2s]_1$ |
| 39 | $C^{2212}_{00,2212}$ | $[[kk]_2s]_2$ | 1 | $[[kk]_2s]_2$ |
| 40 | $C^{2213}_{00,2213}$ | $[[kk]_2s]_3$ | 1 | $[[kk]_2s]_3$ |

TABLE XII: Same as in Table X but for terms that are built from the scalar nonlocal density $\rho(\boldsymbol{r},\boldsymbol{r}')$ and vector nonlocal density $\boldsymbol{s}(\boldsymbol{r},\boldsymbol{r}')$.

| No. | $C^{n'L'v'J'}_{mI,nLvJ}$ | $\rho_{n'L'v'J'}$ | $D_{mI}$ | $\rho_{nLvJ}$ |
|---|---|---|---|---|
| 41 | $C^{1101}_{31,0011}$ | $[k\rho]_1$ | $[\nabla\nabla]_0\nabla$ | $[s]_1$ |
| 42 | $C^{1101}_{11,2011}$ | $[k\rho]_1$ | $\nabla$ | $[[kk]_0s]_1$ |
| 43 | $C^{1101}_{11,2211}$ | $[k\rho]_1$ | $\nabla$ | $[[kk]_2s]_1$ |
| 44 | $C^{1101}_{11,2212}$ | $[k\rho]_1$ | $\nabla$ | $[[kk]_2s]_2$ |
| 45 | $C^{3101}_{11,0011}$ | $[[kk]_0k\rho]_1$ | $\nabla$ | $[s]_1$ |

TABLE XIII: Terms in the EDF (30) that are of sixth order, depend on time-even densities, and are built from the scalar nonlocal density $\rho(\boldsymbol{r},\boldsymbol{r}')$. Coupling constants corresponding to terms that depend on time-even densities are marked by using the bold-face font. Bullets (•) mark coupling constants corresponding to terms that do not vanish for conserved spherical, space-inversion, and time-reversal symmetries, see Sec. IV.

| No. | $C^{n'L'v'J'}_{mI,nLvJ}$ | $\rho_{n'L'v'J'}$ | $D_{mI}$ | $\rho_{nLvJ}$ |
|---|---|---|---|---|
| 1 • | $\boldsymbol{C^{0000}_{60,0000}}$ | $[\rho]_0$ | $[\nabla\nabla]_0^3$ | $[\rho]_0$ |
| 2 • | $\boldsymbol{C^{0000}_{40,2000}}$ | $[\rho]_0$ | $[\nabla\nabla]_0^2$ | $[[kk]_0\rho]_0$ |
| 3 • | $\boldsymbol{C^{0000}_{42,2202}}$ | $[\rho]_0$ | $[\nabla\nabla]_0[\nabla\nabla]_2$ | $[[kk]_2\rho]_2$ |
| 4 • | $\boldsymbol{C^{0000}_{20,4000}}$ | $[\rho]_0$ | $[\nabla\nabla]_0$ | $[[kk]_0^2\rho]_0$ |
| 5 • | $\boldsymbol{C^{0000}_{22,4202}}$ | $[\rho]_0$ | $[\nabla\nabla]_2$ | $[[kk]_0[kk]_2\rho]_2$ |
| 6 • | $\boldsymbol{C^{0000}_{00,6000}}$ | $[\rho]_0$ | 1 | $[[kk]_0^3\rho]_0$ |
| 7 • | $\boldsymbol{C^{2000}_{20,2000}}$ | $[[kk]_0\rho]_0$ | $[\nabla\nabla]_0$ | $[[kk]_0\rho]_0$ |
| 8 • | $\boldsymbol{C^{2000}_{22,2202}}$ | $[[kk]_0\rho]_0$ | $[\nabla\nabla]_2$ | $[[kk]_2\rho]_2$ |
| 9 • | $\boldsymbol{C^{2000}_{00,4000}}$ | $[[kk]_0\rho]_0$ | 1 | $[[kk]_0^2\rho]_0$ |
| 10 • | $\boldsymbol{C^{2202}_{20,2202}}$ | $[[kk]_2\rho]_2$ | $[\nabla\nabla]_0$ | $[[kk]_2\rho]_2$ |
| 11 • | $\boldsymbol{C^{2202}_{22,2202}}$ | $[[kk]_2\rho]_2$ | $[\nabla\nabla]_2$ | $[[kk]_2\rho]_2$ |
| 12 • | $\boldsymbol{C^{2202}_{00,4202}}$ | $[[kk]_2\rho]_2$ | 1 | $[[kk]_0[kk]_2\rho]_2$ |

### B. Galilean and gauge invariance

In the previous Sec. III A, the functional has been required to be consistent with time reversal invariance, invariance under space reflections and rotational invariance. These constraints arise from symmetries of the NN interaction, see e.g. Refs. [31, 32]. Derivations above were much easier to perform in a general form, without imposing any other additional symmetry conditions. In this section, we treat such additional constraints coming from imposing the Galilean and gauge invariance.

Assumption of the Galilean instead of Lorentz invariance goes hand in hand with using the Schrödinger equation as a starting point and relies on the assumption that relativistic effects are negligible. This symmetry ensures that the collective translational mass, calculated within the time-dependent HF or random-phase approximations, is correctly equal to the total mass, $M = Am$.

TABLE XIV: Same as in Table XIII but for terms that are built from the vector nonlocal density $\boldsymbol{s}(\boldsymbol{r},\boldsymbol{r}')$.

| No. | $C^{n'L'v'J'}_{mI,nLvJ}$ | $\rho_{n'L'v'J'}$ | $D_{mI}$ | $\rho_{nLvJ}$ |
|---|---|---|---|---|
| 13 | $\boldsymbol{C^{1110}_{40,1110}}$ | $[ks]_0$ | $[\nabla\nabla]_0^2$ | $[ks]_0$ |
| 14 | $\boldsymbol{C^{1110}_{42,1112}}$ | $[ks]_0$ | $[\nabla\nabla]_0[\nabla\nabla]_2$ | $[ks]_2$ |
| 15 | $\boldsymbol{C^{1110}_{20,3110}}$ | $[ks]_0$ | $[\nabla\nabla]_0$ | $[[kk]_0ks]_0$ |
| 16 | $\boldsymbol{C^{1110}_{22,3112}}$ | $[ks]_0$ | $[\nabla\nabla]_2$ | $[[kk]_0ks]_2$ |
| 17 | $\boldsymbol{C^{1110}_{22,3312}}$ | $[ks]_0$ | $[\nabla\nabla]_2$ | $[[k[kk]_2]_3s]_2$ |
| 18 | $\boldsymbol{C^{1110}_{00,5110}}$ | $[ks]_0$ | 1 | $[[kk]_0^2ks]_0$ |
| 19 • | $\boldsymbol{C^{1111}_{40,1111}}$ | $[ks]_1$ | $[\nabla\nabla]_0^2$ | $[ks]_1$ |
| 20 • | $\boldsymbol{C^{1111}_{42,1111}}$ | $[ks]_1$ | $[\nabla\nabla]_0[\nabla\nabla]_2$ | $[ks]_1$ |
| 21 | $\boldsymbol{C^{1111}_{42,1112}}$ | $[ks]_1$ | $[\nabla\nabla]_0[\nabla\nabla]_2$ | $[ks]_2$ |
| 22 • | $\boldsymbol{C^{1111}_{20,3111}}$ | $[ks]_1$ | $[\nabla\nabla]_0$ | $[[kk]_0ks]_1$ |
| 23 • | $\boldsymbol{C^{1111}_{22,3111}}$ | $[ks]_1$ | $[\nabla\nabla]_2$ | $[[kk]_0ks]_1$ |
| 24 | $\boldsymbol{C^{1111}_{22,3112}}$ | $[ks]_1$ | $[\nabla\nabla]_2$ | $[[kk]_0ks]_2$ |
| 25 | $\boldsymbol{C^{1111}_{22,3312}}$ | $[ks]_1$ | $[\nabla\nabla]_2$ | $[[k[kk]_2]_3s]_2$ |
| 26 • | $\boldsymbol{C^{1111}_{22,3313}}$ | $[ks]_1$ | $[\nabla\nabla]_2$ | $[[k[kk]_2]_3s]_3$ |
| 27 • | $\boldsymbol{C^{1111}_{00,5111}}$ | $[ks]_1$ | 1 | $[[kk]_0^2ks]_1$ |

Therefore, in principle, the Galilean invariance should always be imposed. However, in many phenomenological approaches, like the non-interacting or interacting shell model, the Galilean symmetry is not considered, because the translational motion is not within the scope of such models. The question of whether the Galilean symmetry must be imposed in phenomenological models is not yet resolved, and in the present study we keep this question open.

For a local interaction, $v(\boldsymbol{r}_1',\boldsymbol{r}_2',\boldsymbol{r}_1,\boldsymbol{r}_2) = \delta(\boldsymbol{r}_1' - \boldsymbol{r}_1)\delta(\boldsymbol{r}_2' - \boldsymbol{r}_2)v(\boldsymbol{r}_1,\boldsymbol{r}_2)$, the HF interaction energy is invariant with respect to the local gauge [36]. Therefore, for the total energy (1) obtained within the EDF method, one may also consider constraints resulting from assuming local gauge invariance. As mentioned, this symmetry is only fulfilled when the forces involved are local. An example of a local approximation is the well known local one pion-exchange (OPE) potential, which is only an approximate representation of the correct nonlocal OPE Feynman amplitude [33]. This approximation is good as long as the relative momenta of interacting particles are about the same in the initial and final states (see Fig. 10 in Ref. [33]). Some of the fitted NN potentials like the Argonne $V_{18}$, Nijm-II, and Reid93 use this local approximation while others (CD-Bonn) use the full nonlocal OPE amplitude [33]. The most important nonlocal term is the two-body spin orbit interaction [32], which violates the assumption of gauge invariance. However, in this case a gauge invariant spin-orbit term (used in the Skyrme and Gogny forces) can be obtained in the short-range limit [32, 34, 36].

For the EDF derived in this work it is, however, the symmetries of effective forces rather than bare forces that should be considered. One of the methods to obtain an

TABLE XIV: continued.

| No. | $C^{n'L'v'J'}_{mI,nLvJ}$ | $\rho_{n'L'v'J'}$ | $D_{mI}$ | $\rho_{nLvJ}$ |
|---|---|---|---|---|
| 28 | $\boldsymbol{C^{1112}_{40,1112}}$ | $[ks]_2$ | $[\nabla\nabla]^2_0$ | $[ks]_2$ |
| 29 | $\boldsymbol{C^{1112}_{42,1112}}$ | $[ks]_2$ | $[\nabla\nabla]_0[\nabla\nabla]_2$ | $[ks]_2$ |
| 30 | $\boldsymbol{C^{1112}_{44,1112}}$ | $[ks]_2$ | $[\nabla[\nabla[\nabla\nabla]_2]_3]_4$ | $[ks]_2$ |
| 31 | $\boldsymbol{C^{1112}_{22,3110}}$ | $[ks]_2$ | $[\nabla\nabla]_2$ | $[[kk]_0ks]_0$ |
| 32 | $\boldsymbol{C^{1112}_{22,3111}}$ | $[ks]_2$ | $[\nabla\nabla]_2$ | $[[kk]_0ks]_1$ |
| 33 | $\boldsymbol{C^{1112}_{20,3112}}$ | $[ks]_2$ | $[\nabla\nabla]_0$ | $[[kk]_0ks]_2$ |
| 34 | $\boldsymbol{C^{1112}_{22,3112}}$ | $[ks]_2$ | $[\nabla\nabla]_2$ | $[[kk]_0ks]_2$ |
| 35 | $\boldsymbol{C^{1112}_{20,3312}}$ | $[ks]_2$ | $[\nabla\nabla]_0$ | $[[k[kk]_2]_3s]_2$ |
| 36 | $\boldsymbol{C^{1112}_{22,3312}}$ | $[ks]_2$ | $[\nabla\nabla]_2$ | $[[k[kk]_2]_3s]_2$ |
| 37 | $\boldsymbol{C^{1112}_{22,3313}}$ | $[ks]_2$ | $[\nabla\nabla]_2$ | $[[k[kk]_2]_3s]_3$ |
| 38 | $\boldsymbol{C^{1112}_{22,3314}}$ | $[ks]_2$ | $[\nabla\nabla]_2$ | $[[k[kk]_2]_3s]_4$ |
| 39 | $\boldsymbol{C^{1112}_{00,5112}}$ | $[ks]_2$ | 1 | $[[kk]^2_0ks]_2$ |
| 40 | $\boldsymbol{C^{1112}_{00,5312}}$ | $[ks]_2$ | 1 | $[[kk]_0[k[kk]_2]_3s]_2$ |
| 41 | $\boldsymbol{C^{3110}_{00,3110}}$ | $[[kk]_0ks]_0$ | 1 | $[[kk]_0ks]_0$ |
| 42 • | $\boldsymbol{C^{3111}_{00,3111}}$ | $[[kk]_0ks]_1$ | 1 | $[[kk]_0ks]_1$ |
| 43 | $\boldsymbol{C^{3112}_{00,3112}}$ | $[[kk]_0ks]_2$ | 1 | $[[kk]_0ks]_2$ |
| 44 | $\boldsymbol{C^{3112}_{00,3312}}$ | $[[kk]_0ks]_2$ | 1 | $[[k[kk]_2]_3s]_2$ |
| 45 | $\boldsymbol{C^{3312}_{00,3312}}$ | $[[k[kk]_2]_3s]_2$ | 1 | $[[k[kk]_2]_3s]_2$ |
| 46 • | $\boldsymbol{C^{3313}_{00,3313}}$ | $[[k[kk]_2]_3s]_3$ | 1 | $[[k[kk]_2]_3s]_3$ |
| 47 | $\boldsymbol{C^{3314}_{00,3314}}$ | $[[k[kk]_2]_3s]_4$ | 1 | $[[k[kk]_2]_3s]_4$ |

TABLE XV: Same as in Table XIII but for terms that are built from the scalar nonlocal density $\rho(\boldsymbol{r},\boldsymbol{r}')$ and vector nonlocal density $\boldsymbol{s}(\boldsymbol{r},\boldsymbol{r}')$.

| No. | $C^{n'L'v'J'}_{mI,nLvJ}$ | $\rho_{n'L'v'J'}$ | $D_{mI}$ | $\rho_{nLvJ}$ |
|---|---|---|---|---|
| 48 • | $\boldsymbol{C^{0000}_{51,1111}}$ | $[\rho]_0$ | $[\nabla\nabla]^2_0\nabla$ | $[ks]_1$ |
| 49 • | $\boldsymbol{C^{0000}_{31,3111}}$ | $[\rho]_0$ | $[\nabla\nabla]_0\nabla$ | $[[kk]_0ks]_1$ |
| 50 • | $\boldsymbol{C^{0000}_{33,3313}}$ | $[\rho]_0$ | $[\nabla[\nabla\nabla]_2]_3$ | $[[k[kk]_2]_3s]_3$ |
| 51 • | $\boldsymbol{C^{0000}_{11,5111}}$ | $[\rho]_0$ | $\nabla$ | $[[kk]^2_0ks]_1$ |
| 52 • | $\boldsymbol{C^{2000}_{31,1111}}$ | $[[kk]_0\rho]_0$ | $[\nabla\nabla]_0\nabla$ | $[ks]_1$ |
| 53 • | $\boldsymbol{C^{2000}_{11,3111}}$ | $[[kk]_0\rho]_0$ | $\nabla$ | $[[kk]_0ks]_1$ |
| 54 • | $\boldsymbol{C^{2202}_{31,1111}}$ | $[[kk]_2\rho]_2$ | $[\nabla\nabla]_0\nabla$ | $[ks]_1$ |
| 55 • | $\boldsymbol{C^{2202}_{33,1111}}$ | $[[kk]_2\rho]_2$ | $[\nabla[\nabla\nabla]_2]_3$ | $[ks]_1$ |
| 56 | $\boldsymbol{C^{2202}_{31,1112}}$ | $[[kk]_2\rho]_2$ | $[\nabla\nabla]_0\nabla$ | $[ks]_2$ |
| 57 | $\boldsymbol{C^{2202}_{33,1112}}$ | $[[kk]_2\rho]_2$ | $[\nabla[\nabla\nabla]_2]_3$ | $[ks]_2$ |
| 58 • | $\boldsymbol{C^{2202}_{11,3111}}$ | $[[kk]_2\rho]_2$ | $\nabla$ | $[[kk]_0ks]_1$ |
| 59 | $\boldsymbol{C^{2202}_{11,3112}}$ | $[[kk]_2\rho]_2$ | $\nabla$ | $[[kk]_0ks]_2$ |
| 60 | $\boldsymbol{C^{2202}_{11,3312}}$ | $[[kk]_2\rho]_2$ | $\nabla$ | $[[k[kk]_2]_3s]_2$ |
| 61 • | $\boldsymbol{C^{2202}_{11,3313}}$ | $[[kk]_2\rho]_2$ | $\nabla$ | $[[k[kk]_2]_3s]_3$ |
| 62 • | $\boldsymbol{C^{4000}_{11,1111}}$ | $[[kk]^2_0\rho]_0$ | $\nabla$ | $[ks]_1$ |
| 63 • | $\boldsymbol{C^{4202}_{11,1111}}$ | $[[kk]_0[kk]_2\rho]_2$ | $\nabla$ | $[ks]_1$ |
| 64 | $\boldsymbol{C^{4202}_{11,1112}}$ | $[[kk]_0[kk]_2\rho]_2$ | $\nabla$ | $[ks]_2$ |

TABLE XVI: Terms in the EDF (30) that are of sixth order, depend on time-odd densities, and are built from the scalar nonlocal density $\rho(\boldsymbol{r},\boldsymbol{r}')$.

| No. | $C^{n'L'v'J'}_{mI,nLvJ}$ | $\rho_{n'L'v'J'}$ | $D_{mI}$ | $\rho_{nLvJ}$ |
|---|---|---|---|---|
| 65 | $C^{1101}_{40,1101}$ | $[k\rho]_1$ | $[\nabla\nabla]^2_0$ | $[k\rho]_1$ |
| 66 | $C^{1101}_{42,1101}$ | $[k\rho]_1$ | $[\nabla\nabla]_0[\nabla\nabla]_2$ | $[k\rho]_1$ |
| 67 | $C^{1101}_{20,3101}$ | $[k\rho]_1$ | $[\nabla\nabla]_0$ | $[[kk]_0k\rho]_1$ |
| 68 | $C^{1101}_{22,3101}$ | $[k\rho]_1$ | $[\nabla\nabla]_2$ | $[[kk]_0k\rho]_1$ |
| 69 | $C^{1101}_{22,3303}$ | $[k\rho]_1$ | $[\nabla\nabla]_2$ | $[[k[kk]_2]_3\rho]_3$ |
| 70 | $C^{1101}_{00,5101}$ | $[k\rho]_1$ | 1 | $[[kk]^2_0k\rho]_1$ |
| 71 | $C^{3101}_{00,3101}$ | $[[kk]_0k\rho]_1$ | 1 | $[[kk]_0k\rho]_1$ |
| 72 | $C^{3303}_{00,3303}$ | $[[k[kk]_2]_3\rho]_3$ | 1 | $[[k[kk]_2]_3\rho]_3$ |

TABLE XVII: Same as in Table XVI but for terms that are built from the vector nonlocal density $\boldsymbol{s}(\boldsymbol{r},\boldsymbol{r}')$.

| No. | $C^{n'L'v'J'}_{mI,nLvJ}$ | $\rho_{n'L'v'J'}$ | $D_{mI}$ | $\rho_{nLvJ}$ |
|---|---|---|---|---|
| 73 | $C^{0011}_{60,0011}$ | $[s]_1$ | $[\nabla\nabla]^3_0$ | $[s]_1$ |
| 74 | $C^{0011}_{62,0011}$ | $[s]_1$ | $[\nabla\nabla]^2_0[\nabla\nabla]_2$ | $[s]_1$ |
| 75 | $C^{0011}_{40,2011}$ | $[s]_1$ | $[\nabla\nabla]^2_0$ | $[[kk]_0s]_1$ |
| 76 | $C^{0011}_{42,2011}$ | $[s]_1$ | $[\nabla\nabla]_0[\nabla\nabla]_2$ | $[[kk]_0s]_1$ |
| 77 | $C^{0011}_{40,2211}$ | $[s]_1$ | $[\nabla\nabla]^2_0$ | $[[kk]_2s]_1$ |
| 78 | $C^{0011}_{42,2211}$ | $[s]_1$ | $[\nabla\nabla]_0[\nabla\nabla]_2$ | $[[kk]_2s]_1$ |
| 79 | $C^{0011}_{42,2212}$ | $[s]_1$ | $[\nabla\nabla]_0[\nabla\nabla]_2$ | $[[kk]_2s]_2$ |
| 80 | $C^{0011}_{42,2213}$ | $[s]_1$ | $[\nabla\nabla]_0[\nabla\nabla]_2$ | $[[kk]_2s]_3$ |
| 81 | $C^{0011}_{44,2213}$ | $[s]_1$ | $[\nabla[\nabla[\nabla\nabla]_2]_3]_4$ | $[[kk]_2s]_3$ |
| 82 | $C^{0011}_{20,4011}$ | $[s]_1$ | $[\nabla\nabla]_0$ | $[[kk]^2_0s]_1$ |
| 83 | $C^{0011}_{22,4011}$ | $[s]_1$ | $[\nabla\nabla]_2$ | $[[kk]^2_0s]_1$ |
| 84 | $C^{0011}_{20,4211}$ | $[s]_1$ | $[\nabla\nabla]_0$ | $[[kk]_0[kk]_2s]_1$ |
| 85 | $C^{0011}_{22,4211}$ | $[s]_1$ | $[\nabla\nabla]_2$ | $[[kk]_0[kk]_2s]_1$ |
| 86 | $C^{0011}_{22,4212}$ | $[s]_1$ | $[\nabla\nabla]_2$ | $[[kk]_0[kk]_2s]_2$ |
| 87 | $C^{0011}_{22,4213}$ | $[s]_1$ | $[\nabla\nabla]_2$ | $[[kk]_0[kk]_2s]_3$ |
| 88 | $C^{0011}_{22,4413}$ | $[s]_1$ | $[\nabla\nabla]_2$ | $[[k[k[kk]_2]_3]_4s]_3$ |
| 89 | $C^{0011}_{00,6011}$ | $[s]_1$ | 1 | $[[kk]^3_0s]_1$ |
| 90 | $C^{0011}_{00,6211}$ | $[s]_1$ | 1 | $[[kk]^2_0[kk]_2s]_1$ |

effective NN force from the bare NN force is the unitary correlation operator method (UCOM) [35]. The use of the UCOM scheme, however, leads to a nonlocal effective interaction even if the bare interaction would have been a local one.

But rather than discussing to which extent gauge symmetry is conserved or broken in nuclei we aim to provide a theoretical framework where different choices can be accommodated. Because several successful phenomenological forces (e.g. Skyrme and Gogny [36]) are invariant under the gauge transformation, this symmetry constitutes a natural starting point in the search of improved EDFs. To which extent gauge symmetry is violated for effective renormalized interactions is a question that can be investigated by comparing models using preserved and broken symmetries (see e.g. Ref. [37]).

TABLE XVII: continued.

| No. | $C^{n'L'v'J'}_{mI,nLvJ}$ | $\rho_{n'L'v'J'}$ | $D_{mI}$ | $\rho_{nLvJ}$ |
|---|---|---|---|---|
| 91 | $C^{2011}_{20,2011}$ | $[[kk]_0s]_1$ | $[\nabla\nabla]_0$ | $[[kk]_0s]_1$ |
| 92 | $C^{2011}_{22,2011}$ | $[[kk]_0s]_1$ | $[\nabla\nabla]_2$ | $[[kk]_0s]_1$ |
| 93 | $C^{2011}_{20,2211}$ | $[[kk]_0s]_1$ | $[\nabla\nabla]_0$ | $[[kk]_2s]_1$ |
| 94 | $C^{2011}_{22,2211}$ | $[[kk]_0s]_1$ | $[\nabla\nabla]_2$ | $[[kk]_2s]_1$ |
| 95 | $C^{2011}_{22,2212}$ | $[[kk]_0s]_1$ | $[\nabla\nabla]_2$ | $[[kk]_2s]_2$ |
| 96 | $C^{2011}_{22,2213}$ | $[[kk]_0s]_1$ | $[\nabla\nabla]_2$ | $[[kk]_2s]_3$ |
| 97 | $C^{2011}_{00,4011}$ | $[[kk]_0s]_1$ | 1 | $[[kk]_0^2s]_1$ |
| 98 | $C^{2011}_{00,4211}$ | $[[kk]_0s]_1$ | 1 | $[[kk]_0[kk]_2s]_1$ |
| 99 | $C^{2211}_{20,2211}$ | $[[kk]_2s]_1$ | $[\nabla\nabla]_0$ | $[[kk]_2s]_1$ |
| 100 | $C^{2211}_{22,2211}$ | $[[kk]_2s]_1$ | $[\nabla\nabla]_2$ | $[[kk]_2s]_1$ |
| 101 | $C^{2211}_{22,2212}$ | $[[kk]_2s]_1$ | $[\nabla\nabla]_2$ | $[[kk]_2s]_2$ |
| 102 | $C^{2211}_{22,2213}$ | $[[kk]_2s]_1$ | $[\nabla\nabla]_2$ | $[[kk]_2s]_3$ |
| 103 | $C^{2211}_{00,4011}$ | $[[kk]_2s]_1$ | 1 | $[[kk]_0^2s]_1$ |
| 104 | $C^{2211}_{00,4211}$ | $[[kk]_2s]_1$ | 1 | $[[kk]_0[kk]_2s]_1$ |
| 105 | $C^{2212}_{20,2212}$ | $[[kk]_2s]_2$ | $[\nabla\nabla]_0$ | $[[kk]_2s]_2$ |
| 106 | $C^{2212}_{22,2212}$ | $[[kk]_2s]_2$ | $[\nabla\nabla]_2$ | $[[kk]_2s]_2$ |
| 107 | $C^{2212}_{22,2213}$ | $[[kk]_2s]_2$ | $[\nabla\nabla]_2$ | $[[kk]_2s]_3$ |
| 108 | $C^{2212}_{00,4212}$ | $[[kk]_2s]_2$ | 1 | $[[kk]_0[kk]_2s]_2$ |
| 109 | $C^{2213}_{20,2213}$ | $[[kk]_2s]_3$ | $[\nabla\nabla]_0$ | $[[kk]_2s]_3$ |
| 110 | $C^{2213}_{22,2213}$ | $[[kk]_2s]_3$ | $[\nabla\nabla]_2$ | $[[kk]_2s]_3$ |
| 111 | $C^{2213}_{00,4213}$ | $[[kk]_2s]_3$ | 1 | $[[kk]_0[kk]_2s]_3$ |
| 112 | $C^{2213}_{00,4413}$ | $[[kk]_2s]_3$ | 1 | $[[k[k[kk]_2]_3]_4s]_3$ |

TABLE XVIII: Same as in Table XVI but for terms that are built from the scalar nonlocal density $\rho(\boldsymbol{r},\boldsymbol{r}')$ and vector nonlocal density $\boldsymbol{s}(\boldsymbol{r},\boldsymbol{r}')$.

| No. | $C^{n'L'v'J'}_{mI,nLvJ}$ | $\rho_{n'L'v'J'}$ | $D_{mI}$ | $\rho_{nLvJ}$ |
|---|---|---|---|---|
| 113 | $C^{1101}_{51,0011}$ | $[k\rho]_1$ | $[\nabla\nabla]_0^2\nabla$ | $[s]_1$ |
| 114 | $C^{1101}_{31,2011}$ | $[k\rho]_1$ | $[\nabla\nabla]_0\nabla$ | $[[kk]_0s]_1$ |
| 115 | $C^{1101}_{31,2211}$ | $[k\rho]_1$ | $[\nabla\nabla]_0\nabla$ | $[[kk]_2s]_1$ |
| 116 | $C^{1101}_{31,2212}$ | $[k\rho]_1$ | $[\nabla\nabla]_0\nabla$ | $[[kk]_2s]_2$ |
| 117 | $C^{1101}_{33,2212}$ | $[k\rho]_1$ | $[\nabla[\nabla\nabla]_2]_3$ | $[[kk]_2s]_2$ |
| 118 | $C^{1101}_{33,2213}$ | $[k\rho]_1$ | $[\nabla[\nabla\nabla]_2]_3$ | $[[kk]_2s]_3$ |
| 119 | $C^{1101}_{11,4011}$ | $[k\rho]_1$ | $\nabla$ | $[[kk]_0^2s]_1$ |
| 120 | $C^{1101}_{11,4211}$ | $[k\rho]_1$ | $\nabla$ | $[[kk]_0[kk]_2s]_1$ |
| 121 | $C^{1101}_{11,4212}$ | $[k\rho]_1$ | $\nabla$ | $[[kk]_0[kk]_2s]_2$ |
| 122 | $C^{3101}_{31,0011}$ | $[[kk]_0k\rho]_1$ | $[\nabla\nabla]_0\nabla$ | $[s]_1$ |
| 123 | $C^{3101}_{11,2011}$ | $[[kk]_0k\rho]_1$ | $\nabla$ | $[[kk]_0s]_1$ |
| 124 | $C^{3101}_{11,2211}$ | $[[kk]_0k\rho]_1$ | $\nabla$ | $[[kk]_2s]_1$ |
| 125 | $C^{3101}_{11,2212}$ | $[[kk]_0k\rho]_1$ | $\nabla$ | $[[kk]_2s]_2$ |
| 126 | $C^{3303}_{33,0011}$ | $[[k[kk]_2]_3\rho]_3$ | $[\nabla[\nabla\nabla]_2]_3$ | $[s]_1$ |
| 127 | $C^{3303}_{11,2212}$ | $[[k[kk]_2]_3\rho]_3$ | $\nabla$ | $[[kk]_2s]_2$ |
| 128 | $C^{3303}_{11,2213}$ | $[[k[kk]_2]_3\rho]_3$ | $\nabla$ | $[[kk]_2s]_3$ |
| 129 | $C^{5101}_{11,0011}$ | $[[kk]_0^2k\rho]_1$ | $\nabla$ | $[s]_1$ |

TABLE XIX: Numbers of local densities $\rho_{nLvJ}$, Eq. (23), of different orders, which enter into the EDF up to N$^3$LO. Numbers of local densities constructed from the scalar $\rho(\boldsymbol{r},\boldsymbol{r}')$ or vector $\boldsymbol{s}(\boldsymbol{r},\boldsymbol{r}')$ nonlocal densities, and numbers of time-even and time-odd local densities, are given separately. In Tables III and IV these densities are marked with stars ($\star$).

| order | from $\rho$ | from $\boldsymbol{s}$ | T-even | T-odd | total |
|---|---|---|---|---|---|
| 0 | 1 | 1 | 1 | 1 | 2 |
| 1 | 1 | 3 | 3 | 1 | 4 |
| 2 | 2 | 4 | 2 | 4 | 6 |
| 3 | 2 | 6 | 6 | 2 | 8 |
| 4 | 2 | 5 | 2 | 5 | 7 |
| 5 | 1 | 4 | 4 | 1 | 5 |
| 6 | 1 | 2 | 1 | 2 | 3 |
| total | 10 | 25 | 19 | 16 | 35 |

### 1. Local gauge transformations of the nonlocal densities

The gauge-transformed nonlocal densities read [24, 25, 36]

$$\rho'(\boldsymbol{r},\boldsymbol{r}') = e^{i\left[\phi(\boldsymbol{r})-\phi(\boldsymbol{r}')\right]}\rho(\boldsymbol{r},\boldsymbol{r}'), \tag{34}$$

$$\boldsymbol{s}'(\boldsymbol{r},\boldsymbol{r}') = e^{i\left[\phi(\boldsymbol{r})-\phi(\boldsymbol{r}')\right]}\boldsymbol{s}(\boldsymbol{r},\boldsymbol{r}'). \tag{35}$$

Since the local gauge transformations form a U(1) group, invariance with respect to transformations that are of the first-order in gauge angles, $[1+iG]\,\rho(\boldsymbol{r},\boldsymbol{r}')$, where

$$G(\boldsymbol{r},\boldsymbol{r}') = \phi(\boldsymbol{r}) - \phi(\boldsymbol{r}'), \tag{36}$$

is enough to ensure full gauge invariance. By Taylor expanding the exponential functions in eq. 34 and 35 after they are inserted in the functional one may, of course, also prove this fact explicitly.

One specific type of gauge transformation is the Galilean transformation, for which the gauge angles depend linearly on positions, i.e., $G(\boldsymbol{r},\boldsymbol{r}') = \boldsymbol{p}\cdot(\boldsymbol{r}-\boldsymbol{r}')/\hbar$, and which corresponds to a transformation to a reference frame that moves with velocity $\boldsymbol{p}/m$. For this transformation, only first-order derivatives of $G$ survive, which makes Galilean invariance less restrictive than the full gauge invariance.

### 2. Local gauge transformations of the local densities

Let indices $\beta,\gamma,\ldots = 1,\ldots,35$ label primary local densities $\rho_{nLvJ}$ (23) listed with stars ($\star$) in Tables III and IV, which enter the EDF at N$^3$LO, as shown in Eqs. (30)

and (31). Using this notation, the linearized gauge transformation of one of the local densities can be written as

$$\begin{aligned} \rho'_\beta(\boldsymbol{r}) &= \left\{\left[K_{nL}\left(1+iG(\boldsymbol{r},\boldsymbol{r}')\right)\rho_v(\boldsymbol{r},\boldsymbol{r}')\right]_J\right\}_{\boldsymbol{r}=\boldsymbol{r}'} \\ &= \rho_\beta(\boldsymbol{r}) + \left\{\left[K_{nL}iG(\boldsymbol{r},\boldsymbol{r}')\rho_v(\boldsymbol{r},\boldsymbol{r}')\right]_J\right\}_{\boldsymbol{r}=\boldsymbol{r}'} \\ &= \rho_\beta(\boldsymbol{r}) + \rho^G_\beta(\boldsymbol{r}), \end{aligned} \tag{37}$$

where the first term is the untransformed local density and the second term is the part affected by the gauge transformation.

As an illustration, let us begin by considering the simpler case of Galilean transformation, and look at the term with $n=2$, where only two relative momentum operators $k$ appear. Operator $k$ can be written as $k_\rho + k_G$, with the first term acting only on $\rho_v(\boldsymbol{r},\boldsymbol{r}')$ and the second term acting only on $G(\boldsymbol{r},\boldsymbol{r}')$. Then we have.

$$K_{2L} = [kk]_L = [k_\rho k_\rho]_L + 2\,[k_\rho k_G]_L + [k_G k_G]_L\,. \tag{38}$$

When this is inserted into the expression for $\rho^G_\beta$, the last term can be dropped since only the first-order derivatives of $G(\boldsymbol{r},\boldsymbol{r}')$ survive for the Galilean transformation, and the first term disappears when one takes the limit of $\boldsymbol{r}=\boldsymbol{r}'$ since $G(\boldsymbol{r},\boldsymbol{r})=0$. Thus in this case we obtain

$$\begin{aligned} \rho^G_\beta(\boldsymbol{r}) &= i\left\{2\left[[k_\rho k_G]_l\, G(\boldsymbol{r},\boldsymbol{r}')\rho_v(\boldsymbol{r},\boldsymbol{r}')\right]_J,\right\}_{\boldsymbol{r}=\boldsymbol{r}'} \\ &= i\sum_{J'} c_{J'}\left[\left\{[k_\rho\rho_v]_{J'}\right\}_{\boldsymbol{r}=\boldsymbol{r}'}(\nabla\phi)(\boldsymbol{r})\right]_J, \end{aligned} \tag{39}$$

where the second equation results from the vector recoupling (note that $G(\boldsymbol{r},\boldsymbol{r}')$ is a scalar and $(\nabla\phi)(\boldsymbol{r})$ is a vector) and $c_{J'}$ are the ensuing numerical coefficients.

This example illustrates the main features of the derivation, namely, (i) for the Galilean transformations only terms with first-order derivatives of $G(\boldsymbol{r},\boldsymbol{r}')$ occur in the final expression for $\rho^G_\beta$, (ii) local densities appearing in the sum are of one order less in derivatives than the density being transformed, and (iii) the tensor order is preserved so that a local density is transformed into a sum of densities which can couple with a vector to the same tensor order. This leads to the ansatz for the Galilean transformation,

$$\rho^G_\beta(\boldsymbol{r}) = i\sum_\gamma c(\beta,\gamma)\left[\rho_\gamma(\boldsymbol{r})(\nabla\phi)(\boldsymbol{r})\right]_J, \tag{40}$$

where $c(\beta,\gamma)$ are numerical coefficients. Similarly, for the full gauge transformation the corresponding ansatz reads

$$\rho^G_\beta(\boldsymbol{r}) = i\sum_{\gamma mI} c_{mI}(\beta,\gamma)\left[\rho_\gamma(\boldsymbol{r})\left[D_{mI}\phi\right]_I(\boldsymbol{r})\right]_J. \tag{41}$$

In both cases, the numerical coefficients can be found by using the method outlined above, combined with a repeated use of the $6j$-symbols.

However, instead of using this method, it turned out to be more efficient to proceed in another way. First, by using symbolic programming [38], we constructed the transformed densities $\rho^G_\beta(\boldsymbol{r})$ explicitly in terms of derivatives of the density matrices and the gauge angle. Then, from the resulting expression the ansatz (40) or (41) was subtracted, which gave equations for the numerical coefficients by requesting that these differences must be identically equal to zero. Because these equations must hold for all density matrices and gauge angles, we could randomly assume arbitrary values for these quantities and their derivatives. In this way, all linearly independent equations for the coefficients could be obtained and solved analytically, again by using symbolic programming. The solutions were then double-checked by using the full forms of the densities.

### 3. Galilean or gauge invariant EDF

A Galilean or gauge-invariant EDF is the one which does not change upon inserting Galilean or gauge-transformed densities (37) into the energy density of Eq. (31). Since terms quadratic in $G(\boldsymbol{r},\boldsymbol{r}')$ can be dropped, the condition for the Galilean or gauge invariance reads

$$\int \mathrm{d}^3\boldsymbol{r}\sum C^\beta_{mI,\gamma}\left(\left[\rho^G_\beta(\boldsymbol{r})\left[D_{mI}\rho_\gamma(\boldsymbol{r})\right]_{J'}\right]_0 + \left[\rho_\beta(\boldsymbol{r})\left[D_{mI}\rho^G_\gamma(\boldsymbol{r})\right]_{J'}\right]_0\right) = 0, \tag{42}$$

where $C^\beta_{mI,\gamma}$ is a short-hand notation for the coupling constants $C^{n'L'v'J'}_{mI,nLvJ}$, and the sum runs over all the terms in the energy density.

The task now is to group together all proportional terms in Eq. (42). In doing so, we do not aim at obtaining an invariant energy density but an invariant EDF and total energy. Therefore, after densities (40) or (41) are inserted into Eq. (42), all terms must be integrated by parts to obtain some standard form, where terms equal through integration by parts become identical.

Finally, Eq. (42) can be transformed into a sum of independent terms using recoupling. In this expression each term is multiplied by a specific linear combination of coupling constants $C^{n'L'v'J'}_{mI,nLvJ}$. The Galilean or gauge invariance of the EDF then means that these linear com-binations must all vanish. This gives a set of linear equations that must be fulfilled for an invariant EDF. On the one hand, if a given coupling constant appears in none of these linear equations, the corresponding term of the EDF is invariant on its own, and the corresponding coupling constant is not restricted by the Galilean or gauge symmetry. On the other hand, for some coupling constants the only solution can be the value of zero, and then the corresponding term cannot appear in the invariant energy density.

Among all the remaining coupling constants, we may always select a subset of those that we will call the dependent ones, and express them as linear combinations of the other ones, which we will call the independent ones. This procedure is highly non-unique and can be realized in very many different ways, However, when the dependent coupling constants in function of independent ones are inserted back into the energy density (31), linear combinations of terms appearing at each independent coupling constant will all be invariant with respect to the Galilean or gauge transformations.

Then, the energy density of Eq. (31) takes the form

$$\mathcal{H}(\boldsymbol{r}) = \sum_{\substack{n'L'v'J' \\ mI,nLvJ,J'}} C^{n'L'v'J'}_{mI,nLvJ} G^{n'L'v'J'}_{mI,nLvJ}(\boldsymbol{r}), \tag{43}$$

where the sum runs over indices that correspond to unrestricted and independent coupling constants, which we jointly call free coupling constants. For a term in Eq. (43) that corresponds to an unrestricted coupling constant, we have $G^{n'L'v'J'}_{mI,nLvJ}(\boldsymbol{r}) = T^{n'L'v'J'}_{mI,nLvJ}(\boldsymbol{r})$, i.e., one term in the energy density of Eq. (31) is Galilean or gauge invariant. For a term in Eq. (43) that corresponds to an independent coupling constant, $G^{n'L'v'J'}_{mI,nLvJ}(\boldsymbol{r})$ is equal to a specific linear combination of terms $T$ from the original energy density (31). These linear combinations are listed in Appendix C.

We performed the analysis along these lines for energy densities of orders 0, 2 , 4, and 6, and the obtained results are listed in Appendix C. Derivations were performed by using symbolic programming [38] and employed the technique of forming linear equations by randomly assigning values to local densities and their derivatives, which we also used above. Numbers of linearly independent Galilean or gauge invariant terms are listed in Tables V and VI, and plotted in Fig. 1.

It turns out that only at orders 0 and 2, i.e., for the standard Skyrme functional, all Galilean invariant combinations of terms are also gauge invariant. At orders 4 and 6, there are only 6 gauge invariant terms available, while the numbers of Galilean invariant terms equal to 15 and 26, respectively. This is much less than the total numbers of terms at these orders, which are equal to 45 and 129, respectively. Altogether, at N$^3$LO we obtain the EDF parametrized in general by 188 coupling constants, and by 50 or 21 coupling constants if Galilean or gauge invariance is assumed. If isoscalar and isovector channels are included, all these numbers must be multiplied by a factor of two.

## IV. ENERGY DENSITY AT N$^3$LO WITH CONSERVED SPHERICAL, SPACE-INVERSION, AND TIME-REVERSAL SYMMETRIES

In this section, we apply the results obtained above to the simplest case of spherical even-even nuclei [28], where one can assume that the spherical symmetry, along with the space inversion and time reversal, are simultaneously conserved symmetries. In this case, all primary densities $\rho_{nLvJ}$ (23), which we listed in Tables III and IV, must have the form [39]:

$$\rho_{nLvJ}(\boldsymbol{r}) = R_{JJ}(\boldsymbol{r})\rho_{nLvJ}(|\boldsymbol{r}|), \tag{44}$$

where

$$R_{JJ}(\boldsymbol{r}) = [r[r\ldots,[rr]_2,\ldots,]_{J-2}]_J \tag{45}$$

is the $J$th-order, rank-$J$ stretched coupled tensor built from the position vector $\boldsymbol{r}$ in exactly the same way as the derivative operators $D_{nL}$ of Table I are built from the derivative $\boldsymbol{\nabla}$ in the spherical representation (9), and $\rho_{nLvJ}(|\boldsymbol{r}|)$ is a scalar function depending only on the length $|\boldsymbol{r}|$ of the position vector $\boldsymbol{r}$.

Indeed, due to the generalized Cayley-Hamilton (GCH) theorem, a rank-$J$ tensor function of a rank-$k$ tensor must be a linear combination of all independent rank-$J$ tensors built from that rank-$k$ tensor, with scalar coefficients. In the GCH theorem, tensors that differ by scalar factors are not independent. In our case, only one independent rank-$J$ function $R_{JJ}(\boldsymbol{r})$ can be built from the rank-1 tensor (position vector $\boldsymbol{r}$), which gives Eq. (44). The spherical symmetry assumed here is essential for this argument to work, because many more independent rank-$J$ tensors can be built when other "material" tensors (like, e.g., the quadrupole deformation tensor) are available.

The spherical form of $\rho_{nLvJ}(\boldsymbol{r})$ (44) requires that the following selection rule is obeyed:

$$P = (-1)^J, \tag{46}$$

where $P = (-1)^n$ is the space-inversion parity defined in Eq. (26). For the time-even densities ($T = 1$), selection rule (46) does not impose any new restriction on local densities built from $\rho\,(\boldsymbol{r},\boldsymbol{r}')$ ($v = 0$), see Table III. On the other hand, for local densities built from $\boldsymbol{s}\,(\boldsymbol{r},\boldsymbol{r}')$ ($v = 1$), see Table IV, only the densities with $L = J$ are allowed.

In Tables III and IV, all densities allowed by the conserved spherical, space-inversion, and time-reversal symmetries are marked with bullets (•). One can see that they correspond to quantum numbers $LvJ$ being equal to 000 or 202 [for densities built from $\rho\,(\boldsymbol{r},\boldsymbol{r}')$] and 111 or 313 [for densities built from $\boldsymbol{s}\,(\boldsymbol{r},\boldsymbol{r}')$]. Then, it is easy to select all allowed terms in the energy density—in Tables VII–XVIII and XXII these are also marked with bullets (•). Numbers of such terms are listed in Table XX

TABLE XX: Numbers of terms defined in Eq. (30) of different orders in the EDF up to N$^3$LO, evaluated for the conserved spherical, space-inversion, and time-reversal symmetries. The last two columns give numbers of terms when the Galilean or gauge invariance is assumed, respectively, see Sec. III B. To take into account both isospin channels, the numbers of terms should be multiplied by a factor of two.

| order | Total | Galilean | Gauge |
|---|---|---|---|
| 0 | 1 | 1 | 1 |
| 2 | 4 | 4 | 4 |
| 4 | 13 | 9 | 3 |
| 6 | 32 | 16 | 3 |
| N$^3$LO | 50 | 30 | 11 |

together with those obtained by imposing, in addition, the Galilean or gauge invariance.

All results for the EDF restricted by the spherical, space-inversion, and time-reversal symmetries can now be extracted from the general results presented in Secs. II and III and Appendices B and C. However, in the remainder of this section we give an example of how these results can be translated into those based on the Cartesian representations of derivative operators (18)–(22). Indeed, in this representation, all non-zero densities can be defined as:

$$R_0 = \rho, \tag{47}$$
$$R_2 = \boldsymbol{k}^2\rho, \tag{48}$$
$$\overleftrightarrow{R}_{2ab} = \boldsymbol{k}_a\boldsymbol{k}_b\rho, \tag{49}$$
$$R_4 = \boldsymbol{k}^4\rho, \tag{50}$$
$$\overleftrightarrow{R}_{4ab} = \boldsymbol{k}^2\boldsymbol{k}_a\boldsymbol{k}_b\rho, \tag{51}$$
$$R_6 = \boldsymbol{k}^6\rho, \tag{52}$$

and

$$\boldsymbol{J}_{1a} = (\boldsymbol{k}\times\boldsymbol{s})_a, \tag{53}$$
$$\boldsymbol{J}_{3a} = \boldsymbol{k}^2(\boldsymbol{k}\times\boldsymbol{s})_a, \tag{54}$$
$$\begin{aligned}\overleftrightarrow{J}_{3abc} &= \boldsymbol{k}_a\boldsymbol{k}_b(\boldsymbol{k}\times\boldsymbol{s})_c + \boldsymbol{k}_b\boldsymbol{k}_c(\boldsymbol{k}\times\boldsymbol{s})_a \\ &+ \boldsymbol{k}_c\boldsymbol{k}_a(\boldsymbol{k}\times\boldsymbol{s})_b,\end{aligned} \tag{55}$$
$$\boldsymbol{J}_{5a} = \boldsymbol{k}^4(\boldsymbol{k}\times\boldsymbol{s})_a, \tag{56}$$

where

$$\boldsymbol{k}^2 = \sum_a \boldsymbol{k}_a\boldsymbol{k}_a, \tag{57}$$

and the Cartesian indices are defined as $a, b, c = x, y, z$. To lighten the notation, in these definitions we have omitted the arguments of local densities, $\boldsymbol{r}$, and limits of $\boldsymbol{r}' = \boldsymbol{r}$.

The six local densities (47)–(52) are the Cartesian analogues of densities marked in Table III with bullets (•), and the four local densities (53)–(56) are analogues of those marked in Table IV. However, one should note that rank-2 densities $\overleftrightarrow{R}_{2ab}$ and $\overleftrightarrow{R}_{4ab}$ are not proportional to $\rho_{2202}$ and $\rho_{4202}$, respectively, and the rank-3 density $\overleftrightarrow{J}_{3abc}$ is not proportional to $\rho_{3313}$. This is so, because they are defined in terms of the derivative operators (18)–(22), where appropriate traces have not been subtracted out. Nevertheless, linear relations between densities (47)–(56) and their spherical-representation counterparts $\rho_{nLvJ}$ can easily be worked out and will not be presented here.

Note also that the scalar densities $R_2$ and $R_4$ can be expressed as the corresponding sums of the rank-2 densities $\overleftrightarrow{R}_{2ab}$ and $\overleftrightarrow{R}_{4ab}$, and the vector density $\boldsymbol{J}_{3a}$ as that of $\overleftrightarrow{J}_{3abc}$. However, based on the results obtained in the spherical representation, we know that they have to be treated separately to give separate terms in the energy density.

Again, based on the results obtained in the spherical representation, we can write the N$^3$LO energy density as a sum of contributions from zero, second, fourth, and sixth orders:

$$\mathcal{H} = \mathcal{H}_0 + \mathcal{H}_2 + \mathcal{H}_4 + \mathcal{H}_6, \tag{58}$$

where

$$\mathcal{H}_0 = C^0_{00}R_0R_0, \tag{59}$$

$$\begin{aligned}\mathcal{H}_2 &= C^0_{20}R_0\Delta R_0 + C^0_{02}R_0R_2 \\ &+ C^0_{11}R_0\boldsymbol{\nabla}\cdot\boldsymbol{J}_1, + C^1_{01}\boldsymbol{J}_1^2,\end{aligned} \tag{60}$$

Energy densities $\mathcal{H}_0$ and $\mathcal{H}_2$ correspond, of course, to the standard Skyrme functional [24, 36] with $C^0_{00} = C^\rho$, $C^0_{20} = C^{\Delta\rho} + \frac{1}{4}C^\tau$, $C^0_{02} = C^\tau$, $C^0_{11} = C^{\nabla J}$, and $C^1_{01} = C^{J1}$. At fourth and sixth orders, these energy densities read

$$\begin{aligned}\mathcal{H}_4 &= C^0_{40}R_0\Delta^2R_0 + C^0_{22}R_0\Delta R_2 \\ &+ C^0_{04}R_0R_4 + C^2_{02}R_2R_2 \\ &+ D^0_{22}R_0\sum_{ab}\boldsymbol{\nabla}_a\boldsymbol{\nabla}_b\overleftrightarrow{R}_{2ab} + D^2_{02}\sum_{ab}\overleftrightarrow{R}_{2ab}\overleftrightarrow{R}_{2ab} \\ &+ C^1_{21}\boldsymbol{J}_1\cdot\Delta\boldsymbol{J}_1 + C^1_{03}\boldsymbol{J}_1\cdot\boldsymbol{J}_3 \\ &+ D^1_{21}\boldsymbol{J}_1\cdot\boldsymbol{\nabla}(\boldsymbol{\nabla}\cdot\boldsymbol{J}_1) \\ &+ C^0_{31}R_0\Delta(\boldsymbol{\nabla}\cdot\boldsymbol{J}_1) + C^0_{13}R_0(\boldsymbol{\nabla}\cdot\boldsymbol{J}_3) \\ &+ C^2_{11}R_2(\boldsymbol{\nabla}\cdot\boldsymbol{J}_1) + D^2_{11}\sum_{ab}\overleftrightarrow{R}_{2ab}\boldsymbol{\nabla}_a\boldsymbol{J}_{1b},\end{aligned} \tag{61}$$

$$
\begin{aligned}
\mathcal{H}_6 &= C^0_{60}R_0\Delta^3R_0 + C^0_{42}R_0\Delta^2R_2 \\
&+ C^0_{24}R_0\Delta R_4 + C^0_{06}R_0R_6 \\
&+ C^2_{22}R_2\Delta R_2 + C^2_{04}R_2R_4 \\
&+ D^0_{42}R_0\Delta\sum_{ab}\boldsymbol{\nabla}_a\boldsymbol{\nabla}_b\overleftrightarrow{R}_{2ab} + D^0_{24}R_0\sum_{ab}\boldsymbol{\nabla}_a\boldsymbol{\nabla}_b\overleftrightarrow{R}_{4ab} \\
&+ D^2_{22}R_2\sum_{ab}\boldsymbol{\nabla}_a\boldsymbol{\nabla}_b\overleftrightarrow{R}_{2ab} + E^2_{22}\sum_{ab}\overleftrightarrow{R}_{2ab}\Delta\overleftrightarrow{R}_{2ab} \\
&+ F^2_{22}\sum_{abc}\overleftrightarrow{R}_{2ab}\boldsymbol{\nabla}_a\boldsymbol{\nabla}_c\overleftrightarrow{R}_{2cb} + E^2_{04}\sum_{ab}\overleftrightarrow{R}_{2ab}\overleftrightarrow{R}_{4ab} \\
&+ C^1_{41}\boldsymbol{J}_1\cdot\Delta^2\boldsymbol{J}_1 + C^1_{23}\boldsymbol{J}_1\cdot\Delta\boldsymbol{J}_3 \\
&+ C^1_{05}\boldsymbol{J}_1\cdot\boldsymbol{J}_5 + C^3_{03}\boldsymbol{J}_3\cdot\boldsymbol{J}_3 \\
&+ D^1_{41}\boldsymbol{J}_1\cdot\Delta\boldsymbol{\nabla}\left(\boldsymbol{\nabla}\cdot\boldsymbol{J}_1\right) + D^1_{23}\boldsymbol{J}_1\cdot\boldsymbol{\nabla}\left(\boldsymbol{\nabla}\cdot\boldsymbol{J}_3\right) \\
&+ E^1_{23}\sum_{abc}\boldsymbol{J}_{1a}\boldsymbol{\nabla}_b\boldsymbol{\nabla}_c\overleftrightarrow{J}_{3abc} + D^3_{03}\sum_{abc}\overleftrightarrow{J}_{3abc}\overleftrightarrow{J}_{3abc} \\
&+ C^0_{51}R_0\Delta^2\left(\boldsymbol{\nabla}\cdot\boldsymbol{J}_1\right) + C^0_{33}R_0\Delta\left(\boldsymbol{\nabla}\cdot\boldsymbol{J}_3\right) \\
&+ C^0_{15}R_0\left(\boldsymbol{\nabla}\cdot\boldsymbol{J}_5\right) + C^2_{31}R_2\Delta\left(\boldsymbol{\nabla}\cdot\boldsymbol{J}_1\right) \\
&+ C^2_{13}R_2\left(\boldsymbol{\nabla}\cdot\boldsymbol{J}_3\right) + C^4_{11}R_4\left(\boldsymbol{\nabla}\cdot\boldsymbol{J}_1\right) \\
&+ D^0_{33}R_0\sum_{abc}\boldsymbol{\nabla}_a\boldsymbol{\nabla}_b\boldsymbol{\nabla}_c\overleftrightarrow{J}_{3abc} + D^2_{13}\sum_{abc}\overleftrightarrow{R}_{2ab}\boldsymbol{\nabla}_c\overleftrightarrow{J}_{3abc} \\
&+ D^2_{31}\sum_{ab}\overleftrightarrow{R}_{2ab}\Delta\boldsymbol{\nabla}_a\boldsymbol{J}_{1b} + E^2_{13}\sum_{ab}\overleftrightarrow{R}_{2ab}\boldsymbol{\nabla}_a\boldsymbol{J}_{3b} \\
&+ D^4_{11}\sum_{ab}\overleftrightarrow{R}_{4ab}\boldsymbol{\nabla}_a\boldsymbol{J}_{1b} \\
&+ E^2_{31}\sum_{ab}\overleftrightarrow{R}_{2ab}\boldsymbol{\nabla}_a\boldsymbol{\nabla}_b\left(\boldsymbol{\nabla}\cdot\boldsymbol{J}_1\right).
\end{aligned}
\tag{62}
$$

The energy densities above are given in terms of the coupling constants $C^{n'}_{mn}$, $D^{n'}_{mn}$, $E^{n'}_{mn}$, and $F^{n'}_{mn}$. The indices correspond to orders of derivatives indicated in the same way as for the spherical-representation coupling constants $C^{n'L'v'J'}_{mI,nLvJ}$. Linear relations between both sets of coupling constants can easily be derived and are not reported here.

## V. CONCLUSIONS

In the present study, we constructed nuclear energy density functionals in terms of derivatives of densities up to sixth order. This constitutes the next-to-next-to-next-to-leading order (N$^3$LO) expansion of the functional, whereby, in this scheme, the contact and standard Skyrme forces provide the zero-order (LO) and second-order (NLO) expansions, respectively. The higher-order terms were built to provide tools for testing convergence properties of methods based on energy density functionals, within the spirit of effective field theories.

At N$^3$LO, depending on several options of using the energy density functionals, the numbers of free coupling constants are as follows. If one would like to include the density dependence of all the coupling constants (the option, which is not advocated here), one would have to use 658 different terms in the functional. Full functional with density-independent coupling constants contains 188 terms, while functionals restricted by Galilean and gauge symmetries contain 50 and 21 terms, respectively. If both isoscalar and isovector channels are included, all these numbers must be multiplied by a factor of two

At the present stage of searching for precise, spectroscopic-quality nuclear functionals, extensions beyond the standard Skyrme NLO form are mandatory, see the analysis in Ref. [40]. These may include richer density dependencies [41, 42], higher-order derivative terms. as constructed in the present study, terms of higher powers in densities, richer forms of functional dependence beyond simple power expansions, and possibly many other modifications.

Further studies of higher-order energy density functionals requires constructing appropriate codes to solve self-consistent equations. Although this is a complicated problem, various techniques have already been developed that can be used here. First of all, expressions for mean fields must be derived by using the standard methods presented, e.g., in Refs. [24, 25, 43]. Obviously, such mean fields will involve derivative operators up to sixth order, so the connection with the one-body Schrödinger equation, discussed, e.g., in Ref. [28], will be lost. Nevertheless, all basis-expansions methods can still be used and their implementation will not be basically different than it was done up to now at NLO. Work along these lines is now in progress.

This work was supported in part by the Academy of Finland and University of Jyväskylä within the FIDIPRO programme, by the Polish Ministry of Science under Contract No. N N202 328234, and by the UNEDF SciDAC Collaboration under the U.S. Department of Energy grant No. DE-FC02-07ER41457.

## APPENDIX A: SYMMETRIES

Within the HF approximation, symmetry properties of the interaction carry over to symmetry properties of the total HF energy. It means that whenever the interaction is invariant with respect to a symmetry group, the total HF energy is invariant with respect to transforming densities (or density matrices, in general) by the same symmetry group. However, the well-known phenomenon of the spontaneous symmetry breaking [32] may render densities and energy density themselves not invariant with respect to the symmetry group in question. We must, therefore, consider energy densities for symmetry-breaking densities.

This is best illustrated by the EDF derived for the Skyrme interaction [25, 28]. The standard derivation treats the time-reversal and isospin symmetries in a different way than space symmetries (space inversion and other point symmetries or space rotation) [24, 25, 36]. Indeed, for the time reversal, the nonlocal densities are first split into the time-even and time-odd parts as

$$\begin{aligned}\rho(\boldsymbol{r},\boldsymbol{r}') &= \rho_+(\boldsymbol{r},\boldsymbol{r}')+\rho_-(\boldsymbol{r},\boldsymbol{r}'), && \text{(A1)}\\ \boldsymbol{s}(\boldsymbol{r},\boldsymbol{r}') &= \boldsymbol{s}_+(\boldsymbol{r},\boldsymbol{r}')+\boldsymbol{s}_-(\boldsymbol{r},\boldsymbol{r}'), && \text{(A2)}\end{aligned}$$

where

$$\begin{aligned}\rho_\pm(\boldsymbol{r},\boldsymbol{r}') &= \tfrac{1}{2}\left[\rho(\boldsymbol{r},\boldsymbol{r}')\pm\rho^T(\boldsymbol{r},\boldsymbol{r}')\right], && \text{(A3)}\\ \boldsymbol{s}_\pm(\boldsymbol{r},\boldsymbol{r}') &= \tfrac{1}{2}\left[\boldsymbol{s}(\boldsymbol{r},\boldsymbol{r}')\pm\boldsymbol{s}^T(\boldsymbol{r},\boldsymbol{r}')\right], && \text{(A4)}\end{aligned}$$

such that

$$\begin{aligned}\rho^T_\pm(\boldsymbol{r},\boldsymbol{r}') &= \pm\rho_\pm(\boldsymbol{r},\boldsymbol{r}'), && \text{(A5)}\\ \boldsymbol{s}^T_\pm(\boldsymbol{r},\boldsymbol{r}') &= \pm\boldsymbol{s}_\pm(\boldsymbol{r},\boldsymbol{r}'). && \text{(A6)}\end{aligned}$$

The superscript $T$ here means that the nonlocal densities are calculated for the time-reversed many-body states.

Then, in the derivation of the HF energy density, only squares of the time-even and time-odd densities appear, because, due to the time-reversal symmetry of the interaction, the cross terms do not contribute. As a consequence, the energy density itself is time-even. Similarly, for the isospin symmetry, the densities are first split into the isoscalar and isovector parts, for which no cross terms contribute, and the obtained energy density is an isoscalar. In what follows, we call this kind of derivation 'derivation after separation of symmetries', which implies that the symmetry-breaking terms are absent in the energy density.

The derivation after separation of symmetries can be illustrated by considering the simplest term of the Skyrme interaction – just the contact force,

$$V_\delta(\boldsymbol{r}_1,\boldsymbol{r}_2)=t_0\delta(\boldsymbol{r}_1-\boldsymbol{r}_2), \quad \text{(A7)}$$

for which the energy density reads (we neglected the isospin degree of freedom, so only one type of particles is considered),

$$\mathcal{H}_\delta(\boldsymbol{r})=\tfrac{1}{2}t_0\rho^2(\boldsymbol{r})-\tfrac{1}{2}t_0\boldsymbol{s}^2(\boldsymbol{r}). \quad \text{(A8)}$$

This energy density is invariant with respect to the time reversal of local densities, $\rho^T(\boldsymbol{r})=\rho(\boldsymbol{r})$ and $\boldsymbol{s}^T(\boldsymbol{r})=-\boldsymbol{s}(\boldsymbol{r})$.

Here, the coupling constant multiplying the time-even density, $C^\rho=\frac{1}{2}t_0$, is not independent of the coupling constant multiplying the time-odd density, $C^s=-\frac{1}{2}t_0$. This fact is not related to the time-reversal symmetry but results from the vanishing range of the contact force. Proper treatment of the finite range corrections render these two coupling constants independent of one another. [15, 16]. Irrespective of zero or finite range, the isovector and isoscalar coupling constants are also independent of one another [24, 36].

For space symmetries, the standard derivation proceeds in another way, namely, the energy density is determined directly for the broken-symmetry HF state. For the space-inversion symmetry, for example, this means that both parity-even and parity-odd densities,

$$\begin{aligned}\rho_{P=\pm1}(\boldsymbol{r}) &= \tfrac{1}{2}\left[\rho(\boldsymbol{r})\pm\rho^P(\boldsymbol{r})\right]=\tfrac{1}{2}\left[\rho(\boldsymbol{r})\pm\rho(-\boldsymbol{r})\right], && \text{(A9)}\\ \boldsymbol{s}_{P=\pm1}(\boldsymbol{r}) &= \tfrac{1}{2}\left[\boldsymbol{s}(\boldsymbol{r})\pm\boldsymbol{s}^P(\boldsymbol{r})\right]=\tfrac{1}{2}\left[\boldsymbol{s}(\boldsymbol{r})\pm\boldsymbol{s}(-\boldsymbol{r})\right], && \text{(A10)}\end{aligned}$$

appear in $\rho(\boldsymbol{r})$ and $\boldsymbol{s}(\boldsymbol{r})$ in the energy density of Eq. (A8),

$$\begin{aligned}\rho(\boldsymbol{r}) &= \rho_{P=+1}(\boldsymbol{r})+\rho_{P=-1}(\boldsymbol{r}), && \text{(A11)}\\ \boldsymbol{s}(\boldsymbol{r}) &= \boldsymbol{s}_{P=+1}(\boldsymbol{r})+\boldsymbol{s}_{P=-1}(\boldsymbol{r}). && \text{(A12)}\end{aligned}$$

The superscript $P$ here means that the nonlocal and local densities are calculated for the space-inversed many-body states. We call this kind of derivation 'derivation before separation of symmetries', which implies that the symmetry-breaking terms are then explicitly present in the energy density. If the symmetry is broken, which in the case of space inversion corresponds to $\rho_{P=-1}(\boldsymbol{r})\neq 0$ or $\boldsymbol{s}_{P=-1}(\boldsymbol{r})\neq 0$, then the energy density is *not* invariant with respect to space inversion.

The total energy, i.e., the integral of the energy density (1) is, of course, invariant with respect to space inversion, because the integration then picks up only the space-inversion-invariant parts of the integrand. Therefore, the energy densities derived before and after separation of symmetries are not equal, but they are equivalent. In the case of the space-inversion symmetry, the energy density (A8) derived after separation of symmetries reads

$$\begin{aligned}\mathcal{H}'_\delta(\boldsymbol{r}) &= \tfrac{1}{2}t_0\rho^2_{P=+1}(\boldsymbol{r})+\tfrac{1}{2}t_0\rho^2_{P=-1}(\boldsymbol{r})\\ &- \tfrac{1}{2}t_0\boldsymbol{s}^2_{P=+1}(\boldsymbol{r})-\tfrac{1}{2}t_0\boldsymbol{s}^2_{P=-1}(\boldsymbol{r}). && \text{(A13)}\end{aligned}$$

This energy density is invariant with respect to space inversion and the coupling constants multiplying densities of opposite parities are not independent of one another (see also discussion in Ref. [30]).

The same principle applies to other broken spatial symmetries. For example, for the broken rotational symmetry, the density can be split into the sum of terms belonging to different irreducible representations of the rotational group, which in this case corresponds to the standard multipole series [44],

$$\rho(\boldsymbol{r})=\sum_{\lambda\mu}\rho_{\lambda\mu}(\boldsymbol{r}), \quad \text{(A14)}$$

for

$$\rho_{\lambda\mu}(\boldsymbol{r})=\rho_{\lambda\mu}(r)Y^*_{\lambda\mu}(\theta,\phi), \quad \text{(A15)}$$

where the multipole densities $\rho_{\lambda\mu}(r)$,

$$\rho_{\lambda\mu}(r)=\int \mathrm{d}\theta\mathrm{d}\phi\rho(\boldsymbol{r})Y_{\lambda\mu}(\theta,\phi), \quad \text{(A16)}$$

depend only on the radial coordinate $r$. Then, the first term in the energy density (A8), which is derived before separation of rotational symmetries, and which is *not* rotationally invariant, is equivalent to the following energy density derived after separation of rotational symmetries,

$$\mathcal{H}_\rho(\boldsymbol{r})=\tfrac{1}{2}t_0\sum_\lambda\sqrt{2\lambda+1}[\rho_\lambda(\boldsymbol{r})\rho_\lambda(\boldsymbol{r})]_0, \quad \text{(A17)}$$

(see Ref. [45] for an example application of this series). This energy density is rotationally invariant and the coupling constants multiplying different multipole densities are again not independent of one another.

We have presented a detailed analysis of the problem to arm ourselves with proper tools for discussing construction of EDFs in situations where there is no underlying interaction. Then, the only consideration is the requirement of invariance of the total energy with respect to all symmetries usually conserved by nuclear interactions. We proceed with such a construction in two different ways as described below.

#### 1. Symmetry-invariant energy density

Based on the derivation after separation of symmetries, which we introduced above, it is clear that we can proceed by separating densities into irreducible representations of all required symmetries and then building the energy density by taking scalar products separately in each of the representations. Such a construction gives a symmetry-invariant energy density,

$$\mathcal{H}^S(\boldsymbol{r}) = \mathcal{H}(\boldsymbol{r}), \tag{A18}$$

where $\mathcal{H}^S(\boldsymbol{r})$ denotes the energy density calculated for a many-body state transformed by the symmetry operator $S$. This guarantees the invariance of the EDF and total energy (1) with respect to all considered symmetries. Such a strategy would also allow for using arbitrary, unrelated to one another coupling constants in each of the irreducible representations.

However, in practical applications, such a strategy was never up to now fully implemented—neither at N$^3$LO, for which the present study is the first attempt in the literature, nor at NLO, which corresponds to the standard Skyrme functionals (see Sec. III A below). Only the time reversal and isospin symmetries were up to now treated in this way, and below we are going to follow the same path.

For the time reversal, all local densities discussed in Sec. II C are either time-even or time-odd. Indeed, this simply follows from the facts [25] that

$$\begin{array}{rlll} \rho^T(\boldsymbol{r},\boldsymbol{r}') &= \phantom{-}\rho^*(\boldsymbol{r},\boldsymbol{r}') &= \phantom{-}\rho(\boldsymbol{r}',\boldsymbol{r}), \\ \boldsymbol{s}^T(\boldsymbol{r},\boldsymbol{r}') &= -\ \boldsymbol{s}^*(\boldsymbol{r},\boldsymbol{r}') &= -\ \boldsymbol{s}(\boldsymbol{r}',\boldsymbol{r}), \end{array} \tag{A19}$$

which give the time-even and time-odd parts in Eqs. (A3) and (A4) as

$$\begin{array}{rlll} \rho_+(\boldsymbol{r},\boldsymbol{r}') &= \phantom{-}\rho_+^*(\boldsymbol{r},\boldsymbol{r}') &= \phantom{-}\rho_+(\boldsymbol{r}',\boldsymbol{r}), \\ \rho_-(\boldsymbol{r},\boldsymbol{r}') &= -\ \rho_-^*(\boldsymbol{r},\boldsymbol{r}') &= -\ \rho_-(\boldsymbol{r}',\boldsymbol{r}), \\ \boldsymbol{s}_+(\boldsymbol{r},\boldsymbol{r}') &= -\ \boldsymbol{s}_+^*(\boldsymbol{r},\boldsymbol{r}') &= -\ \boldsymbol{s}_+(\boldsymbol{r}',\boldsymbol{r}), \\ \boldsymbol{s}_-(\boldsymbol{r},\boldsymbol{r}') &= \phantom{-}\boldsymbol{s}_-^*(\boldsymbol{r},\boldsymbol{r}') &= \phantom{-}\boldsymbol{s}_-(\boldsymbol{r}',\boldsymbol{r}), \end{array} \tag{A20}$$

i.e., $\rho_+(\boldsymbol{r},\boldsymbol{r}')$ and $\boldsymbol{s}_-(\boldsymbol{r},\boldsymbol{r}')$ are real symmetric functions and $\rho_-(\boldsymbol{r},\boldsymbol{r}')$ and $\boldsymbol{s}_+(\boldsymbol{r},\boldsymbol{r}')$ are imaginary antisymmetric functions. Moreover, the relative momentum operator $\boldsymbol{k}$ (6), which defines derivative operators $K_{nL}$, is imaginary and antisymmetric with respect to exchanging variables $\boldsymbol{r}$ and $\boldsymbol{r}'$. Altogether, it is easy to see that $T$-parities of primary densities $\rho_{nLvJ}(\boldsymbol{r})$ (23) are equal to $(-1)^{n+v}$, see Eq. (25) and columns denoted by $T$ in Tables III and IV. Similarly, $T$-parities of secondary densities $\rho_{mI,nLvJ,Q}(\boldsymbol{r})$ (24) are also equal to $(-1)^{n+v}$. Construction of the $T$-invariant energy density (A18) can now be realized by multiplying densities that have identical $T$-parities.

#### 2. Symmetry-covariant energy density

Treatment of space symmetries in the construction of EDFs is another kettle of fish. Here, we base our considerations on the derivation before separation of symmetries, which we introduced above, and on the fact that invariance of the energy density itself is not a prerequisite for the invariance of the EDF. In fact, the EDF and total energy (1) are invariant with respect to symmetry $S$ also when the energy density is covariant with $S$, i.e.,

$$\mathcal{H}^S(\boldsymbol{r}) = \mathcal{H}(S^+\boldsymbol{r}S), \tag{A21}$$

where $S^+\boldsymbol{r}S$ denotes the space point transformed by symmetry $S$ (see also discussion in Ref. [32]). Indeed, due to the fact that space integrals are invariant, we have

$$\int \mathrm{d}^3\boldsymbol{r}\mathcal{H}(S^+\boldsymbol{r}S) = \int \mathrm{d}^3r\mathcal{H}(\boldsymbol{r}), \tag{A22}$$

which guarantees invariance of the EDF and total energy.

For the space-inversion symmetry, we have

$$\begin{aligned} \mathcal{H}^P(\boldsymbol{r}) &\equiv \mathcal{H}\left[\rho^P(\boldsymbol{r}\sigma,\boldsymbol{r}'\sigma')\right] \\ &= \mathcal{H}\left[\rho(-\boldsymbol{r}\sigma,-\boldsymbol{r}'\sigma')\right] \qquad \text{(A23)} \\ \mathcal{H}(P^+\boldsymbol{r}P) &\equiv \mathcal{H}(-\boldsymbol{r}), \qquad \text{(A24)} \end{aligned}$$

and the covariance condition (A21) reads

$$\mathcal{H}\left[\rho(-\boldsymbol{r}\sigma,-\boldsymbol{r}'\sigma')\right] = \mathcal{H}(-\boldsymbol{r}). \tag{A25}$$

It is now essential to realize that the arguments of the density matrix, $\rho(-\boldsymbol{r}\sigma,-\boldsymbol{r}'\sigma')$ on which the energy densities in Eq. (A25) depends, are *the same* on both sides of Eq. (A25). The covariance condition then tests only the parity of all other operators that may appear in the definition of local densities. Therefore, to each primary density $\rho_{nLvJ}(\boldsymbol{r})$ (23) we may attribute $P$-parity corresponding to the $P$-parity of the operator $K_{nL}$ only, which is equal to $(-1)^n$, see Eq. (26) and columns denoted by $P$ in Tables III and IV. This attribution is performed *regardless* of space-inversion properties of the nonlocal densities, i.e., *regardless* of whether the parity of the many-body state is conserved or broken. Similarly, $P$-parities of secondary densities $\rho_{mI,nLvJ,Q}(\boldsymbol{r})$ (24) are equal to $(-1)^{n+m}$. Construction of the $P$-covariant energy density (A21) can now be realized by multiplying densities that have identical $P$-parities.

Construction of a rotationally covariant energy density can be performed in an entirely analogous way. We must only ensure, that all tensor operators used in constructing all terms of the energy density are always coupled to total angular momentum (rank) zero. This coupling proceeds *regardless* of any transformation properties of nonlocal densities with respect to rotation, because again, their rotated space arguments appear on *both* sides of the covariance condition (A21).

It is obvious that this is the correct procedure to follow when the rotational symmetry is not broken, and nonlocal densities $\rho(\boldsymbol{r},\boldsymbol{r}')$ and $\boldsymbol{s}(\boldsymbol{r},\boldsymbol{r}')$ are scalar and vector functions of their arguments, respectively. In fact, this is how we refer to these densities throughout the entire paper, seemingly forgetting that the rotational symmetry can be broken, and that these functions can then have no good tensor properties with respect to rotation. Nevertheless, in view of the covariance condition (A21), these rotational properties of broken-symmetry nonlocal densities are irrelevant for the construction of the energy density.

## APPENDIX B: PHASE CONVENTIONS

In the present study, we use four elementary building blocks to construct the EDF, namely, the scalar and vector nonlocal densities, $\rho(\boldsymbol{r},\boldsymbol{r}')$ and $\boldsymbol{s}(\boldsymbol{r},\boldsymbol{r}')$, along with the total derivative $\boldsymbol{\nabla}$ and relative momentum $\boldsymbol{k}$ (6). Spherical representations of the building blocks can be defined by using standard convention of spherical tensors [26] as

$$\rho_{00}(\boldsymbol{r},\boldsymbol{r}') = p_\rho \rho(\boldsymbol{r},\boldsymbol{r}'), \tag{B1}$$

$$s_{1,\mu=\{-1,0,1\}}(\boldsymbol{r},\boldsymbol{r}') = p_s \left\{ \tfrac{1}{\sqrt{2}}\left(s_x(\boldsymbol{r},\boldsymbol{r}') - i s_y(\boldsymbol{r},\boldsymbol{r}')\right), s_z(\boldsymbol{r},\boldsymbol{r}'), \tfrac{-1}{\sqrt{2}}\left(s_x(\boldsymbol{r},\boldsymbol{r}') + i s_y(\boldsymbol{r},\boldsymbol{r}')\right) \right\}, \tag{B2}$$

$$\nabla_{1,\mu=\{-1,0,1\}} = p_\nabla \left\{ \tfrac{1}{\sqrt{2}}\left(\nabla_x - i\nabla_y\right), \nabla_z, \tfrac{-1}{\sqrt{2}}\left(\nabla_x + i\nabla_y\right) \right\}, \tag{B3}$$

$$k_{1,\mu=\{-1,0,1\}} = p_k \left\{ \tfrac{1}{\sqrt{2}}\left(k_x - ik_y\right), k_z, \tfrac{-1}{\sqrt{2}}\left(k_x + ik_y\right) \right\}, \tag{B4}$$

where $p_\rho$, $p_s$, $p_\nabla$, and $p_k$, are arbitrary phase factors, $|p_\rho| = |p_s| = |p_\nabla| = |p_k| = 1$. These phase factors define the phase convention of the building blocks, and can be used to achieve specific phase properties of densities and terms in the EDF, as discussed in this Appendix.

In order to motivate the best suitable choice of the phase convention, in Tables XXI and XXII we present relations between the spherical and Cartesian representations of densities and terms in the EDF, respectively. All NLO densities in the Cartesian representation, which are listed in Table XXI, are real. It is then clear that the phase convention, which would render all NLO densities in the spherical representation real does not exist. However, for the phase factors $p_\rho$, $p_s$, $p_\nabla$, and $p_k$ equal to $\pm1$ or $\pm i$, in the spherical representation all NLO densities and terms in the EDF are either real or imaginary.

Among many options of choosing the phase convention, in the present study we set

$$p_\rho = +1, \quad p_s = -i, \quad p_\nabla = -i, \quad \text{and } p_k = -i. \tag{B5}$$

This choice is unique in the fact that all scalar densities and all terms in the EDF are then characterized by phase factors +1 connecting the spherical and Cartesian representations, see the last columns in Tables XXI and XXII. This allows for the closest possible relationships between both representations, which may facilitate the use of the spherical representation as it is introduced in the present study. In particular, relations between coupling constants up to NLO (Table XXII) and standard coupling constants in the Cartesian representation [24] then read, for terms depending on time-even densities:

$$\boldsymbol{C}^{\mathbf{0000}}_{\mathbf{00,0000}} = C^\rho, \tag{B6}$$

$$\boldsymbol{C}^{\mathbf{0000}}_{\mathbf{20,0000}} = \sqrt{3}\left(C^{\Delta\rho} + \tfrac{1}{4}C^\tau\right), \tag{B7}$$

$$\boldsymbol{C}^{\mathbf{0000}}_{\mathbf{00,2000}} = \sqrt{3}C^\tau, \tag{B8}$$

$$\boldsymbol{C}^{\mathbf{1110}}_{\mathbf{00,1110}} = 3C^{J0}, \tag{B9}$$

$$\boldsymbol{C}^{\mathbf{1111}}_{\mathbf{00,1111}} = \sqrt{12}C^{J1}, \tag{B10}$$

$$\boldsymbol{C}^{\mathbf{1112}}_{\mathbf{00,1112}} = \sqrt{5}C^{J2}, \tag{B11}$$

$$\boldsymbol{C}^{\mathbf{0000}}_{\mathbf{11,1111}} = \sqrt{6}C^{\nabla J}, \tag{B12}$$

and for terms depending on time-odd densities:

$$C^{0011}_{00,0011} = \sqrt{3}C^s, \tag{B13}$$

$$C^{1101}_{00,1101} = \sqrt{3}C^j, \tag{B14}$$

$$C^{0011}_{20,0011} = 3\left(C^{\Delta s} + \tfrac{1}{4}C^T\right) + \tfrac{1}{4}C^F - C^{\nabla s}, \tag{B15}$$

$$C^{0011}_{22,0011} = \sqrt{5}\left(\tfrac{1}{4}C^F - C^{\nabla s}\right), \tag{B16}$$

$$C^{0011}_{00,2011} = 3C^T + C^F, \tag{B17}$$

$$C^{0011}_{00,2211} = \sqrt{5}C^F, \tag{B18}$$

$$C^{1101}_{00,0011} = \sqrt{6}C^{\nabla j}. \tag{B19}$$

At the same time, all vector densities in Table XXI and vector operators in Eqs. (B2)–(B4) are consistently characterized by phase factors $-i$ connecting the spherical and Cartesian representations.

TABLE XXI: Spherical and Cartesian representations of local densities (24) up to NLO. Only scalar densities and the $\mu = 0$ components of vector densities are shown. Numbers in the first column refer to numbers of primary densities (23) shown in Tables III and IV. The last column shows factors preceding densities in the Cartesian representation evaluated for the phase conventions of Eq. (B5). Time-even densities are marked by using the bold-face font.

| No. | $\rho_{mI,nLvJ,Q\mu}$ | | | Cartesian representation [24, 25, 30] | Phase |
|---|---|---|---|---|---|
| 1 | $\boldsymbol{\rho_{00,0000,00}}$ | $= [\rho]_{00}$ | $= p_\rho$ | $\rho$ | $+1$ |
| | $\boldsymbol{\rho_{11,0000,10}}$ | $= [\nabla\rho]_{10}$ | $= p_\nabla p_\rho$ | $\nabla_z\rho$ | $-i$ |
| | $\boldsymbol{\rho_{20,0000,00}}$ | $= [[\nabla\nabla]_0\rho]_{00}$ | $= -\, p_\nabla^2 p_\rho$ | $\frac{1}{\sqrt{3}}\Delta\rho$ | $+1$ |
| 2 | $\rho_{00,1101,10}$ | $= [k\rho]_{10}$ | $= p_k p_\rho$ | $j_z$ | $-i$ |
| | $\rho_{11,1101,00}$ | $= [\nabla[k\rho]_1]_{00}$ | $= -\, p_\nabla p_k p_\rho$ | $\frac{1}{\sqrt{3}}\boldsymbol{\nabla}\cdot\boldsymbol{j}$ | $+1$ |
| | $\rho_{11,1101,10}$ | $= [\nabla[k\rho]_1]_{10}$ | $= i p_\nabla p_k p_\rho$ | $\frac{1}{\sqrt{2}}\left(\boldsymbol{\nabla}\times\boldsymbol{j}\right)_z$ | $-i$ |
| 3 | $\boldsymbol{\rho_{00,2000,00}}$ | $= [[kk]_0\rho]_{00}$ | $= -\, p_k^2 p_\rho$ | $\frac{1}{\sqrt{3}}\left(\tau - \frac{1}{4}\Delta\rho\right)$ | $+1$ |
| 17 | $\rho_{00,0011,10}$ | $= [s]_{10}$ | $= p_s$ | $s_z$ | $-i$ |
| | $\rho_{11,0011,00}$ | $= [\nabla s]_{00}$ | $= -\, p_\nabla p_s$ | $\frac{1}{\sqrt{3}}\boldsymbol{\nabla}\cdot\boldsymbol{s}$ | $+1$ |
| | $\rho_{11,0011,10}$ | $= [\nabla s]_{10}$ | $= i p_\nabla p_s$ | $\frac{1}{\sqrt{2}}\left(\boldsymbol{\nabla}\times\boldsymbol{s}\right)_z$ | $-i$ |
| | $\rho_{20,0011,10}$ | $= [[\nabla\nabla]_0 s]_{10}$ | $= -\, p_\nabla^2 p_s$ | $\frac{1}{\sqrt{3}}\Delta s_z$ | $-i$ |
| | $\rho_{22,0011,10}$ | $= [[\nabla\nabla]_2 s]_{10}$ | $= -\, p_\nabla^2 p_s$ | $\frac{1}{\sqrt{15}}\left(3\nabla_z\boldsymbol{\nabla}\cdot\boldsymbol{s} - \Delta s_z\right)$ | $-i$ |
| 18 | $\boldsymbol{\rho_{00,1110,00}}$ | $= [ks]_{00}$ | $= -\, p_k p_s$ | $\frac{1}{\sqrt{3}}J^{(0)}$ | $+1$ |
| | $\boldsymbol{\rho_{11,1110,10}}$ | $= [\nabla[ks]_0]_{10}$ | $= -\, p_\nabla p_k p_s$ | $\frac{1}{\sqrt{3}}\nabla_z J^{(0)}$ | $-i$ |
| 19 | $\boldsymbol{\rho_{00,1111,10}}$ | $= [ks]_{10}$ | $= i p_k p_s$ | $\frac{1}{\sqrt{2}}J_z$ | $-i$ |
| | $\boldsymbol{\rho_{11,1111,00}}$ | $= [\nabla[ks]_1]_{00}$ | $= -i p_\nabla p_k p_s$ | $\frac{1}{\sqrt{6}}\boldsymbol{\nabla}\cdot\boldsymbol{J}$ | $+1$ |
| | $\boldsymbol{\rho_{11,1111,10}}$ | $= [\nabla[ks]_1]_{10}$ | $= -\, p_\nabla p_k p_s$ | $\frac{1}{2}\left(\boldsymbol{\nabla}\times\boldsymbol{J}\right)_z$ | $-i$ |
| 21 | $\rho_{00,2011,10}$ | $= [[kk]_0 s]_{10}$ | $= -\, p_k^2 p_s$ | $\frac{1}{\sqrt{3}}\left(T_z - \frac{1}{4}\Delta s_z\right)$ | $-i$ |
| 22 | $\rho_{00,2211,10}$ | $= [[kk]_2 s]_{10}$ | $= -\, p_k^2 p_s$ | $\frac{1}{\sqrt{15}}\left(3F_z - \frac{3}{4}\nabla_z\boldsymbol{\nabla}\cdot\boldsymbol{s} - T_z + \frac{1}{4}\Delta s_z\right)$ | $-i$ |

TABLE XXII: Spherical and Cartesian representations of terms in the EDF (30) up to NLO. The last column shows factors preceding terms in the Cartesian representation evaluated using the phase conventions of Eq. (B5). Integration by parts was used to transform $\boldsymbol{s}\cdot\boldsymbol{\nabla\nabla}\cdot\boldsymbol{s}$ into $-(\boldsymbol{\nabla}\cdot\boldsymbol{s})^2$, which is the term used previously in Refs. [24, 30]. Coupling constants corresponding to terms that depend on time-even densities are marked by using the bold-face font. Bullets (•) mark coupling constants corresponding to terms that do not vanish for conserved spherical, space-inversion, and time-reversal symmetries, see Sec. IV.

| No. | $C_{mI,nLvJ}^{n'L'v'J'}$ | $\left[\rho_{n'L'v'J'}[D_{mI}\rho_{nLvJ}]_{J'}\right]_0$ | | Cartesian representation | Phase |
|---|---|---|---|---|---|
| 1 | • $\boldsymbol{C_{00,0000}^{0000}}$ | $[\rho\rho]_0$ | $= p_\rho^2$ | $\rho^2$ | $+1$ |
| 2 | $C_{00,0011}^{0011}$ | $[ss]_0$ | $= -\, p_s^2$ | $\frac{1}{\sqrt{3}}\boldsymbol{s}^2$ | $+1$ |
| 3 | • $\boldsymbol{C_{20,0000}^{0000}}$ | $[\rho[[\nabla\nabla]_0\rho]_0]_0$ | $= -\, p_\nabla^2 p_\rho^2$ | $\frac{1}{\sqrt{3}}\rho\Delta\rho$ | $+1$ |
| 4 | • $\boldsymbol{C_{00,2000}^{0000}}$ | $[\rho[[kk]_0\rho]_0]_0$ | $= -\, p_k^2 p_\rho^2$ | $\frac{1}{\sqrt{3}}\left(\rho\tau - \frac{1}{4}\rho\Delta\rho\right)$ | $+1$ |
| 5 | $\boldsymbol{C_{00,1110}^{1110}}$ | $[[ks]_0[ks]_0]_0$ | $= p_k^2 p_s^2$ | $\frac{1}{3}\left(J^{(0)}\right)^2$ | $+1$ |
| 6 | • $\boldsymbol{C_{00,1111}^{1111}}$ | $[[ks]_1[ks]_1]_0$ | $= p_k^2 p_s^2$ | $\frac{1}{\sqrt{12}}\boldsymbol{J}^2$ | $+1$ |
| 7 | $\boldsymbol{C_{00,1112}^{1112}}$ | $[[ks]_2[ks]_2]_0$ | $= p_k^2 p_s^2$ | $\frac{1}{\sqrt{5}}\sum_{ab} J_{ab}^{(2)} J_{ab}^{(2)}$ | $+1$ |
| 8 | • $\boldsymbol{C_{11,1111}^{0000}}$ | $[\rho[\nabla[ks]_1]_0]_0$ | $= -i p_\nabla p_k p_s p_\rho$ | $\frac{1}{\sqrt{6}}\rho\boldsymbol{\nabla}\cdot\boldsymbol{J}$ | $+1$ |
| 9 | $C_{00,1101}^{1101}$ | $[[k\rho]_1[k\rho]_1]_0$ | $= -\, p_k^2 p_\rho^2$ | $\frac{1}{\sqrt{3}}\boldsymbol{j}^2$ | $+1$ |
| 10 | $C_{20,0011}^{0011}$ | $[s[[\nabla\nabla]_0 s]_1]_0$ | $= p_\nabla^2 p_s^2$ | $\frac{1}{3}\boldsymbol{s}\Delta\boldsymbol{s}$ | $+1$ |
| 11 | $C_{22,0011}^{0011}$ | $[s[[\nabla\nabla]_2 s]_1]_0$ | $= p_\nabla^2 p_s^2$ | $\frac{-1}{\sqrt{5}}\left((\boldsymbol{\nabla}\cdot\boldsymbol{s})^2 + \frac{1}{3}\boldsymbol{s}\Delta\boldsymbol{s}\right)$ | $+1$ |
| 12 | $C_{00,2011}^{0011}$ | $[s[[kk]_0 s]_1]_0$ | $= p_k^2 p_s^2$ | $\frac{1}{3}\left(\boldsymbol{s}\cdot\boldsymbol{T} - \frac{1}{4}\boldsymbol{s}\Delta\boldsymbol{s}\right)$ | $+1$ |
| 13 | $C_{00,2211}^{0011}$ | $[s[[kk]_2 s]_1]_0$ | $= p_k^2 p_s^2$ | $\frac{1}{\sqrt{5}}\Big(\boldsymbol{s}\cdot\boldsymbol{F} + \frac{1}{4}\left(\boldsymbol{\nabla}\cdot\boldsymbol{s}\right)^2$ $-\frac{1}{3}\boldsymbol{s}\cdot\boldsymbol{T} + \frac{1}{12}\boldsymbol{s}\Delta\boldsymbol{s}\Big)$ | $+1$ |
| 14 | $C_{11,0011}^{1101}$ | $[[k\rho]_1[\nabla s]_1]_0$ | $= -i p_\nabla p_k p_s p_\rho$ | $\frac{1}{\sqrt{6}}\boldsymbol{j}\cdot\boldsymbol{\nabla}\times\boldsymbol{s}$ | $+1$ |

Phase conventions (B5) also lead to very simple phase properties, which our spherical tensors have with respect to complex conjugation. Indeed, spherical tensors (B1)–(B4) obey standard transformation rules under complex conjugation [26],

$$A^*_{\lambda\mu} = P_A(-1)^{\lambda-\mu}A_{\lambda,-\mu}, \tag{B20}$$

where $P_A = \pm 1$. For nonlocal densities (B1)–(B2), Eq. (B20) holds separately for their time-even and time-odd parts, split as in Eqs. (A1) and (A2). Using Eqs. (A20) we then have

$$\begin{aligned} P_{\rho_+} &= +p^2_\rho \ \ , \ \ P_{\rho_-} = -p^2_\rho, \\ P_{s_+} &= +p^2_s \ \ , \ \ P_{s_-} = -p^2_s, \\ P_\nabla &= -p^2_\nabla \ \ , \ \ P_k = p^2_k, \end{aligned} \tag{B21}$$

which for the phase convention of Eq. (B5) reads

$$\begin{aligned} P_{\rho_+} &= +1 \ \ , \ \ P_{\rho_-} = -1, \\ P_{s_+} &= -1 \ \ , \ \ P_{s_-} = +1, \\ P_\nabla &= +1 \ \ , \ \ P_k = -1. \end{aligned} \tag{B22}$$

Standard rule (B20) propagates through the angular momentum coupling, i.e., if signs $P_A$ and $P_{A'}$ characterize tensors $A_\lambda$ and $A'_{\lambda'}$, respectively, then the coupled tensor,

$$A''_{\lambda''\mu''} = [A_\lambda A'_{\lambda'}]_{\lambda''\mu''} = \sum_{\mu\mu'} C^{\lambda''\mu''}_{\lambda\mu\lambda'\mu'} A_{\lambda\mu}A'_{\lambda'\mu'}, \tag{B23}$$

is characterized by the product of signs $P_{A''} = P_A P_{A'}$. Therefore, coupled higher-order densities (24) are characterized by signs,

$$P_{\rho_{mI,nLvJ,Q}} = P^m_\nabla P^n_k P_{vT}, \tag{B24}$$

where $v = 0$ or 1 denotes the scalar or vector density, $\rho$ or $\boldsymbol{s}$, respectively, and $T = +1$ or $T = -1$ denotes the time-even or time-odd density. However, symmetry conditions (A20) require that powers of the $k$ derivative determine the time-reversal symmetry of each local density, so that $T = (-1)^{n+v}$. From Eqs. (B21) we then obtain

$$P_{\rho_{mI,nLvJ,Q}} = (-1)^{m+n+v} p^{2m}_\nabla p^{2n}_k p^2_v. \tag{B25}$$

which for the phase convention of Eq. (B5) reads

$$P_{\rho_{mI,nLvJ,Q}} = +1 \tag{B26}$$

for all densities. Therefore, the phase convention of Eq. (B5) ensures that scalar densities and all terms in the EDF are always real.

## APPENDIX C: RESULTS FOR THE GALILEAN OR GAUGE INVARIANT ENERGY DENSITY FUNCTIONAL

As discussed in Sec. III B 3, when the Galilean or gauge invariance is imposed on the EDF, this induces specific constraints on the coupling constants and terms of the functional. We pointed out that there can be three disconnected classes of terms in the EDF with related properties of the coupling constants:

1. Terms that are invariant with respect to the Galilean or gauge transformation, and, therefore, the corresponding coupling constants are not restricted by the imposed symmetries.

2. Terms that cannot appear in the energy density when the Galilean or gauge symmetry is imposed, and, therefore, the corresponding coupling constants must be equal to zero.

3. Terms that can appear in the energy density only in certain specific linear combinations with other terms. This means that the coupling constants corresponding to these terms must obey specific linear conditions. We then distinguish:

   (a) *independent* coupling constants, which multiply invariant combinations of terms and, therefore, their values are not restricted by the imposed symmetries and

   (b) *dependent* coupling constants, which are equal to specific linear combinations of independent coupling constants, and, therefore, their values are in this way restricted by the imposed symmetries.

   Division into the sets of independent and dependent coupling constants is not unique, and below, in each case, we present only one specific choice thereof.

TABLE XXIII: Numbers of unrestricted, vanishing, independent, and dependent coupling constants in the EDF at zero, second, fourth, and sixth orders. Left and right columns correspond to the Galilean and gauge symmetries imposed, respectively.

| | Galilean | | | | Gauge | | | |
|---|---|---|---|---|---|---|---|---|
| order | 0 | 2 | 4 | 6 | 0 | 2 | 4 | 6 |
| unrestricted | 2 | 3 | 3 | 3 | 2 | 3 | 3 | 3 |
| vanishing | 0 | 0 | 0 | 0 | 0 | 0 | 27 | 100 |
| independent | 0 | 4 | 12 | 23 | 0 | 4 | 3 | 3 |
| dependent | 0 | 5 | 30 | 103 | 0 | 5 | 12 | 23 |

In Table XXIII we show numbers of unrestricted, vanishing, independent, and dependent coupling constants that appear at a given order when either Galilean or gauge symmetry is imposed. In what follows, we use the name of a free coupling constant to denote either the unrestricted or independent one. Indeed, in the Galilean or gauge invariant energy density (43), these two groups of coupling constants become free parameters.

Below we simultaneously discuss the Galilean and gauge symmetries. In doing so, we use the fact that the Galilean symmetry is a special case of the gauge symmetry, and, therefore, the latter may impose more restrictions on the EDF than the former. At NLO, this is not the case, and the Galilean and gauge symmetries impose, in fact, identical restrictions on the EDF [25, 36]. However, at higher orders, restrictions imposed by the Galilean and gauge symmetries are very different.

Both zero-order terms in the EDF, which correspond to the contact interaction, are Galilean and gauge invariant, i.e., these symmetries do not restrict the form of the EDF at LO. In the three following sections we give results for second, fourth, and sixth orders, respectively.

### 1. Second order

At second order, we obtain the same restrictions of the EDF as those already identified for the Skyrme functional, see Ref. [24] for a complete list thereof. Then, 5 dependent coupling constants are equal to specific linear combinations of 4 independent ones:

$$\boldsymbol{C^{0000}_{00,2000}} = -C^{1101}_{00,1101}, \tag{C1}$$
$$\boldsymbol{C^{1110}_{00,1110}} = -\tfrac{1}{3}C^{0011}_{00,2011} - \tfrac{1}{3}\sqrt{5}C^{0011}_{00,2211}, \tag{C2}$$
$$\boldsymbol{C^{1111}_{00,1111}} = \tfrac{1}{2}\sqrt{\tfrac{5}{3}}C^{0011}_{00,2211} - \tfrac{1}{\sqrt{3}}C^{0011}_{00,2011}, \tag{C3}$$
$$\boldsymbol{C^{1112}_{00,1112}} = -\tfrac{1}{3}\sqrt{5}C^{0011}_{00,2011} - \tfrac{1}{6}C^{0011}_{00,2211}, \tag{C4}$$
$$\boldsymbol{C^{0000}_{11,1111}} = C^{1101}_{11,0011}. \tag{C5}$$

These relations are obtained by imposing either Galilean or gauge invariance. In Eqs. (C1)–(C5), coupling constants corresponding to terms that depend on time-even densities are marked by using the bold-face font. The same convention also applies below.

At this order, the Galilean or gauge invariant energy density of Eq. (43) is composed of three terms corresponding to unrestricted coupling constants:

$$G^{0000}_{20,0000} = \boldsymbol{T^{0000}_{20,0000}}, \tag{C6}$$
$$G^{0011}_{20,0011} = T^{0011}_{20,0011}, \tag{C7}$$
$$G^{0011}_{22,0011} = T^{0011}_{22,0011}, \tag{C8}$$

and of four terms corresponding to the independent coupling constants:

$$G^{1101}_{00,1101} = T^{1101}_{00,1101} - \boldsymbol{T^{0000}_{00,2000}}, \tag{C9}$$
$$G^{1101}_{11,0011} = T^{1101}_{11,0011} + \boldsymbol{T^{0000}_{11,1111}}, \tag{C10}$$
$$G^{0011}_{00,2011} = -\tfrac{1}{3}\boldsymbol{T^{1110}_{00,1110}} - \tfrac{1}{\sqrt{3}}\boldsymbol{T^{1111}_{00,1111}} - \tfrac{1}{3}\sqrt{5}\boldsymbol{T^{1112}_{00,1112}} + T^{0011}_{00,2011}, \tag{C11}$$
$$G^{0011}_{00,2211} = -\tfrac{1}{3}\sqrt{5}\boldsymbol{T^{1110}_{00,1110}} + \tfrac{1}{2}\sqrt{\tfrac{5}{3}}\boldsymbol{T^{1111}_{00,1111}} - \tfrac{1}{6}\boldsymbol{T^{1112}_{00,1112}} + T^{0011}_{00,2211}. \tag{C12}$$

Again, terms that depend on time-even densities are marked by using the bold-face font. Altogether, 7 free coupling constants (3 unrestricted and 4 independent) define the Galilean or gauge invariant EDF at second order, cf. Table VI.

### 2. Fourth order

At fourth order, imposing either Galilean or gauge symmetry forces 12 dependent coupling constants to be specific linear combinations of 3 independent ones:

$$\boldsymbol{C^{0000}_{00,4000}} = \tfrac{3}{2\sqrt{5}}\boldsymbol{C^{2202}_{00,2202}}, \tag{C13}$$
$$\boldsymbol{C^{2000}_{00,2000}} = \tfrac{1}{2}\sqrt{5}\boldsymbol{C^{2202}_{00,2202}}, \tag{C14}$$
$$C^{1101}_{00,3101} = -\tfrac{6}{\sqrt{5}}\boldsymbol{C^{2202}_{00,2202}}, \tag{C15}$$
$$\boldsymbol{C^{1110}_{00,3110}} = -2\sqrt{\tfrac{3}{5}}C^{2212}_{00,2212} - \tfrac{7}{\sqrt{5}}C^{0011}_{00,4211}, \tag{C16}$$
$$\boldsymbol{C^{1111}_{00,3111}} = -\tfrac{6}{\sqrt{5}}C^{2212}_{00,2212}, \tag{C17}$$
$$\boldsymbol{C^{1112}_{00,3112}} = -2\sqrt{3}C^{2212}_{00,2212} - \tfrac{14}{5}C^{0011}_{00,4211}, \tag{C18}$$
$$\boldsymbol{C^{1112}_{00,3312}} = -2\sqrt{\tfrac{7}{15}}C^{0011}_{00,4211}, \tag{C19}$$
$$C^{0011}_{00,4011} = \tfrac{3}{2}\sqrt{\tfrac{3}{5}}C^{2212}_{00,2212} + \tfrac{7}{4\sqrt{5}}C^{0011}_{00,4211}, \tag{C20}$$
$$C^{2011}_{00,2011} = \tfrac{1}{2}\sqrt{15}C^{2212}_{00,2212} + \tfrac{7}{12}\sqrt{5}C^{0011}_{00,4211}, \tag{C21}$$
$$C^{2011}_{00,2211} = \tfrac{7}{3}C^{0011}_{00,4211}, \tag{C22}$$
$$C^{2211}_{00,2211} = \sqrt{\tfrac{3}{5}}C^{2212}_{00,2212} + \tfrac{7}{3\sqrt{5}}C^{0011}_{00,4211}, \tag{C23}$$
$$C^{2213}_{00,2213} = \sqrt{\tfrac{7}{5}}C^{2212}_{00,2212} + \tfrac{1}{2}\sqrt{\tfrac{21}{5}}C^{0011}_{00,4211}. \tag{C24}$$

At this order, there are 3 stand-alone Galilean and gauge invariant terms:

$$G^{0000}_{40,0000} = \boldsymbol{T^{0000}_{40,0000}}, \tag{C25}$$
$$G^{0011}_{40,0011} = T^{0011}_{40,0011}, \tag{C26}$$
$$G^{0011}_{42,0011} = T^{0011}_{42,0011}, \tag{C27}$$

and 3 Galilean and gauge invariant linear combinations of terms, corresponding to the 3 independent coupling constants:

$$G^{2202}_{00,2202} = \tfrac{1}{2}\sqrt{5}\boldsymbol{T^{2000}_{00,2000}} + \boldsymbol{T^{2202}_{00,2202}} - \tfrac{6}{\sqrt{5}}T^{1101}_{00,3101} + \tfrac{3}{2\sqrt{5}}\boldsymbol{T^{0000}_{00,4000}}, \tag{C28}$$

$$G^{0011}_{00,4211} = \tfrac{7}{12}\sqrt{5}T^{2011}_{00,2011} - \tfrac{7}{\sqrt{5}}\boldsymbol{T^{1110}_{00,3110}} - \tfrac{14}{5}\boldsymbol{T^{1112}_{00,3112}} - 2\sqrt{\tfrac{7}{15}}\boldsymbol{T^{1112}_{00,3312}} + \tfrac{7}{4\sqrt{5}}T^{0011}_{00,4011} + T^{0011}_{00,4211} + \tfrac{7}{3}T^{2011}_{00,2211} + \tfrac{7}{3\sqrt{5}}T^{2211}_{00,2211} + \tfrac{1}{2}\sqrt{\tfrac{21}{5}}T^{2213}_{00,2213}, \tag{C29}$$

$$G^{2212}_{00,2212} = \tfrac{1}{2}\sqrt{15}T^{2011}_{00,2011} + \sqrt{\tfrac{3}{5}}T^{2211}_{00,2211} - 2\sqrt{\tfrac{3}{5}}\boldsymbol{T^{1110}_{00,3110}} - \tfrac{6}{\sqrt{5}}\boldsymbol{T^{1111}_{00,3111}} - 2\sqrt{3}\boldsymbol{T^{1112}_{00,3112}} + \tfrac{3}{2}\sqrt{\tfrac{3}{5}}T^{0011}_{00,4011} + T^{2212}_{00,2212} + \sqrt{\tfrac{7}{5}}T^{2213}_{00,2213}, \tag{C30}$$

Altogether, these 6 free coupling constants (3 unrestricted and 3 independent) occur in both the Galilean and gauge invariant EDF at fourth order, cf. Table VI.

Apart from these 6 free and 12 dependent coupling constants, the gauge invariance requires that all the remaining 27 coupling constants are equal to zero. These 27 constants are allowed to be non-zero if the Galilean symmetry is imposed instead of the full gauge invariance. Then, there are 18 dependent coupling constants that are forced to be linear combinations of 9 independent ones:

$$\boldsymbol{C^{0000}_{20,2000}} = -C^{1101}_{20,1101}, \tag{C31}$$

$$\boldsymbol{C^{0000}_{22,2202}} = -C^{1101}_{22,1101}, \tag{C32}$$

$$\boldsymbol{C^{1110}_{20,1110}} = -\tfrac{1}{3}C^{0011}_{20,2011} - \tfrac{1}{3}\sqrt{5}C^{0011}_{20,2211}, \tag{C33}$$

$$\boldsymbol{C^{1110}_{22,1112}} = -\boldsymbol{C^{1111}_{22,1112}} - \tfrac{2}{\sqrt{7}}\boldsymbol{C^{1112}_{22,1112}} - 2\sqrt{\tfrac{15}{7}}C^{0011}_{22,2213}, \tag{C34}$$

$$\boldsymbol{C^{1111}_{20,1111}} = \tfrac{1}{2}\sqrt{\tfrac{5}{3}}C^{0011}_{20,2211} - \tfrac{1}{\sqrt{3}}C^{0011}_{20,2011}, \tag{C35}$$

$$\boldsymbol{C^{1111}_{22,1111}} = -\tfrac{1}{\sqrt{3}}\boldsymbol{C^{1111}_{22,1112}} - 3\sqrt{\tfrac{3}{7}}\boldsymbol{C^{1112}_{22,1112}} - 2\sqrt{\tfrac{5}{7}}C^{0011}_{22,2213}, \tag{C36}$$

$$\boldsymbol{C^{1112}_{20,1112}} = -\tfrac{1}{3}\sqrt{5}C^{0011}_{20,2011} - \tfrac{1}{6}C^{0011}_{20,2211}, \tag{C37}$$

$$C^{0011}_{22,2011} = \tfrac{2}{3}\boldsymbol{C^{1111}_{22,1112}} - \tfrac{2}{\sqrt{7}}\boldsymbol{C^{1112}_{22,1112}} + \sqrt{\tfrac{5}{21}}C^{0011}_{22,2213}, \tag{C38}$$

$$C^{0011}_{22,2211} = \tfrac{1}{6}\sqrt{5}\boldsymbol{C^{1111}_{22,1112}} + \sqrt{\tfrac{5}{7}}\boldsymbol{C^{1112}_{22,1112}} + \tfrac{8}{\sqrt{21}}C^{0011}_{22,2213}, \tag{C39}$$

$$C^{0011}_{22,2212} = \tfrac{1}{2}\sqrt{3}\boldsymbol{C^{1111}_{22,1112}} + 3\sqrt{\tfrac{3}{7}}\boldsymbol{C^{1112}_{22,1112}} + 2\sqrt{\tfrac{5}{7}}C^{0011}_{22,2213}, \tag{C40}$$

$$\boldsymbol{C^{0000}_{31,1111}} = C^{1101}_{31,0011}, \tag{C41}$$

$$\boldsymbol{C^{0000}_{11,3111}} = -\sqrt{\tfrac{3}{5}}C^{1101}_{11,2212}, \tag{C42}$$

$$\boldsymbol{C^{2000}_{11,1111}} = -\sqrt{\tfrac{5}{3}}C^{1101}_{11,2212}, \tag{C43}$$

$$\boldsymbol{C^{2202}_{11,1111}} = \tfrac{1}{\sqrt{3}}C^{1101}_{11,2212}, \tag{C44}$$

$$\boldsymbol{C^{2202}_{11,1112}} = C^{1101}_{11,2212}, \tag{C45}$$

$$C^{1101}_{11,2011} = -\sqrt{\tfrac{5}{3}}C^{1101}_{11,2212}, \tag{C46}$$

$$C^{1101}_{11,2211} = \tfrac{1}{\sqrt{3}}C^{1101}_{11,2212}, \tag{C47}$$

$$C^{3101}_{11,0011} = -\sqrt{\tfrac{3}{5}}C^{1101}_{11,2212}. \tag{C48}$$

Finally, we list 9 combinations of terms that are invariant with respect to the Galilean symmetry and correspond to the independent coupling constants:

$$G^{1111}_{22,1112} = -\frac{1}{\sqrt{3}}\boldsymbol{T^{1111}_{22,1111}} - \boldsymbol{T^{1110}_{22,1112}} + \boldsymbol{T^{1111}_{22,1112}} + \frac{2}{3}T^{0011}_{22,2011} + \frac{1}{6}\sqrt{5}T^{0011}_{22,2211} + \frac{1}{2}\sqrt{3}T^{0011}_{22,2212}, \quad (C49)$$

$$G^{1112}_{22,1112} = -3\sqrt{\frac{3}{7}}\boldsymbol{T^{1111}_{22,1111}} - \frac{2}{\sqrt{7}}\boldsymbol{T^{1110}_{22,1112}} + \boldsymbol{T^{1112}_{22,1112}} - \frac{2}{\sqrt{7}}T^{0011}_{22,2011} + \sqrt{\frac{5}{7}}T^{0011}_{22,2211} + 3\sqrt{\frac{3}{7}}T^{0011}_{22,2212}, \quad (C50)$$

$$G^{1101}_{20,1101} = T^{1101}_{20,1101} - \boldsymbol{T^{0000}_{20,2000}}, \quad (C51)$$

$$G^{1101}_{22,1101} = T^{1101}_{22,1101} - \boldsymbol{T^{0000}_{22,2202}}, \quad (C52)$$

$$G^{1101}_{31,0011} = T^{1101}_{31,0011} + \boldsymbol{T^{0000}_{31,1111}}, \quad (C53)$$

$$G^{1101}_{11,2212} = -\sqrt{\frac{5}{3}}\boldsymbol{T^{2000}_{11,1111}} + \frac{1}{\sqrt{3}}\boldsymbol{T^{2202}_{11,1111}} + \boldsymbol{T^{2202}_{11,1112}} - \sqrt{\frac{5}{3}}T^{1101}_{11,2011} + \frac{1}{\sqrt{3}}T^{1101}_{11,2211} - \sqrt{\frac{3}{5}}\boldsymbol{T^{0000}_{11,3111}} + T^{1101}_{11,2212} - \sqrt{\frac{3}{5}}T^{3101}_{11,0011}, \quad (C54)$$

$$G^{0011}_{20,2011} = -\frac{1}{3}\boldsymbol{T^{1110}_{20,1110}} - \frac{1}{\sqrt{3}}\boldsymbol{T^{1111}_{20,1111}} - \frac{1}{3}\sqrt{5}\boldsymbol{T^{1112}_{20,1112}} + T^{0011}_{20,2011}, \quad (C55)$$

$$G^{0011}_{20,2211} = -\frac{1}{3}\sqrt{5}\boldsymbol{T^{1110}_{20,1110}} + \frac{1}{2}\sqrt{\frac{5}{3}}\boldsymbol{T^{1111}_{20,1111}} - \frac{1}{6}\boldsymbol{T^{1112}_{20,1112}} + T^{0011}_{20,2211}, \quad (C56)$$

$$G^{0011}_{22,2213} = -2\sqrt{\frac{5}{7}}\boldsymbol{T^{1111}_{22,1111}} - 2\sqrt{\frac{15}{7}}\boldsymbol{T^{1110}_{22,1112}} + \sqrt{\frac{5}{21}}T^{0011}_{22,2011} + \frac{8}{\sqrt{21}}T^{0011}_{22,2211} + 2\sqrt{\frac{5}{7}}T^{0011}_{22,2212} + T^{0011}_{22,2213}. \quad (C57)$$

### 3. Sixth order

An entirely analogous pattern of terms and coupling constants appears at sixth order. Imposing either Galilean or gauge symmetry forces 23 dependent coupling constants to be specific linear combinations of 3 independent ones:.

$$\boldsymbol{C^{0000}_{00,6000}} = -\frac{3}{4}\sqrt{\frac{3}{7}}C^{3303}_{00,3303}, \quad (C58)$$

$$\boldsymbol{C^{2000}_{00,4000}} = -\frac{3}{4}\sqrt{21}C^{3303}_{00,3303}, \quad (C59)$$

$$\boldsymbol{C^{2202}_{00,4202}} = -3\sqrt{\frac{15}{7}}C^{3303}_{00,3303}, \quad (C60)$$

$$C^{1101}_{00,5101} = \frac{9}{2}\sqrt{\frac{3}{7}}C^{3303}_{00,3303}, \quad (C61)$$

$$C^{3101}_{00,3101} = \frac{9}{10}\sqrt{21}C^{3303}_{00,3303}, \quad (C62)$$

$$\boldsymbol{C^{1110}_{00,5110}} = -\frac{1}{2}\sqrt{\frac{3}{5}}C^{2212}_{00,4212} - \frac{9}{\sqrt{5}}C^{0011}_{00,6211}, \quad (C63)$$

$$\boldsymbol{C^{1111}_{00,5111}} = -\frac{3}{2\sqrt{5}}C^{2212}_{00,4212}, \quad (C64)$$

$$\boldsymbol{C^{1112}_{00,5112}} = -\frac{1}{2}\sqrt{3}C^{2212}_{00,4212} - \frac{18}{5}C^{0011}_{00,6211}, \quad (C65)$$

$$\boldsymbol{C^{1112}_{00,5312}} = -4\sqrt{\frac{7}{15}}C^{0011}_{00,6211}, \quad (C66)$$

$$\boldsymbol{C^{3110}_{00,3110}} = -\frac{7}{10}\sqrt{\frac{3}{5}}C^{2212}_{00,4212} - \frac{63}{5\sqrt{5}}C^{0011}_{00,6211}, \quad (C67)$$

$$\boldsymbol{C^{3111}_{00,3111}} = -\frac{21}{10\sqrt{5}}C^{2212}_{00,4212}, \quad (C68)$$

$$\boldsymbol{C^{3112}_{00,3112}} = -\frac{7}{10}\sqrt{3}C^{2212}_{00,4212} - \frac{126}{25}C^{0011}_{00,6211}, \quad (C69)$$

$$\boldsymbol{C^{3112}_{00,3312}} = -\frac{12}{5}\sqrt{\frac{21}{5}}C^{0011}_{00,6211}, \quad (C70)$$

$$\boldsymbol{C^{3312}_{00,3312}} = -\frac{1}{3\sqrt{3}}C^{2212}_{00,4212} - \frac{6}{5}C^{0011}_{00,6211}, \quad (C71)$$

$$\boldsymbol{C^{3313}_{00,3313}} = -\frac{1}{3}\sqrt{\frac{7}{15}}C^{2212}_{00,4212}, \quad (C72)$$

$$\boldsymbol{C^{3314}_{00,3314}} = -\frac{1}{\sqrt{15}}C^{2212}_{00,4212} - \frac{8}{3\sqrt{5}}C^{0011}_{00,6211}, \quad (C73)$$

$$C^{0011}_{00,6011} = \frac{1}{4}\sqrt{\frac{3}{5}}C^{2212}_{00,4212} + \frac{3}{2\sqrt{5}}C^{0011}_{00,6211}, \quad (C74)$$

$$C^{2011}_{00,4011} = \frac{7}{4}\sqrt{\frac{3}{5}}C^{2212}_{00,4212} + \frac{21}{2\sqrt{5}}C^{0011}_{00,6211}, \quad (C75)$$

$$C^{2011}_{00,4211} = 6C^{0011}_{00,6211}, \quad (C76)$$

$$C^{2211}_{00,4011} = \frac{21}{5}C^{0011}_{00,6211}, \quad (C77)$$

$$C^{2211}_{00,4211} = \sqrt{\frac{3}{5}}C^{2212}_{00,4212} + \frac{12}{\sqrt{5}}C^{0011}_{00,6211}, \quad (C78)$$

$$C^{2213}_{00,4213} = \sqrt{\frac{7}{5}}C^{2212}_{00,4212} + 18\sqrt{\frac{3}{35}}C^{0011}_{00,6211}, \quad (C79)$$

$$C^{2213}_{00,4413} = \frac{4}{\sqrt{5}}C^{0011}_{00,6211}. \quad (C80)$$

At this order, there are 3 stand-alone Galilean and gauge invariant terms:

$$G^{0000}_{60,0000} = \boldsymbol{T^{0000}_{60,0000}}, \quad (C81)$$

$$G^{0011}_{60,0011} = T^{0011}_{60,0011}, \quad (C82)$$

$$G^{0011}_{62,0011} = T^{0011}_{62,0011}, \quad (C83)$$

and 3 Galilean and gauge invariant linear combinations of terms, corresponding to the 3 independent coupling constants:

$$
\begin{aligned}
G^{3303}_{00,3303} &= \tfrac{9}{10}\sqrt{21}T^{3101}_{00,3101} + T^{3303}_{00,3303} - \tfrac{3}{4}\sqrt{21}\boldsymbol{T^{2000}_{00,4000}} - 3\sqrt{\tfrac{15}{7}}\boldsymbol{T^{2202}_{00,4202}} \\
&\quad + \tfrac{9}{2}\sqrt{\tfrac{3}{7}}T^{1101}_{00,5101} - \tfrac{3}{4}\sqrt{\tfrac{3}{7}}\boldsymbol{T^{0000}_{00,6000}}, \qquad \text{(C84)}\\
G^{0011}_{00,6211} &= -\tfrac{63}{5\sqrt{5}}\boldsymbol{T^{3110}_{00,3110}} - \tfrac{126}{25}\boldsymbol{T^{3112}_{00,3112}} - \tfrac{12}{5}\sqrt{\tfrac{21}{5}}\boldsymbol{T^{3112}_{00,3312}} - \tfrac{9}{\sqrt{5}}\boldsymbol{T^{1110}_{00,5110}} - \tfrac{18}{5}\boldsymbol{T^{1112}_{00,5112}} \\
&\quad -4\sqrt{\tfrac{7}{15}}\boldsymbol{T^{1112}_{00,5312}} - \tfrac{6}{5}\boldsymbol{T^{3312}_{00,3312}} - \tfrac{8}{3\sqrt{5}}\boldsymbol{T^{3314}_{00,3314}} + \tfrac{21}{2\sqrt{5}}T^{2011}_{00,4011} + 6T^{2011}_{00,4211} + \tfrac{3}{2\sqrt{5}}T^{0011}_{00,6011} \\
&\quad + T^{0011}_{00,6211} + \tfrac{21}{5}T^{2211}_{00,4011} + \tfrac{12}{\sqrt{5}}T^{2211}_{00,4211} + 18\sqrt{\tfrac{3}{35}}T^{2213}_{00,4213} + \tfrac{4}{\sqrt{5}}T^{2213}_{00,4413}, \qquad \text{(C85)}\\
G^{2212}_{00,4212} &= -\tfrac{7}{10}\sqrt{\tfrac{3}{5}}\boldsymbol{T^{3110}_{00,3110}} - \tfrac{21}{10\sqrt{5}}\boldsymbol{T^{3111}_{00,3111}} - \tfrac{7}{10}\sqrt{3}\boldsymbol{T^{3112}_{00,3112}} - \tfrac{1}{2}\sqrt{\tfrac{3}{5}}\boldsymbol{T^{1110}_{00,5110}} - \tfrac{3}{2\sqrt{5}}\boldsymbol{T^{1111}_{00,5111}} \\
&\quad -\tfrac{1}{2}\sqrt{3}\boldsymbol{T^{1112}_{00,5112}} - \tfrac{1}{3\sqrt{3}}\boldsymbol{T^{3312}_{00,3312}} - \tfrac{1}{3}\sqrt{\tfrac{7}{15}}\boldsymbol{T^{3313}_{00,3313}} - \tfrac{1}{\sqrt{15}}\boldsymbol{T^{3314}_{00,3314}} + \tfrac{7}{4}\sqrt{\tfrac{3}{5}}T^{2011}_{00,4011} \\
&\quad + \sqrt{\tfrac{3}{5}}T^{2211}_{00,4211} + \tfrac{1}{4}\sqrt{\tfrac{3}{5}}T^{0011}_{00,6011} + T^{2212}_{00,4212} + \sqrt{\tfrac{7}{5}}T^{2213}_{00,4213}. \qquad \text{(C86)}
\end{aligned}
$$

Altogether, 6 free coupling constants (3 unrestricted and 3 independent) occur in both the Galilean and gauge invariant EDF at sixth order, cf. Table VI.

Apart from the 6 free and 23 dependent coupling constants, at sixth order the gauge invariance requires that all the remaining 100 coupling constants are equal to zero. These 100 constants are allowed to be non-zero if the Galilean symmetry is imposed instead of the full gauge invariance. Then, there are 80 dependent coupling constants that are forced to be linear combinations of 20 independent ones:

$$
\begin{aligned}
\boldsymbol{C^{0000}_{40,2000}} &= -C^{1101}_{40,1101}, \qquad \text{(C87)}\\
\boldsymbol{C^{0000}_{42,2202}} &= -C^{1101}_{42,1101}, \qquad \text{(C88)}\\
\boldsymbol{C^{0000}_{20,4000}} &= \tfrac{3}{2\sqrt{5}}\boldsymbol{C^{2202}_{20,2202}}, \qquad \text{(C89)}\\
\boldsymbol{C^{0000}_{22,4202}} &= -\tfrac{1}{2}\sqrt{\tfrac{15}{7}}C^{1101}_{22,3303}, \qquad \text{(C90)}\\
\boldsymbol{C^{2000}_{20,2000}} &= \tfrac{1}{2}\sqrt{5}\boldsymbol{C^{2202}_{20,2202}}, \qquad \text{(C91)}\\
\boldsymbol{C^{2000}_{22,2202}} &= -\tfrac{1}{2}\sqrt{\tfrac{35}{3}}C^{1101}_{22,3303}, \qquad \text{(C92)}\\
\boldsymbol{C^{2202}_{22,2202}} &= -\tfrac{1}{2}\sqrt{\tfrac{5}{3}}C^{1101}_{22,3303}, \qquad \text{(C93)}\\
C^{1101}_{20,3101} &= -\tfrac{6}{\sqrt{5}}\boldsymbol{C^{2202}_{20,2202}}, \qquad \text{(C94)}\\
C^{1101}_{22,3101} &= \sqrt{\tfrac{21}{5}}C^{1101}_{22,3303}, \qquad \text{(C95)}\\
\boldsymbol{C^{1110}_{40,1110}} &= \tfrac{1}{\sqrt{5}}\boldsymbol{C^{1112}_{40,1112}} - \tfrac{3}{2\sqrt{5}}C^{0011}_{40,2211}, \qquad \text{(C96)}\\
\boldsymbol{C^{1110}_{20,3110}} &= -\tfrac{4}{5}C^{2011}_{20,2011} - \tfrac{2}{\sqrt{5}}C^{2011}_{20,2211}, \qquad \text{(C97)}\\
\boldsymbol{C^{1110}_{22,3312}} &= \tfrac{6}{\sqrt{35}}C^{2011}_{22,2212} + \sqrt{\tfrac{5}{21}}\boldsymbol{C^{1110}_{22,3112}} \\
&\quad + \tfrac{2}{3}\sqrt{\tfrac{5}{21}}\boldsymbol{C^{1111}_{22,3112}} - 2\sqrt{2}\boldsymbol{C^{1111}_{22,3313}}. \qquad \text{(C98)}\\
\boldsymbol{C^{1111}_{40,1111}} &= \sqrt{\tfrac{3}{5}}\boldsymbol{C^{1112}_{40,1112}} + \sqrt{\tfrac{3}{5}}C^{0011}_{40,2211}, \qquad \text{(C99)}\\
\boldsymbol{C^{1111}_{42,1112}} &= 2\sqrt{3}\boldsymbol{C^{1111}_{42,1111}} + 2\sqrt{3}C^{0011}_{42,2212}, \qquad \text{(C100)}\\
\boldsymbol{C^{1111}_{20,3111}} &= \sqrt{\tfrac{3}{5}}C^{2011}_{20,2211} - \tfrac{4}{5}\sqrt{3}C^{2011}_{20,2011}, \qquad \text{(C101)}\\
\boldsymbol{C^{1111}_{22,3111}} &= \tfrac{1}{\sqrt{3}}\boldsymbol{C^{1111}_{22,3112}} - \tfrac{6}{5}C^{2011}_{22,2212}, \qquad \text{(C102)}\\
\boldsymbol{C^{1111}_{22,3312}} &= \tfrac{6}{\sqrt{35}}C^{2011}_{22,2212} - 2\sqrt{2}\boldsymbol{C^{1111}_{22,3313}}, \qquad \text{(C103)}\\
\boldsymbol{C^{1112}_{42,1112}} &= \tfrac{1}{\sqrt{7}}\boldsymbol{C^{1110}_{42,1112}} - \sqrt{\tfrac{3}{7}}\boldsymbol{C^{1111}_{42,1111}}, \qquad \text{(C104)}\\
\boldsymbol{C^{1112}_{44,1112}} &= -C^{0011}_{44,2213}, \qquad \text{(C105)}\\
\boldsymbol{C^{1112}_{22,3110}} &= \boldsymbol{C^{1110}_{22,3112}}, \qquad \text{(C106)}\\
\boldsymbol{C^{1112}_{22,3111}} &= -\boldsymbol{C^{1111}_{22,3112}}, \qquad \text{(C107)}\\
\boldsymbol{C^{1112}_{20,3112}} &= -\tfrac{4}{\sqrt{5}}C^{2011}_{20,2011} - \tfrac{1}{5}C^{2011}_{20,2211}, \qquad \text{(C108)}\\
\boldsymbol{C^{1112}_{22,3112}} &= \tfrac{6}{5}\sqrt{\tfrac{3}{7}}C^{2011}_{22,2212} + \tfrac{2}{\sqrt{7}}\boldsymbol{C^{1110}_{22,3112}} \\
&\quad - \tfrac{1}{\sqrt{7}}\boldsymbol{C^{1111}_{22,3112}}, \qquad \text{(C109)}\\
\boldsymbol{C^{1112}_{20,3312}} &= -2\sqrt{\tfrac{3}{35}}C^{2011}_{20,2211}, \qquad \text{(C110)}\\
\boldsymbol{C^{1112}_{22,3312}} &= \tfrac{2}{7\sqrt{5}}C^{2011}_{22,2212} + \tfrac{2}{7}\sqrt{\tfrac{5}{3}}\boldsymbol{C^{1110}_{22,3112}} \\
&\quad + \tfrac{4}{21}\sqrt{\tfrac{5}{3}}\boldsymbol{C^{1111}_{22,3112}} - 2\sqrt{\tfrac{2}{7}}\boldsymbol{C^{1111}_{22,3313}}, \qquad \text{(C111)}\\
\boldsymbol{C^{1112}_{22,3313}} &= \tfrac{1}{\sqrt{2}}\boldsymbol{C^{1111}_{22,3313}} - \tfrac{4}{\sqrt{35}}C^{2011}_{22,2212}, \qquad \text{(C112)}\\
\boldsymbol{C^{1112}_{22,3314}} &= \tfrac{4}{7\sqrt{15}}C^{2011}_{22,2212} + \tfrac{4}{21}\sqrt{5}\boldsymbol{C^{1110}_{22,3112}} \\
&\quad + \tfrac{8}{63}\sqrt{5}\boldsymbol{C^{1111}_{22,3112}} - \tfrac{1}{\sqrt{42}}\boldsymbol{C^{1111}_{22,3313}}, \qquad \text{(C113)}\\
C^{0011}_{40,2011} &= -\tfrac{3}{\sqrt{5}}\boldsymbol{C^{1112}_{40,1112}} - \tfrac{1}{2\sqrt{5}}C^{0011}_{40,2211}, \qquad \text{(C114)}
\end{aligned}
$$

$$C^{0011}_{42,2011} = \frac{4}{\sqrt{3}}\boldsymbol{C}^{\mathbf{1111}}_{\mathbf{42,1111}} - \frac{1}{2}\boldsymbol{C}^{\mathbf{1110}}_{\mathbf{42,1112}} + \sqrt{3}C^{0011}_{42,2212}, \tag{C115}$$

$$C^{0011}_{42,2211} = -\frac{4}{\sqrt{15}}\boldsymbol{C}^{\mathbf{1111}}_{\mathbf{42,1111}} - \frac{1}{\sqrt{5}}\boldsymbol{C}^{\mathbf{1110}}_{\mathbf{42,1112}} - \sqrt{\frac{3}{5}}C^{0011}_{42,2212}, \tag{C116}$$

$$C^{0011}_{42,2213} = -\frac{6}{\sqrt{35}}\boldsymbol{C}^{\mathbf{1111}}_{\mathbf{42,1111}} - \frac{3}{2}\sqrt{\frac{3}{35}}\boldsymbol{C}^{\mathbf{1110}}_{\mathbf{42,1112}} - \sqrt{\frac{7}{5}}C^{0011}_{42,2212}, \tag{C117}$$

$$C^{0011}_{20,4011} = \frac{3}{5}C^{2011}_{20,2011}, \tag{C118}$$

$$C^{0011}_{22,4011} = -\frac{1}{10}\sqrt{3}C^{2011}_{22,2212} - \frac{1}{4}\boldsymbol{C}^{\mathbf{1110}}_{\mathbf{22,3112}} + \frac{1}{3}\boldsymbol{C}^{\mathbf{1111}}_{\mathbf{22,3112}}, \tag{C119}$$

$$C^{0011}_{20,4211} = \frac{3}{7}C^{2011}_{20,2211}, \tag{C120}$$

$$C^{0011}_{22,4211} = \frac{1}{7}\sqrt{\frac{3}{5}}C^{2011}_{22,2212} - \frac{1}{7}\sqrt{5}\boldsymbol{C}^{\mathbf{1110}}_{\mathbf{22,3112}} - \frac{2}{21}\sqrt{5}\boldsymbol{C}^{\mathbf{1111}}_{\mathbf{22,3112}}, \tag{C121}$$

$$C^{0011}_{22,4212} = \frac{3}{7}C^{2011}_{22,2212}, \tag{C122}$$

$$C^{0011}_{22,4213} = -\frac{3}{7\sqrt{35}}C^{2011}_{22,2212} - \frac{3}{14}\sqrt{\frac{15}{7}}\boldsymbol{C}^{\mathbf{1110}}_{\mathbf{22,3112}} - \frac{1}{7}\sqrt{\frac{15}{7}}\boldsymbol{C}^{\mathbf{1111}}_{\mathbf{22,3112}}, \tag{C123}$$

$$C^{0011}_{22,4413} = -\frac{2}{7}\sqrt{\frac{5}{3}}C^{2011}_{22,2212} - \frac{1}{21}\sqrt{5}\boldsymbol{C}^{\mathbf{1110}}_{\mathbf{22,3112}} - \frac{2}{63}\sqrt{5}\boldsymbol{C}^{\mathbf{1111}}_{\mathbf{22,3112}} + \sqrt{\frac{7}{6}}\boldsymbol{C}^{\mathbf{1111}}_{\mathbf{22,3313}}, \tag{C124}$$

$$C^{2011}_{22,2011} = -\frac{1}{2\sqrt{3}}C^{2011}_{22,2212} - \frac{5}{12}\boldsymbol{C}^{\mathbf{1110}}_{\mathbf{22,3112}} + \frac{5}{9}\boldsymbol{C}^{\mathbf{1111}}_{\mathbf{22,3112}}, \tag{C125}$$

$$C^{2011}_{22,2211} = \frac{1}{\sqrt{15}}C^{2011}_{22,2212} - \frac{1}{3}\sqrt{5}\boldsymbol{C}^{\mathbf{1110}}_{\mathbf{22,3112}} - \frac{2}{9}\sqrt{5}\boldsymbol{C}^{\mathbf{1111}}_{\mathbf{22,3112}}, \tag{C126}$$

$$C^{2011}_{22,2213} = -\frac{1}{\sqrt{35}}C^{2011}_{22,2212} - \frac{1}{2}\sqrt{\frac{15}{7}}\boldsymbol{C}^{\mathbf{1110}}_{\mathbf{22,3112}} - \sqrt{\frac{5}{21}}\boldsymbol{C}^{\mathbf{1111}}_{\mathbf{22,3112}}, \tag{C127}$$

$$C^{2211}_{20,2211} = \frac{2}{5}C^{2011}_{20,2011} + \frac{1}{2\sqrt{5}}C^{2011}_{20,2211}, \tag{C128}$$

$$C^{2211}_{22,2211} = -\frac{43}{25\sqrt{3}}C^{2011}_{22,2212} - \frac{1}{3}\boldsymbol{C}^{\mathbf{1110}}_{\mathbf{22,3112}} - \frac{17}{90}\boldsymbol{C}^{\mathbf{1111}}_{\mathbf{22,3112}} + \frac{9}{5}\sqrt{\frac{21}{10}}\boldsymbol{C}^{\mathbf{1111}}_{\mathbf{22,3313}}, \tag{C129}$$

$$C^{2211}_{22,2212} = \frac{4}{5\sqrt{5}}C^{2011}_{22,2212} + \frac{1}{\sqrt{15}}\boldsymbol{C}^{\mathbf{1111}}_{\mathbf{22,3112}} - \frac{3}{5}\sqrt{14}\boldsymbol{C}^{\mathbf{1111}}_{\mathbf{22,3313}}, \tag{C130}$$

$$C^{2211}_{22,2213} = -\frac{4}{25\sqrt{7}}C^{2011}_{22,2212} - \sqrt{\frac{3}{7}}\boldsymbol{C}^{\mathbf{1110}}_{\mathbf{22,3112}} + \frac{4}{5\sqrt{21}}\boldsymbol{C}^{\mathbf{1111}}_{\mathbf{22,3112}} + \frac{3}{5}\sqrt{\frac{2}{5}}\boldsymbol{C}^{\mathbf{1111}}_{\mathbf{22,3313}}, \tag{C131}$$

$$C^{2212}_{20,2212} = \frac{2}{\sqrt{15}}C^{2011}_{20,2011} - \frac{1}{2\sqrt{3}}C^{2011}_{20,2211}, \tag{C132}$$

$$C^{2212}_{22,2212} = \frac{1}{5}\sqrt{7}C^{2011}_{22,2212} - \frac{1}{6}\sqrt{\frac{7}{3}}\boldsymbol{C}^{\mathbf{1111}}_{\mathbf{22,3112}} - \frac{3}{\sqrt{10}}\boldsymbol{C}^{\mathbf{1111}}_{\mathbf{22,3313}}, \tag{C133}$$

$$C^{2212}_{22,2213} = -\frac{2}{5}\sqrt{\frac{2}{35}}C^{2011}_{22,2212} - \frac{2}{3}\sqrt{\frac{14}{15}}\boldsymbol{C}^{\mathbf{1111}}_{\mathbf{22,3112}} + \frac{3}{5}\boldsymbol{C}^{\mathbf{1111}}_{\mathbf{22,3313}}, \tag{C134}$$

$$C^{2213}_{20,2213} = \frac{2}{5}\sqrt{\frac{7}{3}}C^{2011}_{20,2011} + \frac{1}{\sqrt{105}}C^{2011}_{20,2211}, \tag{C135}$$

$$C^{2213}_{22,2213} = -\frac{9}{25}\sqrt{\frac{6}{7}}C^{2011}_{22,2212} - \frac{1}{\sqrt{14}}\boldsymbol{C}^{\mathbf{1110}}_{\mathbf{22,3112}} + \frac{2}{15}\sqrt{\frac{2}{7}}\boldsymbol{C}^{\mathbf{1111}}_{\mathbf{22,3112}} + \frac{1}{5}\sqrt{\frac{3}{5}}\boldsymbol{C}^{\mathbf{1111}}_{\mathbf{22,3313}}, \tag{C136}$$

$$\boldsymbol{C}^{\mathbf{0000}}_{\mathbf{51,1111}} = C^{1101}_{51,0011}, \tag{C137}$$

$$\boldsymbol{C}^{\mathbf{0000}}_{\mathbf{31,3111}} = -\sqrt{\frac{3}{5}}C^{1101}_{31,2212}, \tag{C138}$$

$$\boldsymbol{C}^{\mathbf{0000}}_{\mathbf{33,3313}} = C^{1101}_{33,2213}, \tag{C139}$$

$$\boldsymbol{C}^{\mathbf{0000}}_{\mathbf{11,5111}} = -\frac{1}{14}\sqrt{15}C^{3101}_{11,2212}, \tag{C140}$$

$$\boldsymbol{C}^{\mathbf{2000}}_{\mathbf{31,1111}} = -\sqrt{\frac{5}{3}}C^{1101}_{31,2212}, \tag{C141}$$

$$\boldsymbol{C}^{\mathbf{2000}}_{\mathbf{11,3111}} = -\sqrt{\frac{5}{3}}C^{3101}_{11,2212}, \tag{C142}$$

$$\boldsymbol{C}^{\mathbf{2202}}_{\mathbf{31,1111}} = \frac{1}{\sqrt{3}}C^{1101}_{31,2212}, \tag{C143}$$

$$\boldsymbol{C}^{\mathbf{2202}}_{\mathbf{33,1111}} = \sqrt{6}C^{1101}_{33,2213}, \tag{C144}$$

$$\boldsymbol{C}^{\mathbf{2202}}_{\mathbf{31,1112}} = C^{1101}_{31,2212}, \tag{C145}$$

$$\boldsymbol{C}^{\mathbf{2202}}_{\mathbf{33,1112}} = -\sqrt{3}C^{1101}_{33,2213}, \tag{C146}$$

$$\boldsymbol{C^{2202}_{11,3111}} = \frac{1}{\sqrt{3}} C^{3101}_{11,2212}, \tag{C147}$$

$$\boldsymbol{C^{2202}_{11,3112}} = C^{3101}_{11,2212}, \tag{C148}$$

$$\boldsymbol{C^{2202}_{11,3312}} = -\frac{2}{3}\sqrt{\frac{5}{21}} C^{3101}_{11,2212}, \tag{C149}$$

$$\boldsymbol{C^{2202}_{11,3313}} = -\frac{2}{3}\sqrt{\frac{10}{21}} C^{3101}_{11,2212}, \tag{C150}$$

$$\boldsymbol{C^{4000}_{11,1111}} = -\frac{1}{2}\sqrt{\frac{5}{3}} C^{3101}_{11,2212}, \tag{C151}$$

$$\boldsymbol{C^{4202}_{11,1111}} = \frac{5}{7\sqrt{3}} C^{3101}_{11,2212}, \tag{C152}$$

$$\boldsymbol{C^{4202}_{11,1112}} = \frac{5}{7} C^{3101}_{11,2212}, \tag{C153}$$

$$C^{1101}_{31,2011} = -\sqrt{\frac{5}{3}} C^{1101}_{31,2212}, \tag{C154}$$

$$C^{1101}_{31,2211} = \frac{1}{\sqrt{3}} C^{1101}_{31,2212}, \tag{C155}$$

$$C^{1101}_{33,2212} = -2\sqrt{2} C^{1101}_{33,2213}, \tag{C156}$$

$$C^{1101}_{11,4011} = -\frac{1}{2}\sqrt{\frac{5}{3}} C^{3101}_{11,2212}, \tag{C157}$$

$$C^{1101}_{11,4211} = \frac{5}{7\sqrt{3}} C^{3101}_{11,2212}, \tag{C158}$$

$$C^{1101}_{11,4212} = \frac{5}{7} C^{3101}_{11,2212}, \tag{C159}$$

$$C^{3101}_{31,0011} = -\sqrt{\frac{3}{5}} C^{1101}_{31,2212}, \tag{C160}$$

$$C^{3101}_{11,2011} = -\sqrt{\frac{5}{3}} C^{3101}_{11,2212}, \tag{C161}$$

$$C^{3101}_{11,2211} = \frac{1}{\sqrt{3}} C^{3101}_{11,2212}, \tag{C162}$$

$$C^{3303}_{33,0011} = C^{1101}_{33,2213}, \tag{C163}$$

$$C^{3303}_{11,2212} = -\frac{2}{3}\sqrt{\frac{5}{21}} C^{3101}_{11,2212}, \tag{C164}$$

$$C^{3303}_{11,2213} = -\frac{2}{3}\sqrt{\frac{10}{21}} C^{3101}_{11,2212}, \tag{C165}$$

$$C^{5101}_{11,0011} = -\frac{1}{14}\sqrt{15} C^{3101}_{11,2212}. \tag{C166}$$

Finally, we list 20 combinations of terms that are invariant with respect to the Galilean symmetry and correspond to the independent coupling constants:

$$G^{2202}_{20,2202} = \tfrac{1}{2}\sqrt{5}\boldsymbol{T}^{\mathbf{2000}}_{\mathbf{20,2000}} + \boldsymbol{T}^{\mathbf{2202}}_{\mathbf{20,2202}} - \tfrac{6}{\sqrt{5}}T^{1101}_{20,3101} + \tfrac{3}{2\sqrt{5}}\boldsymbol{T}^{\mathbf{0000}}_{\mathbf{20,4000}}, \quad \text{(C167)}$$

$$G^{1110}_{42,1112} = \boldsymbol{T}^{\mathbf{1110}}_{\mathbf{42,1112}} + \tfrac{1}{\sqrt{7}}\boldsymbol{T}^{\mathbf{1112}}_{\mathbf{42,1112}} - \tfrac{1}{2}T^{0011}_{42,2011} - \tfrac{1}{\sqrt{5}}T^{0011}_{42,2211} - \tfrac{3}{2}\sqrt{\tfrac{3}{35}}T^{0011}_{42,2213}, \quad \text{(C168)}$$

$$\begin{aligned} G^{1110}_{22,3112} = {} & \boldsymbol{T}^{\mathbf{1112}}_{\mathbf{22,3110}} + \boldsymbol{T}^{\mathbf{1110}}_{\mathbf{22,3112}} + \tfrac{2}{\sqrt{7}}\boldsymbol{T}^{\mathbf{1112}}_{\mathbf{22,3112}} + \sqrt{\tfrac{5}{21}}\boldsymbol{T}^{\mathbf{1110}}_{\mathbf{22,3312}} + \tfrac{2}{7}\sqrt{\tfrac{5}{3}}\boldsymbol{T}^{\mathbf{1112}}_{\mathbf{22,3312}} + \tfrac{4}{21}\sqrt{5}\boldsymbol{T}^{\mathbf{1112}}_{\mathbf{22,3314}} \\ & - \tfrac{5}{12}T^{2011}_{22,2011} - \tfrac{1}{3}\sqrt{5}T^{2011}_{22,2211} - \tfrac{1}{4}T^{0011}_{22,4011} - \tfrac{1}{7}\sqrt{5}T^{0011}_{22,4211} - \tfrac{3}{14}\sqrt{\tfrac{15}{7}}T^{0011}_{22,4213} \\ & - \tfrac{1}{21}\sqrt{5}T^{0011}_{22,4413} - \tfrac{1}{3}T^{2211}_{22,2211} - \tfrac{1}{2}\sqrt{\tfrac{15}{7}}T^{2011}_{22,2213} - \sqrt{\tfrac{3}{7}}T^{2211}_{22,2213} - \tfrac{1}{\sqrt{14}}T^{2213}_{22,2213}, \end{aligned} \quad \text{(C169)}$$

$$G^{1111}_{42,1111} = \boldsymbol{T}^{\mathbf{1111}}_{\mathbf{42,1111}} + 2\sqrt{3}\boldsymbol{T}^{\mathbf{1111}}_{\mathbf{42,1112}} - \sqrt{\tfrac{3}{7}}\boldsymbol{T}^{\mathbf{1112}}_{\mathbf{42,1112}} + \tfrac{4}{\sqrt{3}}T^{0011}_{42,2011} - \tfrac{4}{\sqrt{15}}T^{0011}_{42,2211} - \tfrac{6}{\sqrt{35}}T^{0011}_{42,2213}, \quad \text{(C170)}$$

$$\begin{aligned} G^{1111}_{22,3112} = {} & \tfrac{1}{\sqrt{3}}\boldsymbol{T}^{\mathbf{1111}}_{\mathbf{22,3111}} - \boldsymbol{T}^{\mathbf{1112}}_{\mathbf{22,3111}} + \boldsymbol{T}^{\mathbf{1111}}_{\mathbf{22,3112}} - \tfrac{1}{\sqrt{7}}\boldsymbol{T}^{\mathbf{1112}}_{\mathbf{22,3112}} + \tfrac{2}{3}\sqrt{\tfrac{5}{21}}\boldsymbol{T}^{\mathbf{1110}}_{\mathbf{22,3312}} + \tfrac{4}{21}\sqrt{\tfrac{5}{3}}\boldsymbol{T}^{\mathbf{1112}}_{\mathbf{22,3312}} \\ & + \tfrac{5}{9}T^{2011}_{22,2011} + \tfrac{8}{63}\sqrt{5}\boldsymbol{T}^{\mathbf{1112}}_{\mathbf{22,3314}} + \tfrac{1}{3}T^{0011}_{22,4011} - \tfrac{2}{21}\sqrt{5}T^{0011}_{22,4211} - \tfrac{1}{7}\sqrt{\tfrac{15}{7}}T^{0011}_{22,4213} \\ & - \tfrac{2}{63}\sqrt{5}T^{0011}_{22,4413} - \tfrac{2}{9}\sqrt{5}T^{2011}_{22,2211} - \tfrac{17}{90}T^{2211}_{22,2211} + \tfrac{1}{\sqrt{15}}T^{2211}_{22,2212} - \tfrac{1}{6}\sqrt{\tfrac{7}{3}}T^{2212}_{22,2212} \\ & - \sqrt{\tfrac{5}{21}}T^{2011}_{22,2213} + \tfrac{4}{5\sqrt{21}}T^{2211}_{22,2213} + \tfrac{2}{15}\sqrt{\tfrac{2}{7}}T^{2213}_{22,2213} - \tfrac{2}{3}\sqrt{\tfrac{14}{15}}T^{2212}_{22,2213}, \end{aligned} \quad \text{(C171)}$$

$$\begin{aligned} G^{1111}_{22,3313} = {} & -2\sqrt{2}\boldsymbol{T}^{\mathbf{1110}}_{\mathbf{22,3312}} - 2\sqrt{2}\boldsymbol{T}^{\mathbf{1111}}_{\mathbf{22,3312}} - 2\sqrt{\tfrac{2}{7}}\boldsymbol{T}^{\mathbf{1112}}_{\mathbf{22,3312}} + \boldsymbol{T}^{\mathbf{1111}}_{\mathbf{22,3313}} + \tfrac{1}{\sqrt{2}}\boldsymbol{T}^{\mathbf{1112}}_{\mathbf{22,3313}} \\ & - \tfrac{1}{\sqrt{42}}\boldsymbol{T}^{\mathbf{1112}}_{\mathbf{22,3314}} + \tfrac{9}{5}\sqrt{\tfrac{21}{10}}T^{2211}_{22,2211} - \tfrac{3}{5}\sqrt{14}T^{2211}_{22,2212} - \tfrac{3}{\sqrt{10}}T^{2212}_{22,2212} + \tfrac{3}{5}\sqrt{\tfrac{2}{5}}T^{2211}_{22,2213} \\ & + \tfrac{3}{5}T^{2212}_{22,2213} + \sqrt{\tfrac{7}{6}}T^{0011}_{22,4413} + \tfrac{1}{5}\sqrt{\tfrac{3}{5}}T^{2213}_{22,2213}, \end{aligned} \quad \text{(C172)}$$

$$G^{1112}_{40,1112} = \tfrac{1}{\sqrt{5}}\boldsymbol{T}^{\mathbf{1110}}_{\mathbf{40,1110}} + \sqrt{\tfrac{3}{5}}\boldsymbol{T}^{\mathbf{1111}}_{\mathbf{40,1111}} + \boldsymbol{T}^{\mathbf{1112}}_{\mathbf{40,1112}} - \tfrac{3}{\sqrt{5}}T^{0011}_{40,2011}, \quad \text{(C173)}$$

$$G^{1101}_{40,1101} = T^{1101}_{40,1101} - \boldsymbol{T}^{\mathbf{0000}}_{\mathbf{40,2000}}, \quad \text{(C174)}$$

$$G^{1101}_{42,1101} = T^{1101}_{42,1101} - \boldsymbol{T}^{\mathbf{0000}}_{\mathbf{42,2202}}, \quad \text{(C175)}$$

$$G^{1101}_{22,3303} = -\tfrac{1}{2}\sqrt{\tfrac{35}{3}}\boldsymbol{T}^{\mathbf{2000}}_{\mathbf{22,2202}} - \tfrac{1}{2}\sqrt{\tfrac{5}{3}}\boldsymbol{T}^{\mathbf{2202}}_{\mathbf{22,2202}} + \sqrt{\tfrac{21}{5}}T^{1101}_{22,3101} + T^{1101}_{22,3303} - \tfrac{1}{2}\sqrt{\tfrac{15}{7}}\boldsymbol{T}^{\mathbf{0000}}_{\mathbf{22,4202}}, \quad \text{(C176)}$$

$$G^{1101}_{51,0011} = T^{1101}_{51,0011} + \boldsymbol{T}^{\mathbf{0000}}_{\mathbf{51,1111}}, \quad \text{(C177)}$$

$$\begin{aligned} G^{2202}_{31,1112} = {} & -\sqrt{\tfrac{5}{3}}\boldsymbol{T}^{\mathbf{2000}}_{\mathbf{31,1111}} + \tfrac{1}{\sqrt{3}}\boldsymbol{T}^{\mathbf{2202}}_{\mathbf{31,1111}} + \boldsymbol{T}^{\mathbf{2202}}_{\mathbf{31,1112}} - \sqrt{\tfrac{5}{3}}T^{1101}_{31,2011} + \tfrac{1}{\sqrt{3}}T^{1101}_{31,2211} - \sqrt{\tfrac{3}{5}}\boldsymbol{T}^{\mathbf{0000}}_{\mathbf{31,3111}} \\ & + T^{1101}_{31,2212} - \sqrt{\tfrac{3}{5}}T^{3101}_{31,0011}, \end{aligned} \quad \text{(C178)}$$

$$G^{1101}_{33,2213} = T^{3303}_{33,0011} + \sqrt{6}\boldsymbol{T}^{\mathbf{2202}}_{\mathbf{33,1111}} - \sqrt{3}\boldsymbol{T}^{\mathbf{2202}}_{\mathbf{33,1112}} - 2\sqrt{2}T^{1101}_{33,2212} + T^{1101}_{33,2213} + \boldsymbol{T}^{\mathbf{0000}}_{\mathbf{33,3313}}, \quad \text{(C179)}$$

$$\begin{aligned} G^{3101}_{11,2212} &= -\sqrt{\tfrac{5}{3}}\boldsymbol{T}^{\mathbf{2000}}_{\mathbf{11,3111}} + \tfrac{1}{\sqrt{3}}\boldsymbol{T}^{\mathbf{2202}}_{\mathbf{11,3111}} + \boldsymbol{T}^{\mathbf{2202}}_{\mathbf{11,3112}} - \tfrac{2}{3}\sqrt{\tfrac{5}{21}}\boldsymbol{T}^{\mathbf{2202}}_{\mathbf{11,3312}} - \tfrac{2}{3}\sqrt{\tfrac{10}{21}}\boldsymbol{T}^{\mathbf{2202}}_{\mathbf{11,3313}} \\ &\quad -\tfrac{1}{14}\sqrt{15}\boldsymbol{T}^{\mathbf{0000}}_{\mathbf{11,5111}} - \tfrac{1}{2}\sqrt{\tfrac{5}{3}}\boldsymbol{T}^{\mathbf{4000}}_{\mathbf{11,1111}} + \tfrac{5}{7\sqrt{3}}\boldsymbol{T}^{\mathbf{4202}}_{\mathbf{11,1111}} + \tfrac{5}{7}\boldsymbol{T}^{\mathbf{4202}}_{\mathbf{11,1112}} - \tfrac{1}{2}\sqrt{\tfrac{5}{3}}T^{1101}_{11,4011} \\ &\quad +\tfrac{5}{7\sqrt{3}}T^{1101}_{11,4211} + \tfrac{5}{7}T^{1101}_{11,4212} - \tfrac{1}{14}\sqrt{15}T^{5101}_{11,0011} - \sqrt{\tfrac{5}{3}}T^{3101}_{11,2011} + \tfrac{1}{\sqrt{3}}T^{3101}_{11,2211} \\ &\quad +T^{3101}_{11,2212} - \tfrac{2}{3}\sqrt{\tfrac{5}{21}}T^{3303}_{11,2212} - \tfrac{2}{3}\sqrt{\tfrac{10}{21}}T^{3303}_{11,2213}, \end{aligned} \tag{C180}$$

$$G^{0011}_{40,2211} = -\tfrac{3}{2\sqrt{5}}\boldsymbol{T}^{\mathbf{1110}}_{\mathbf{40,1110}} + \sqrt{\tfrac{3}{5}}\boldsymbol{T}^{\mathbf{1111}}_{\mathbf{40,1111}} - \tfrac{1}{2\sqrt{5}}T^{0011}_{40,2011} + T^{0011}_{40,2211}, \tag{C181}$$

$$G^{0011}_{42,2212} = 2\sqrt{3}\boldsymbol{T}^{\mathbf{1111}}_{\mathbf{42,1112}} + \sqrt{3}T^{0011}_{42,2011} - \sqrt{\tfrac{3}{5}}T^{0011}_{42,2211} + T^{0011}_{42,2212} - \sqrt{\tfrac{7}{5}}T^{0011}_{42,2213}, \tag{C182}$$

$$G^{0011}_{44,2213} = T^{0011}_{44,2213} - \boldsymbol{T}^{\mathbf{1112}}_{\mathbf{44,1112}}, \tag{C183}$$

$$\begin{aligned} G^{2011}_{20,2011} &= T^{2011}_{20,2011} + \tfrac{2}{5}T^{2211}_{20,2211} - \tfrac{4}{5}\boldsymbol{T}^{\mathbf{1110}}_{\mathbf{20,3110}} - \tfrac{4}{5}\sqrt{3}\boldsymbol{T}^{\mathbf{1111}}_{\mathbf{20,3111}} - \tfrac{4}{\sqrt{5}}\boldsymbol{T}^{\mathbf{1112}}_{\mathbf{20,3112}} + \tfrac{3}{5}T^{0011}_{20,4011} \\ &\quad +\tfrac{2}{\sqrt{15}}T^{2212}_{20,2212} + \tfrac{2}{5}\sqrt{\tfrac{7}{3}}T^{2213}_{20,2213}, \end{aligned} \tag{C184}$$

$$\begin{aligned} G^{2011}_{20,2211} &= T^{2011}_{20,2211} - \tfrac{2}{\sqrt{5}}\boldsymbol{T}^{\mathbf{1110}}_{\mathbf{20,3110}} + \sqrt{\tfrac{3}{5}}\boldsymbol{T}^{\mathbf{1111}}_{\mathbf{20,3111}} - \tfrac{1}{5}\boldsymbol{T}^{\mathbf{1112}}_{\mathbf{20,3112}} - 2\sqrt{\tfrac{3}{35}}\boldsymbol{T}^{\mathbf{1112}}_{\mathbf{20,3312}} + \tfrac{3}{7}T^{0011}_{20,4211} \\ &\quad +\tfrac{1}{2\sqrt{5}}T^{2211}_{20,2211} - \tfrac{1}{2\sqrt{3}}T^{2212}_{20,2212} + \tfrac{1}{\sqrt{105}}T^{2213}_{20,2213}, \end{aligned} \tag{C185}$$

$$\begin{aligned} G^{2011}_{22,2212} &= -\tfrac{6}{5}\boldsymbol{T}^{\mathbf{1111}}_{\mathbf{22,3111}} + \tfrac{6}{5}\sqrt{\tfrac{3}{7}}\boldsymbol{T}^{\mathbf{1112}}_{\mathbf{22,3112}} + \tfrac{6}{\sqrt{35}}\boldsymbol{T}^{\mathbf{1110}}_{\mathbf{22,3312}} + \tfrac{6}{\sqrt{35}}\boldsymbol{T}^{\mathbf{1111}}_{\mathbf{22,3312}} + \tfrac{2}{7\sqrt{5}}\boldsymbol{T}^{\mathbf{1112}}_{\mathbf{22,3312}} \\ &\quad -\tfrac{4}{\sqrt{35}}\boldsymbol{T}^{\mathbf{1112}}_{\mathbf{22,3313}} + \tfrac{4}{7\sqrt{15}}\boldsymbol{T}^{\mathbf{1112}}_{\mathbf{22,3314}} - \tfrac{1}{10}\sqrt{3}T^{0011}_{22,4011} + \tfrac{1}{7}\sqrt{\tfrac{3}{5}}T^{0011}_{22,4211} + \tfrac{3}{7}T^{0011}_{22,4212} \\ &\quad -\tfrac{3}{7\sqrt{35}}T^{0011}_{22,4213} - \tfrac{2}{7}\sqrt{\tfrac{5}{3}}T^{0011}_{22,4413} - \tfrac{1}{2\sqrt{3}}T^{2011}_{22,2011} + \tfrac{1}{\sqrt{15}}T^{2011}_{22,2211} - \tfrac{43}{25\sqrt{3}}T^{2211}_{22,2211} \\ &\quad +T^{2011}_{22,2212} + \tfrac{4}{5\sqrt{5}}T^{2211}_{22,2212} - \tfrac{1}{\sqrt{35}}T^{2011}_{22,2213} + \tfrac{1}{5}\sqrt{7}T^{2212}_{22,2212} - \tfrac{4}{25\sqrt{7}}T^{2211}_{22,2213} \\ &\quad -\tfrac{2}{5}\sqrt{\tfrac{2}{35}}T^{2212}_{22,2213} - \tfrac{9}{25}\sqrt{\tfrac{6}{7}}T^{2213}_{22,2213}. \end{aligned} \tag{C186}$$